\documentclass[prl,amsfonts,twocolumn,nofootinbib,showpacs,floatfix]{revtex4}
\usepackage{graphicx,epsfig}
\usepackage[usenames]{color}
\newcommand{\be}{\begin{equation}}
\newcommand{\ee}{\end{equation}}
\newcommand{\bea}{\begin{eqnarray}}
\newcommand{\eea}{\end{eqnarray}}

\newcommand{\gapp}{\mathrel{\raise.3ex\hbox{$>$}\mkern-14mu 
\lower0.6ex\hbox{$\sim$}}}
\newcommand{\lapp}{\mathrel{\raise.3ex\hbox{$<$}\mkern-14mu 
\lower0.6ex\hbox{$\sim$}}}
\newcommand{\maxa}{\mathrel{\raise.1ex\hbox{${\rm max}$}\mkern-14mu 
\lower0.7ex\hbox{$a$}}}
\newcommand{\mina}{\mathrel{\raise.1ex\hbox{${\rm min}$}\mkern-14mu 
\lower0.7ex\hbox{$a$}}}
\def\bbox{{\,\lower0.9pt\vbox{\hrule \hbox{\vrule height 0.2 cm
\hskip 0.2 cm \vrule height 0.2 cm}\hrule}\,}}
\newcommand{\etc}{{\it etc.}}
\newcommand{\ECM}{E_{\rm CM}}
\begin{document}

\title{BlackMax: A black-hole event generator with rotation, recoil, split branes and brane tension.}
\author{De-Chang Dai$^1$, Glenn Starkman$^1$, Dejan Stojkovic$^2$, Cigdem 
Issever$^3$, Eram Rizvi$^4$, Jeff Tseng$^3$} 

\affiliation{$^1$Case Western Reserve University, Cleveland OH 44106-7079, USA}
\affiliation{$^2$Department of Physics, SUNY at Buffalo, Buffalo NY 14260-1500, 
USA}
\affiliation{$^3$University of Oxford, Oxford,  UK} 
\affiliation{$^4$Queen Mary, University of London, London, UK}

\begin{abstract}

\widetext
We present a comprehensive black-hole event generator, 
BlackMax, which simulates the experimental signatures of microscopic and 
Planckian black-hole production and evolution at the LHC 
in the context of brane world models with low-scale quantum gravity. 
The generator is based on phenomenologically realistic models free of 
serious problems that plague low-scale gravity, thus offering 
more realistic predictions for hadron-hadron colliders. 
The generator includes all of the black-hole gray-body factors known to date 
and incorporates the effects of black-hole rotation, splitting between 
the fermions, non-zero brane tension and black-hole recoil due to Hawking radiation
(although not all simultaneously).

The generator can be interfaced with Herwig and Pythia.

The main code can be downloaded from \cite{code}.
\end{abstract}


\pacs{04.50.Gh, 04.70.Dy}

\maketitle

\section{Introduction}
\label{sec:introduction} 
\indent Models with TeV-scale quantum gravity \cite{ADD,RS,CHMADD,UED}
offer very rich collider phenomenology. Most of them assume the existence 
of a three-plus-one-dimensional hypersurface,
which is referred as ``the brane,'' where Standard-Model particles are confined,
while only gravity and possibly other particles that carry no
gauge quantum numbers, such as right handed neutrinos 
can propagate in the full space, the so called ``bulk''.
Under certain assumptions, this setup allows the fundamental 
quantum-gravity energy scale, $M_*$, 
to be close to the electroweak scale. 
The observed weakness of gravity compared to other forces 
on the brane ({\it i.e.} in the laboratory) is a consequence 
of the large volume of the bulk which dilutes the strength 
of gravity.

In the context of these models of TeV-scale quantum gravity, 
probably the most exciting new physics
is the production of micro-black-holes in near-future accelerators 
like the Large Hadron Collider (LHC) \cite{BHacc}. 
According to the ``hoop conjecture" \cite{hoop},
if the impact parameter of two colliding particles is less than two times the
gravitational radius, $r_h$, corresponding to their center-of-mass energy 
($\ECM$),
a black-hole with a mass of the order of $\ECM$ and horizon radius, $r_h$, will 
form.
Typically, this gravitational radius is approximately $\ECM/M_*^2$
Thus, when particles collide at center-of-mass energies above $M_*$,
the probability of black-hole formation is high.

Strictly speaking, there exist no complete calculation (including 
radiation during the process of formation and back-reaction) which 
proves that a black-hole really forms. It may happen that a true 
event horizon and singularity never forms, and that Hawking (or rather 
Hawking-like) radiation is never quite thermal. In \cite{Vachaspati:2006ki} this 
question was analyzed in detail from a point of view of an asymptotic 
observer, who is in the context of the LHC the most relevant observer. 
It was shown that though such observers never observe the formation of 
an event horizon even in the full quantum treatment, they do
register pre-Hawking quantum radiation that takes away energy from a 
collapsing system. Pre-Hawking radiation is non-thermal and becomes 
thermal only in the limit when the horizon is formed. Since a collapsing 
system has only a finite amount of energy, it disappears before the 
horizon is seen to be formed.  While these results 
have important implications for theoretical issues like the 
information loss paradox, in a practical sense very little will
change. The characteristic time for gravitational collapse in the 
context of collisions of particles at the LHC is very short. 
This implies that pre-Hawking radiation will be quickly experimentally 
indistinguishable from Hawking radiation calculated for a real black 
hole. Also, calculations in \cite{Vachaspati:2006ki} indicate that 
the characteristic time in which a collapsing system losses all 
of its energy is very similar to a life time of a real black-hole. 
Thus, one may proceed with a standard theory of black-holes.

Once a black-hole is formed, it is believed to decay via Hawking radiation. 
This Hawking radiation will consist of two parts: 
radiation of Standard-Model particles into the brane and radiation 
of gravitons and any other bulk modes into the bulk. 
The relative probability for the emission of each particle type
is given by the gray-body factor for that mode. 
This gray-body factor depends on the properties of the particle 
(charge, spin, mass, momentum), of the black-hole (mass, spin, charge) and, 
in the context of TeV-scale quantum gravity, on environmental properties --
the number of extra dimensions, the location of the black-hole relative
to the brane (or branes), \etc. In order to properly describe
the experimental signatures of black-hole production and decay
one must therefore calculate the gray-body factors 
for all of the relevant degrees of freedom.

There are several black-hole event generators available 
in the literature \cite{othergenerators}
based on particular, simplified models of low-scale quantum gravity
and incorporating limited aspects of micro-black-hole physics.
Unfortunately, low-scale gravity is plagued with many phenomenological
challenges like fast proton decay, large $n{\bar n}$ oscillations, 
flavor-changing neutral currents and large mixing between leptons 
\cite{Dai:2006dz,Stojkovic:2005zq}.
For a realistic understanding of the experimental signature of black
hole production and decay, 
one needs calculations based on phenomenologically viable gravity models,
and incorporating all necessary aspects of the production and
evolution of the black-holes.

One low-scale gravity model in which the above mentioned phenomenological
challenges can be addressed is the split-fermion model \cite{splitfermions}. 
In this model, the Standard Model fields are confined to a 
``thick brane'', much thicker than $M_*^{-1}$. 
Within this thick brane, quarks and leptons are stuck on different 
three-dimensional slices (or on different branes), 
which are separated by much more than $M_*^{-1}$. 
This separation causes an exponential suppression of all
direct couplings between quarks and leptons, 
because of exponentially small overlaps between their wave-functions.
The proton decay rate will be safely suppressed if the spatial separation 
between quarks and leptons
is greater by a factor of at least $10$ than the widths of their wave functions. 
Since $\Delta{B} = 2$ processes, like $n{\bar n}$ oscillations, 
are mediated by operators of the type uddudd, suppressing them 
requires a further splitting between up-type and down-type quarks. 
Since the experimental limits on $\Delta{B}=2$ operators 
are much less stringent than those on $\Delta{B}=1$ operators, 
the u and d-type quarks need only be separated by a few times the width 
of their wave functions \cite{splitfermions}.

Current black-hole generators assume that the black-holes that are formed
are Schwarzschild-like. However, most of the black-holes that
would be formed at the LHC would be highly rotating,
due to the non-zero impact parameter of the colliding partons. 
Due to the existence of an ergosphere 
(a region between the infinite redshift surface and the event horizon), 
a rotating black-hole exhibits super-radiance: 
some modes of radiation get amplified compared to others.
The effect of super-radiance \cite{superradiance} is strongly spin-dependent, 
with emission of higher-spin particles strongly favored. 
In particular the emission of gravitons is enhanced over
lower-spin Standard-Model particles. Since graviton emission
appears in detectors as missing energy,
the effects of black-hole rotation cannot be ignored.
Similarly, black-holes may be formed with non-zero gauge charge,
or acquire charge during their decay. This again may alter
the decay properties of the black-hole and should be included.

Another effect neglected in other generators is the recoil of the
black-hole. A small black-hole attached to a brane in a 
higher-dimensional space emitting quanta into the bulk 
could leave the brane as a result of a recoil\footnote{Although if the 
black-hole carries gauge charge it will be prevented from leaving the brane.}.
In this case, visible black-hole radiation would cease.
Alternately, in a split-brane model, as a black-hole traverses
the thick brane the Standard-Model particles that it is able
to emit will change depending on which fermionic branes are nearby.

It is also the case that virtually all the work in this field has been done for
the idealized case where the brane tension is negligible. 
However, one generically expects the brane tension
to be of the order of the fundamental energy scale,
being determined by the vacuum energy contributions 
of brane-localized matter fields\cite{NKD}. 
As shown in \cite{DKSS}, 
finite brane tension modifies the standard gray-body factors.

Finally, it has been suggested \cite{Meade} that more common than the formation
and evaporation of black-holes will be gravitational
scattering of parton pairs into a two-body final state.
We include this possibility.

Here we present a comprehensive black-hole event generator, BlackMax,
that takes into account practically all of the above mentioned 
issues\footnote{although not necessarily simultaneously}, 
and includes almost all the necessary 
gray-body factors\footnote{Except in the one case of the graviton gray-body 
factor for a rotating black-hole, where the calculation has yet to be 
achieved.}. 
Preliminary studies
show how the signatures of black-hole production and decay change 
when one includes splitting between the fermions, 
black-hole rotation, positive brane tension and black-hole recoil.
Future papers will explore the implications of these changes
in greater detail.

In section \ref{sec:bhproduction} and \ref{sec:graybodyfactors} we discuss the 
production of black-holes and 
the gray-body factors respectively. The 
evaporation process and final burst of the black-holes is discussed in section 
\ref{sec:bhevolution} and \ref{sec:finalburst}. Sections \ref{sec:input_output} and 
describe the input and output of the generator. Section 
\ref{sec:results} shows some characteristic distributions of black-holes for 
different extra dimension scenarios. The reference list is extensive, reflecting 
the great interest in the topic 
\cite{BHacc,hoop,othergenerators,Dai:2006dz,Stojkovic:2005zq,NKD,DKSS,Cardoso-gravit
on,SM-brane,SM-brane-1,kant-ro-s,kant-ro-s-1,kant-ro-b,kant-ro-fer,Ida1,Creek:20
07sy,Meade,Stojkovic:2004hp,Anchordoqui,recoil,flux,ida,1,2,3,4,5,6,7,9,10,11,12
,13,14,15,16,17,18,19,20,21,22,23,24,25,26,27,28,29,30,31,32,33,34,35,36,37,38,3
9,40,41,42,43,44,45,46,47,48,49,50,51,52,53,54,55,56,57,58,59,60,61,62,63,64,65,
66,67,68,69,70,71,72,73,74,75,76}, but by no means 
complete.

\section{black-hole production}
\label{sec:bhproduction}
\indent
We assume that the fundamental quantum-gravity energy scale $M_*$ 
is not too far above the electroweak scale. 
Consider two particles colliding with a center-of-mass energy $E_{CM}$. 
They will also have an angular momentum $J$ in their center-of-mass (CM) frame.
By the hoop conjecture, if the impact parameter, $b$, 
between the two colliding particles is smaller than 
the diameter of the horizon of a $(d+1)$-dimensional black-hole 
(where $d$ is the total number of space-like dimensions) of mass $M=E_{CM}$ and 
angular momentum $J$,
\be
b < 2 r_h(d,M,J) ,
\ee
then a black-hole with $r_{h}$ will form.
The cross section for this process is approximately equal to the
interaction area $\pi (2r_h)^2$.

In Boyer-Lindquist coordinates,
the metric for a $(d+1)$-dimensional rotating black-hole
(with angular momentum parallel to the $\hat{\omega}$ in the rest frame of the 
black-hole) is:
\begin{eqnarray}
\label{eqn:bhmetric}
ds^2 &=&\left( 1-\frac{\mu r^{4-d}}{\Sigma (r,\theta)}\right)dt^{2}
\nonumber \\
&-& \sin^{2}\theta \left(r^{2}+a^{2}\left(
+ \sin^{2}\theta \frac{\mu r^{4-d}}{\Sigma (r,\theta)}\right)\right)
d\phi ^{2} \nonumber \\
&+& 2 a \sin^{2}\theta \frac{\mu r^{4-d}}{\Sigma (r,\theta )} dt d\phi
- \frac{\Sigma(r,\theta)}{\Delta} dr^{2} \nonumber \\
&-& \Sigma(r,\theta) d\theta ^{2} - r^{2}cos^2\theta
d^{d-3}\Omega
\end{eqnarray}
where $\mu$ is a parameter related to mass of the black-hole, while 
\begin{eqnarray}
&&\Sigma =r^2+a^{2}cos^{2}\theta \\
{\rm and}&& \nonumber \\
&& \Delta=r^2+a^{2}-\mu r^{4-d} .
\end{eqnarray}
The mass of the black-hole is
\begin{equation}
M=\frac{(d-1) A_{d-1}}{16\pi G_d}\mu ,
\end{equation}
and
\begin{equation}
J=\frac{2Ma}{d-1}
\end{equation}
is its angular momentum. Here,
\be
A_{d-1} = \frac{2 \pi^{d/2}}{\Gamma(d/2)}
\ee
is the hyper-surface area of a $(d-1)$-dimensional unit sphere.
The higher-dimensional gravitational constant $G_d$ is defined as
\begin {equation}
G_{d}=\frac{\pi^{d-4}}{4 M_{\star}^{d-1}} .
\end{equation}

The horizon occurs when $\Delta =0$. That is at a radius given
implicitly by
\begin{equation}
\label{eqn:horizon}
r^{(d)}_h=\left[\frac{\mu}{1+(a/r^{(d)}_h)^{2}}\right]^{\frac{1}{d-2}}
=\frac{r^{(d)}_s}{\left[1+(a/r^{(d)}_h)^{2}\right]^{\frac{1}{d-2}}}.
\end{equation}
Here
\be
\label{eqn:defrSchwarzshild}
r^{(d)}_s\equiv\mu ^{1/(d-2)}
\ee
is the Schwarzschild radius of a $(d+1)$-dimensional black-hole,
{\it i.e.} the horizon radius of a non-rotating black-hole.
Equation \ref{eqn:defrSchwarzshild} can be rewritten as:
\begin{equation}
\label{eqn:horizon2}
r^{(d)}_s(\ECM,d,M_*)=k(d)M_{*}^{-1}[\ECM/M_{*}]^{1/(d-2)},
\end{equation}
where
\begin{equation}
\label{eqn:horizon3}
k(d)\equiv\left[2^{d-3}{\pi}^{(d-6)/2}\frac{\Gamma[d/2]}{d-1}\right]^{1/(d-2)} .
\end{equation}

\begin{figure}
\centering{
\includegraphics[width=3.2in]{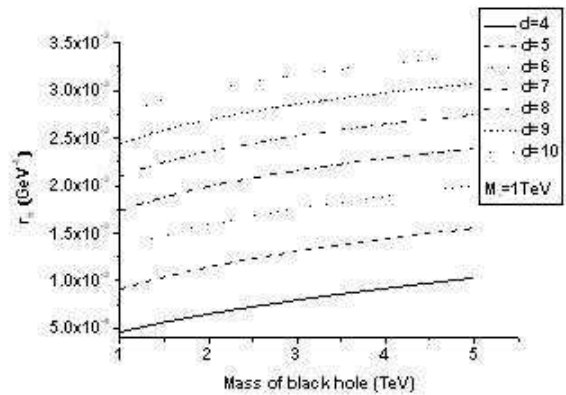} }
\caption{Horizon radius (in GeV$^{-1}$) of a non-rotating black-hole as a 
function of mass 
for 4-10 spatial dimensions.}
\label{fig:rhorizon}
\end{figure}

\begin{figure}
\centering{
\includegraphics[width=3.2in]{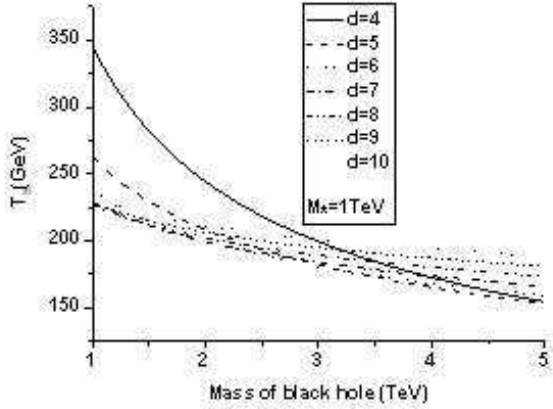} }
\caption{Hawking temperature(in GeV) of a non-rotating black-hole as a function 
of mass for 4-10 spatial dimensions.}
\label{fig:THawk}
\end{figure}

Figure \ref{fig:rhorizon} shows the horizon radius as a function of black-hole 
mass for $d$ from $4$ to $10$. We see that the
horizon radius increases with mass; it also increases with $d$.
Figure \ref{fig:THawk} shows the Hawking temperature of a black-hole
\be
\label{eqn:THawking}
T_{H} = \frac{d-2}{4\pi r_h}
\ee

as a function of the black-hole mass for $d$ from $4$ to $10$. 
The Hawking temperature is a measure of the characteristic
energies of the particles emitted by the black-hole.
$T_H$ decreases with increasing mass.
However, the behaviour of $T_H$ with changing $d$ is
complicated, reflecting the competing effect of an increasing
horizon radius and an increasing $d-2$ in equation \ref{eqn:THawking}.


For the model with with non-zero tension brane, the radius of the black-hole is 
defined as 
\begin{equation}
\label{eqn:r-tension}
r_{h}^{(t)}=\frac{r_{s}}{B^{1/3}},
\end{equation}
with $B$ the deficit-angle parameter which is inverse proportional to the 
tension of the brane.

\begin{figure}
\centering{
\includegraphics[width=3.2in]{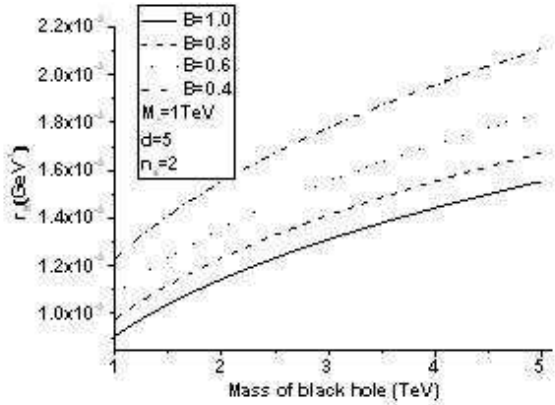} }
\caption{Horizon radius (in GeV$^{-1}$) of a black-hole as a function of mass 
for different B in d=5 spatial dimensions. }
\label{fig:rhorizon-t}
\end{figure}

Figure \ref{fig:rhorizon-t} shows the horizon radius as a function of black-hole 
mass 
for the model with non-zero tension brane. As the deficit-angle parameter 
increases, 
the size of the black-hole increases. 

\begin{figure}
\centering{
\includegraphics[width=3.2in]{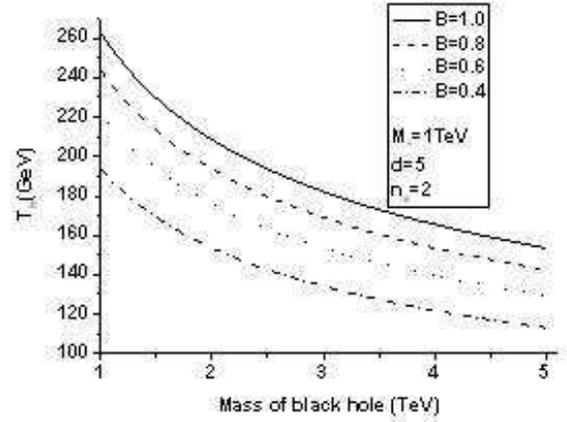} }
\caption{Hawking temperature(in GeV) of a black-hole as a function of mass for 
different B in d=5 spatial dimensions.}
\label{fig:THawk-t}
\end{figure}

Figure \ref{fig:THawk-t} shows the Hawking temperature of a black-hole for the 
model with non-zero tension brane. 
The Hawking temperature decreases as the deficit angle decreases.

\begin{figure}
\centering{
\includegraphics[width=3in]{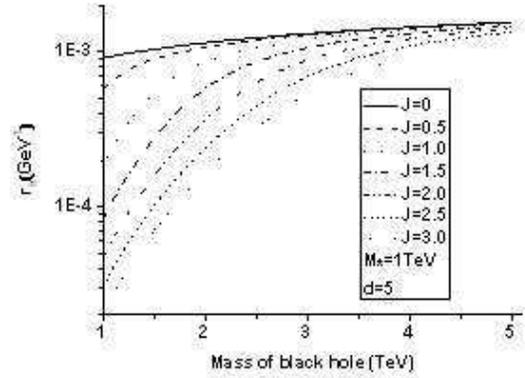} }
\caption{Horizon radius (in GeV$^{-1}$) of a rotating black-hole as a function 
of mass 
for different angular momentum in d=5 spatial dimensions. Angular momentum J is 
in unit of $\hbar$.}
\label{fig:rhorizon-an}
\end{figure}

\begin{figure}
\centering{
\includegraphics[width=3in]{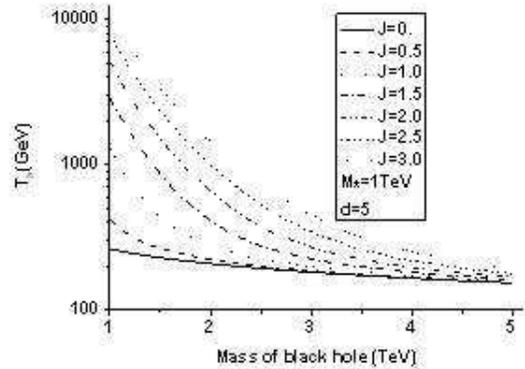} }
\caption{Hawking temperature(in GeV) of a rotating black-hole as a function 
of mass for different angular momentum in d=5 spatial dimensions. Angular 
momentum J is in unit of $\hbar$.}
\label{fig:THawk-an}
\end{figure}

Figure \ref{fig:rhorizon-an} shows the horizon radius as a function of 
black-hole mass 
for a rotating black-hole. The angular momentum decreases the size of the 
horizon 
and increases the Hawking temperature (see figures \ref{fig:rhorizon-an} and 
\ref{fig:THawk-an}).

If two highly relativistic particles collide with center-of-mass energy
$E_{CM}$, and impact parameter $b$, then their angular momentum
in the center-of-mass frame before the collision is $L_{in}=b E_{CM}/2$. 
Suppose for now that the black-hole that is formed 
retains all this energy and angular momentum.
Then the mass and angular momentum of the black-hole will be
$M_{in}=\ECM$ and $J_{in}=L_{in}$.
A black-hole will form if:
\begin{equation}
\label{bmax1}
b < b_{\rm max} \equiv 2r^{(d)}_h (E_{CM}, b_{max}E_{CM}/2) \, .
\end{equation}
We see that $b_{max}$ is a function of both $E_{CM}$ and the number of
extra dimensions.

We can rewrite condition (\ref{bmax1}) as
\begin{equation}
\label{eqn:bmaxfinalform}
b_{max}(E_{CM};d)=
2\frac{r^{(d)}_s(E_{CM})}{\left[1+\left(\frac{d-1}{2}\right)^{2}
\right]^{1 \over {d-2}}} \, .
\label{bmax2}
\end{equation}

There is one exception to this condition. In the case 
where we are including the effects of the brane tension,
the metric (and hence gray-body factors) for a rotating
black-hole are not known. In this case we
consider only non-rotating black-holes. Therefore,
for branes with tension 
\be
b^{tension}_{max}(E_{CM},d) = 2 r^{(d)}_s(\ECM) .
\ee
Also, for branes with tension only the $d=5$ metric is known.

At the LHC, each proton will have $E=7$\,TeV in the CM frame.
Therefore, the total proton-proton center of mass energy
will be $\sqrt{s} = 14$\,TeV. However, it is not the protons
that collide to make the black-holes, but the partons of which
the protons are made. If two partons have energy $vE$ and $\frac{uE}{v}$,
much greater than their respective masses, then the parton-parton
collision will have
\begin{equation}
s'=|p_{i}+p_{j}|^{2}=|v(E,E)+\frac{u}{v} (E,-E)|^{2}=4uE^{2}=us \, .
\end{equation}
We define a quantity $Q'$
\begin{equation}
\label{QandQ'}
Q'= \ECM = \sqrt{s'}=\sqrt{us}
\end{equation}
The center-of-mass energy for the two colliding partons will be $\sqrt{us}$,
as will be the 4-momentum transfer $Q'^{2}$.
The largest impact parameter between the two partons that can form a
black-hole with this mass will therefore be 
$b_{max}(\sqrt{us};d)$, as given by equation \ref{eqn:bmaxfinalform}.

The total proton-proton cross section for black-hole production is
therefore
\begin{eqnarray} \label{partcross}
&&\sigma^{pp\to BH}(s;d,M_\star) =
\int_{M_\star^2/s}^{1}\!\!\!\!du \int_{u}^{1}\frac{dv}{v}
\pi \left[b_{max}(\sqrt{us};d)\right]^2 \nonumber \\
&&\quad\quad \times \sum_{ij}
f_{i}(v,Q')f_{j}(u/v,Q') \, .
\end{eqnarray}
Here $f_i(v,Q')$ is the i-th parton distribution function.
Loosely this is the expected number of partons of type i and momentum
$vE$ to be found in the proton in a collision at momentum transfer $Q'$.

In \cite{Meade} it is argued that strong gravity effects at energies close to 
the Planck scale will lead to an increase in the $2 \rightarrow 2$ cross 
section via the exchange of Planckian ``black-holes'' (by which 
any quantum gravity effect or resonance is meant).  
Final states with high multiplicities are predicted to be suppressed.
 Although the intermediate state is created in 
 the strong gravity regime,
it is not a conventional microscopic black hole. 
The state is not stable.  Thermal Hawking radiation does not take place. 
Especially since inelastic 
collisions increase the energy loss, the threshold for creating stable black-holes 
shifts to even higher values. Thus $2\rightarrow 2$ scattering may be the most 
important signal in the LHC instead of black-holes evaporating via Hawking 
radiation. One should find that the cross section for two-body final states 
suddenly jumps to a larger value, as the energy reaches the 
quantum-gravity scale. We calculate the cross section of two-body final states
by replacing $\pi b_{max}^{2}$ in equation (\ref{partcross}) with 
\begin{eqnarray}
\pi b_{max}(\sqrt{s'}>M_{min})^{2}& \approx & \pi r_{s}^{2}P_{2}
\end{eqnarray}
where
\begin{eqnarray}
P_{2}& = &e^{-<N>}\sum^{2}_{i=0}\frac{<N>^{i}}{i!}\\
<N>& = &\rho (\frac{4\pi k(d)}{d-1}\frac{M_{BH}}{M_{*}})^{(d-1)/(d-2)}\\
\rho &=&\frac{\sum c_{i}g_{i}\Gamma_{i}\zeta (3)\Gamma (3)}{\sum 
c_{i}f_{i}\Phi_{i}\zeta (4)\Gamma (4)}\\
\Gamma_{i}&=&\frac{1}{4\pi r^{2}}\int 
\frac{\sigma_{i}(\omega)\omega^{2}d\omega}{e^{\omega/T}\pm 1}[\int 
\frac{\omega^{2}d\omega}{e^{\omega/T}\pm 1}]^{-1}\\
\Phi_{i}&=&\frac{1}{4\pi r^{2}}\int 
\frac{\sigma_{i}(\omega)\omega^{3}d\omega}{e^{\omega/T}\pm 1}[\int 
\frac{\omega^{3}d\omega}{e^{\omega/T}\pm 1}]^{-1}
\end{eqnarray}

Here $c_{i}$ is the number of internal degrees of freedom of particle species 
$i$, $g_{i}=1$ and $f_{i}=1$ for bosons, and $g_{i}=3/4$ and $f_{i}=7/8$ for 
fermions \cite{Anchordoqui}. 

\begin{figure}
\centering{
\includegraphics[width=3in]{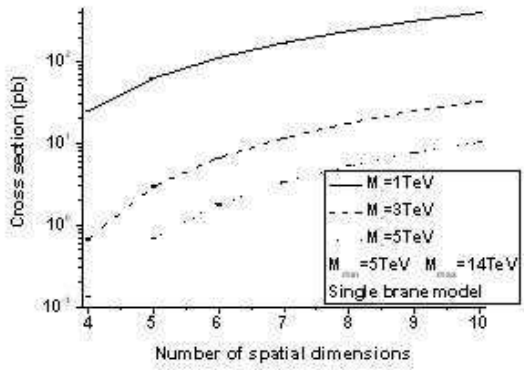} }
\caption{Cross section for production of a black-hole (rotating or non-rotating)
as a function of the number of spatial dimensions,
for a tensionless brane, with no fermion brane-splitting.}
\label{fig:cross-d}
\end{figure}

Figure \ref{fig:cross-d} shows the cross section for non-rotating black-holes on 
a tensionless brane as a function of mass for different numbers of spatial 
dimensions. 
The cross section increases with the number of spatial dimensions.

\begin{figure}
\centering{
\includegraphics[width=3in]{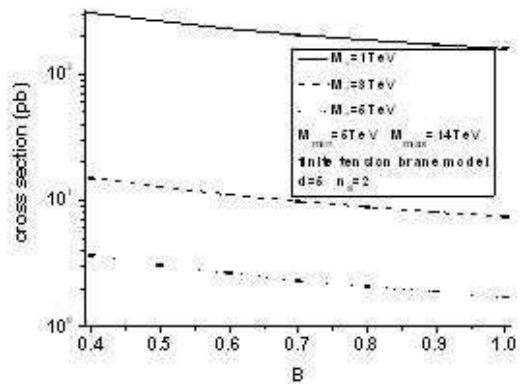} }
\caption{Cross section for production of a non-rotating black-hole 
as a function of the deficit angle parameter for $d=5$ and $n_s = 2$.}
\label{fig:cross-tension}
\end{figure}

Figure \ref{fig:cross-tension} shows the cross section for non-rotating black 
holes 
on a brane with positive tension as a function of mass for various deficit angle 
parameter, $B$. The cross section increases as the tension increases (as $B$ 
decreases).

\begin{figure}
\centering{
\includegraphics[width=3in]{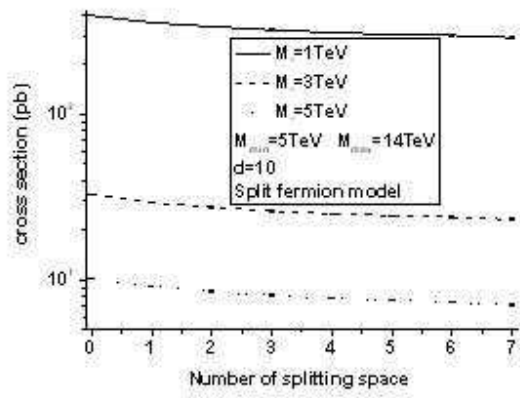} }
\caption{Cross section for production of a non-rotating black-hole 
as a function of the number of fermion brane-splitting dimensions for $d=10$.}
\label{fig:cross-split}
\end{figure}

Figure \ref{fig:cross-split} shows the cross section for non-rotating black 
holes as function of the number of split-fermion space dimensions, $n_{s}$. 
When a pair of partons are separated in the extra-dimensions they must approach 
more
closely in the ordinary dimensions in order to form a black-hole. Thus
the effective cross-section for black-hole formation in collisions is decreased.
This effect become more severe as $n_s$ increases because the partons
are more likely to be more widely separated in the extra dimensions, 
therefore the cross section decreases with increasing $n_{s}$.

\begin{figure}
\centering{
\includegraphics[width=3in]{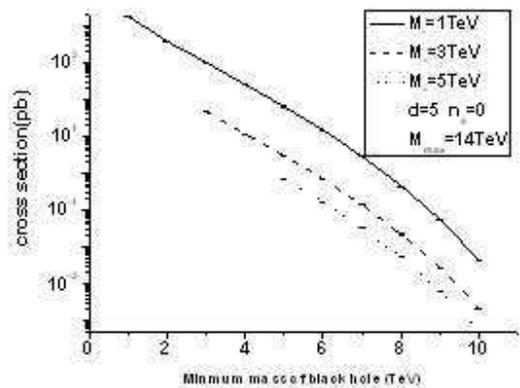} }
\caption{Cross section for formation of a black-hole (rotating or non-rotating)
as function of the minimum mass of black-hole,
for a zero-tension brane, with no fermion brane-splitting. 
The vertical lines are the error bars. }
\label{fig:cross-mass}
\end{figure}

Figure \ref{fig:cross-mass} shows the cross section as a function of the chosen 
minimum black-hole mass. The parton distribution functions strongly suppress the 
events with high black-hole masses.


\begin{figure}[ht]
\centering
\includegraphics[width=3.2in]{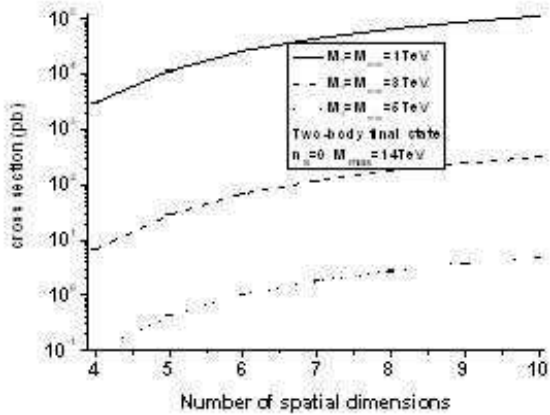}
\caption{Cross section for the two-body final-state scenario as a 
function of number of spatial dimensions where $M_{min}=M_{*}=1\,\rm{TeV}$, $M_{min}=M_{*}=3\,\rm{TeV}$ and 
$M_{min}=5\,\rm{TeV}$.}
\label{fig:cross-lisa}
\end{figure}

Figure \ref{fig:cross-lisa} shows the cross section for the two-body final-state 
scenario as a function of the number of spatial dimensions, for 
$M_{min}=M_{*}=1\,\rm{TeV}$, $M_{min}=M_{*}=3\,\rm{TeV}$ and $M_{min}=M_{*}=5\,\rm{TeV}$. It increases with 
the number of spatial dimensions. 

\subsection{Black-Hole Formation in BlackMax}
\label{subsec:formation}

Within BlackMax, the probability of creating a black-hole of center-of-mass
energy $\sqrt{us}$, in the collision of two protons of center-of-mass-energy 
$\sqrt{s}$,
is given by
\begin{eqnarray} \label{prob}
&&P(Q') =
\int_{u}^{1}\frac{dv}{v}
\sum_{ij} f_{i}(v,Q')f_{j}(\frac{u}{v},Q') \, .
\end{eqnarray}

According to the theory, there will be some minimum mass for a black-hole.
We expect $M_{min} \sim M_*$, but leave $M_{min}\geq M_*$ as a free parameter.
Therefore, a black-hole will only form if $u>(M_{min}/Q)^{2}$. 
The type of partons from which a black-hole is formed 
determines the gauge charges of the black-hole. 
Clearly, the probability to create a black-hole in the collision of any 
two particular partons $i$ and $j$ with energies (momenta) 
$vE$ and $\frac{uE}{v}$, is given by
\begin{equation}
P(vE,\frac{uE}{v},i,j)=f_{i}(v,Q')f_{j}(u/v,Q')
\end{equation}


The energies and types of the two colliding partons 
determine their momenta and affect their locations within the
ordinary and extra dimensions.
For protons moving in the $z$-direction, 
we arbitrarily put one of the partons at the origin 
and locate the second parton randomly within a disk in the xy-plane of radius
$b_{max}(\ECM=\sqrt{us};d)$. 

We must, however, also take into account that the partons will 
be separated in the extra dimensions as well. Each parton type 
is given a wave function in the extra dimension. 
For fermions, these wave functions are parametrized by their 
centers and widths which are input parameters (cf. Fig \ref{fig:input}). 
In the split-fermion case, the centers of these wave functions may
be widely separated;
but even in the non-split case, the wave functions have non-zero widths.
For gauge bosons, the wave functions are taken to be constant across the 
(thick) brane.

The output from the generator (described in greater detail below)
includes the energies, momenta, and types of partons that yielded black-holes. 
The locations in time and space of the black-holes are also output.

The formation of the black-hole is a very non-linear and complicated process. 
We assume that, 
before settling down to a stationary phase, 
a black-hole loses some fraction of its energy, linear and angular momentum. 
We parameterize these losses by three parameters: 
$1 - f_{E}$, $1-f_{P}$ and $1-f_{L}$. 
Thus the black-hole initial state that we actually evolve is characterized by 
\begin{eqnarray}
\label{eqn:initlossfactors}
E&=&E_{in}f_{E};\nonumber \\
P_z&=&P_{z\, in}f_{P};\\
J'&=&L_{in}f_{L}\nonumber ;
\end{eqnarray}
where $E_{in}$, $P_{z\, in}$ and $L_{in}$ are initial energy, 
momentum and angular momentum of colliding partons, 
while $f_{E}$, $f_{P}$ and $f_{L}$ are the fractions of 
the initial energy, momentum and angular momentum
that are retained by the stationary black-hole. 
We expect that most of the energy lost in the non-linear regime is 
radiated in the form of gravitational waves and thus represents missing energy. 
Yoshino and Rychkov \cite{15} have calculated the 
energy losses by numerical simulation of collisions.
Their results will be incorporated in a future upgrade of BlackMax.

For a small black-hole, the numerical value for the angular momentum 
is of the order of several $\hbar$. In that range of values, 
angular momentum is quantized. Therefore a black-hole cannot 
have arbitrary values of angular momentum. We keep the 
actual angular momentum of the black-hole, $J$, to be the nearest 
half-integer, i.e. $2J =\left [2J' + \frac{1}{2} \right ]$. 

\begin{figure}[ht]
\centering{
\includegraphics[width=3.2in]{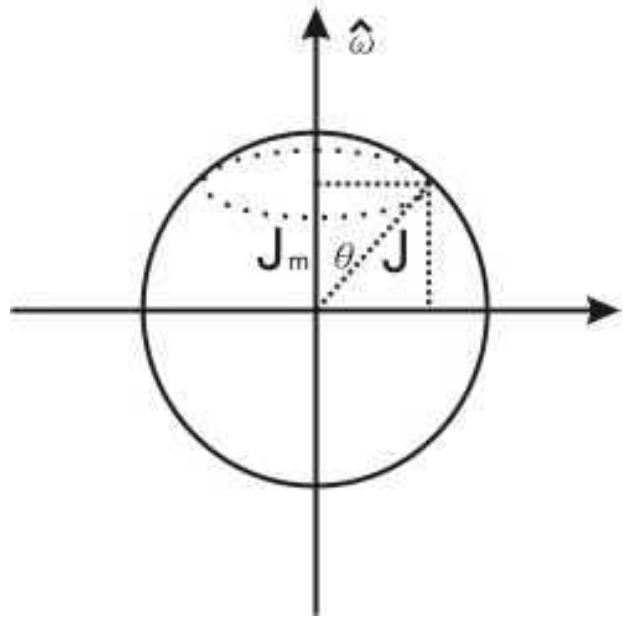} }
\caption{Angular momentum tilt geometry.} \label{fig:anglevsJ}
\end{figure}

The loss of the initial angular momentum in the non-linear regime 
has as a consequence that the black-hole angular momentum 
is no longer in the transverse plane of the colliding protons. 
We therefore introduce a tilt in the angular-momentum
\begin{equation}
\theta \equiv cos^{-1} (\frac{J}{\sqrt{J(J+1)}}) .
\end{equation}
Figure \ref{fig:anglevsJ} illustrates this geometry.

In this version of the generator, 
we have assumed that the angular-momentum quantum numbers 
of the black-hole were $(J,J_m=J)$\footnote{Future versions of the generator 
may randomize the choice of $J_m$.}.
We next randomly choose an angle $\phi$, 
and then reset the angular-momentum axis to $(\theta,\phi)$.

\section{Gray-body Factors} \label{sec:graybodyfactors} 
\indent 

Once the black-hole settles down to its stationary configuration, 
it is expected to emit semi-classical Hawking radiation. 
The emission spectra of different particles
from a given black-hole depend in principle on the mass, 
spin and charge of the black-hole, on the ``environment''\footnote{
dimensionality and geometry of the bulk, brane tension, 
location of fermionic branes} and on the mass and spin of
the particular particle. Wherever possible we have made use of the
correct emission spectrum often phrased in terms of the
gray-body factor for black-holes in 3+1-dimensional space-time.
In most cases, these were extant in the literature, but
we have calculated the spectra for the split-fermion model ourselves,
and reproduced existing spectra independently.
The sources of the gray-body factors are summarized in 
Table \ref{tab:spectra}.


\begin{itemize}
\item {\it Non-rotating black-hole on a tensionless brane:} 
For a non-rotating black-hole, we used previously known gray-body factors 
for spin $0,1/2$ and $1$ fields in the brane, and for spin $2$ fields 
({\it i.e.} gravitons) in the bulk.

\item {\it Rotating black-hole on a tensionless brane:} 
For rotating black-holes, we used known gray-body factors for spin 
$0,1/2$ and $1$ fields on the brane. The correct emission spectrum
for spin $2$ bulk fields is not yet known for rotating black-holes,
we currently do not allow for the emission of bulk gravitons from
rotating black-holes. {\em As discussed below, this remains a serious 
shortcoming of current micro-black-hole phenomenology, since
super-radiance might be expected to significantly increase graviton emission 
from rotating black-holes, and thus increase the missing energy in a detector.}

\item {\it Non-rotating black-holes on a tensionless brane with fermion brane 
splitting:}
In the split-fermion models, gauge fields can propagate through the bulk
as well as on the brane, so we have calculated gray-body factors for 
spin $0$ and $1$ fields propagating through the bulk,
but only for a non-rotating black-hole for the split-fermion model. 
These are shown in Figures \ref{spectra:s-2}-\ref{spectra:ga-7}.

\item {\it Non-rotating black-holes on a non-zero tension brane:} 
The bulk gray-body factors for a brane with non-zero tension are affected by
non-zero tension because of the modified bulk geometry (deficit angle).
We have calculated gray-body factors for spin $0,1$ and $2$ fields 
propagating through the bulk, again only for the non-rotating black-hole for a 
brane with non-zero tension and $d=5$.

\item {\it Two particle final states:}
We use the same gray-body factors as a non-rotating black-hole to calculate 
the cross section of two-particle final states (excluding gravitons).
\end{itemize}

In all cases, the relevant emission spectra are loaded 
into a data base as described in appendices A.

\section{Black-Hole Evolution }
\label{sec:bhevolution}
\indent

The Hawking radiation spectra are calculated for the black-hole at rest in the 
center-of-mass frame of the colliding partons. 
The spectra are then transformed to the laboratory frame as needed. 
In all cases we have not (yet) taken the charge of the black-hole into account 
in calculating the emission spectrum, but have included phenomenological 
factors to account for it as explained below.

The degrees of freedom of the Standard-Model particles are given 
in Table \ref{tab:degenecy1}. Using the calculated Hawking spectrum 
and the number of degrees of freedom per particle, we determine
the expected radiated flux of each type of particle as a 
function of black-hole and environmental properties. 
For each particle type $i$ we assign to it a specific energy, 
$\hbar\omega_i$ 
with a probability determined by that particle's emission spectrum.
(The particle ``types'' are listed in Table \ref{tab:degenecy1}.)

Assume a black-hole with mass $M_{bh}$ emits a massless particle with 
energy $\hbar \omega_i$. The remaining black-hole will have energy and momentum 
like 
\begin{equation}
\label{eqn:M-max1}
(M_{bh}-\hbar \omega_{i},-\hbar \omega_{i})
\end{equation}
Here we ignore the other dimensions. 
We use a classical model to simulate the events. 
The mass of the remaining black-hole should remain positive.
So from equation (\ref{eqn:M-max1}) one gets 
\begin{equation}
\label{eqn:M-max}
\hbar \omega_{i} < M_{bh}/2 .
\end{equation}
Combining this with the observation that energy of a particle is larger than its 
mass, 
leads us to require that 
\begin{equation}
M_{i}<\hbar\omega_{i} \end{equation}

\begin{figure}[ht]
\centering
\includegraphics[width=3.2in]{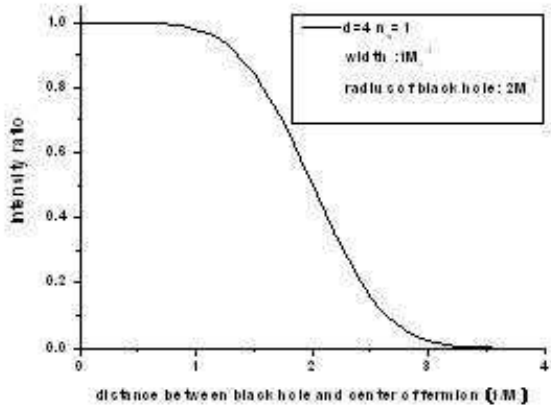}
\caption{The emitted fermion intensity (normalized to one) as a function of 
the distance between the black-hole and the center of the gaussian distribution 
of a fermionic brane. As a black-hole 
increases the distance from a fermionic brane due to recoil the intensity of the 
emitted fermions of that type falls down quickly. The radius of black-hole is 
set to be $2M_{*}^{-1}$. The width of the fermionic brane is $M_{*}^{-1}$. The 
plot is for the case of one extra dimension.}
\label{fig:decay-factor}
\end{figure}

We next need to determine whether that particle with that
energy is actually emitted within one generator time-step $\Delta t$.
The time-step itself is an input parameter (cf. figure \ref{fig:input}).
We choose a random number $N_{r}$ from the interval $\left[0,1\right]$. 
Given $L_{Fi}$, the total number flux of particles of type $i$, 
and $N_{i}$, the number of degrees of freedom of that particle type, 
the particle will be emitted if
\begin{equation}
\label{eqn:emit-poss}
 L_{Fi}N_{i}\Delta t >N_{r}.
\end{equation}

In the single-brane model, 
$L_{Fi}$ is derived from the power spectrum of the Hawking radiation.
In the split-fermion model, 
we include a suppression factor for fermions. 
The factor depends on the overlap between the 
particular fermion brane and the black-hole when the
black-hole is not located on that fermion's brane.
Fig. \ref{fig:decay-factor} shows 
how the spectrum of emitted particles changes 
as the black-hole drifts away from the center of the fermion brane. 
As a black-hole increases its distance from a fermion brane 
due to recoil, 
the intensity of the emitted fermions of that type declines quickly.

If the particle is to be emitted, 
we choose its angular-momentum quantum numbers $(l,m)$ according
to:
\begin{equation}
P_{em}(i,l,m,E)=L_{i,l,m}(E)/\sum_{l',m'} L_{i,l',m'}(E) .
\end{equation}
Here $P_{em}(i,l,m,E)$ is the probability that a type $i$ particle 
with quantum numbers $(l,m)$ will be emitted. 
$L_{i,l,m}(E)$ is the emission spectrum of a particle of type $i$ 
with quantum numbers $(l,m)$.  
This step is omitted in the case of non-rotating black-holes since
we do not follow the angular-momentum evolution of the black-hole.

Once the quantum numbers of the emitted particle are determined, 
we calculate the direction of emission according to the corresponding 
spheroidal wave function:
\begin{equation}
\label{eq:p_ang}
P_{em}(\theta,\phi)= \vert\Psi_{lm}(\theta,\phi)\vert^2 \sin\theta\Delta \theta \Delta \phi
\end{equation} 
\noindent
Here $P_{em}(\theta,\phi)$ is the probability of emission 
in the $(\theta, \phi)$ direction for 
the angular quantum numbers $(\ell, m)$.
$\Psi_{lm}(\theta,\phi)$ is the (properly normalized) 
spheroidal wave function
of the mode with those angular quantum numbers.

Once the energy and angular-momentum quantum numbers 
are determined for the i-th particle type, 
then, if that particle type carries SU(3) color 
we assign the color randomly.
The color is treated as a three-dimensional vector 
$\vec{c}_i=(r,b,g)$, in which
a quark's color-vector is either $(1,0,0)$, $(0,1,0)$, or $(0,0,1)$. Similarly,
an anti-quark has $-1$ entries in its color-vector. A gluon's color-vector has
one $+1$ entry, and one $-1$ entry.

The black-hole gray-body factors which quantify the relative emission
probabilities of particles with different spin are calculated for 
a fixed background, i.e. assuming that the black-hole metric
does not change during the emission process.
However, as the black-hole emits particles, 
the spin and charge of the black-hole do change. 

\subsection{Electric and Color Charge Suppression}
\label{subsec:qcsuppression}

A charged and highly rotating black-hole will tend 
to shed its charge and angular momentum. 
Thus, emission of particles with charges 
of the same sign as that of the black-hole 
and angular momentum parallel to the black-hole's will be preferred. 
Emission of particles that increase the black-hole's charge or 
angular momentum should be suppressed. 
The precise calculation of these effects has not as yet been accomplished.
Therefore, to account for these effects 
we allow optional phenomenological suppression factors 
for both charge and angular momentum. 

The following charge-suppression factors can currently be used
by setting parameter 19 (cf. section \ref{sec:input_output}) equal to 2.
\begin{eqnarray}
\label{eqn:highqandcsuppression}
F^{Q}&=& \exp(\zeta_Q Q^{bh}Q^{em})\\
F^{3}_a&=& \exp(\zeta_3 c^{bh}_{a}c^{em}_{a}) \quad \quad {\rm a=r,b,g} .
\end{eqnarray}
$Q^{bh}$ is the electromagnetic charge of the black-hole, 
$Q^{em}$ is the charge of the emitted particle; 
$c^{bh}_{a}$, is the color value for the color $a$, with ${\rm a=r,b,g}$, of 
the black-hole, 
and
$c^{em}_{a}$, is the color value for the color $a$, with ${\rm a=r,b,g}$, of the 
emitted particle. 
$\zeta_Q$ and $\zeta_3$ are phenomenological suppression parameters that are set 
as input parameters of the generator.

We estimate 
$\zeta_{Q} = {\cal O}(\alpha_{em})$ and 
$\zeta_{3} = {\cal O}(\alpha_{s})$,
where $\alpha_{em}$ and $\alpha_{s}$ 
are the values of the electromagnetic and
strong couplings at the Hawking temperature of the black-hole.
Note that we currently neglect the possible restoration 
of the electroweak symmetry in the vicinity 
of the black-hole when its Hawking temperature
is above the electroweak scale.
Clearly, since $\alpha_{em}\simeq 10^{-2}$ we do not expect
electromagnetic (or more correctly) electroweak
charge suppression to be a significant effect.
However, since $\alpha_s(1\,\rm{TeV})\simeq0.1$,
color suppression may well play a role in the
evolution of the black-hole.

Once we have determined the type of particle
to be emitted by the black-hole, we draw a random number $N_r$
between $0$ and $1$ from a uniform distribution.
If $N_r > F^Q$ then the emission process is allowed to occur,
if $N_r < F^Q$ then the emission process is aborted. We repeat 
the same procedure for color suppression factor, $F^3_a$.
Thus, particle emission which decreases the magnitude
of the charge or color of the black-hole is unsuppressed;
this suppression prevents the black-hole from acquiring
a large charge/color, and gives preference to particle
emission which reduces the charge/color of the black-hole.

\subsection{Movement of the Black-Hole during Evaporation}
\label{subsec:bhmovement}

We choose the direction of the momentum of the emitted particle 
($\hat{P}_{e}$) according to equation (\ref{eq:p_ang}) in the center-of-mass frame and then
transform the energy and momentum to their laboratory frame
values $\hbar\omega'$ and $\vec{P}'_{e}$.
The black-hole properties (energy, momentum, mass, colors, and charge)
are then accordingly updated for the next time step:
\begin{eqnarray}
E(t+\Delta t)&=&E(t)-\hbar\omega'\\
\vec{P}(t+\Delta t)&=&\vec {P}(t)-\vec {P'_{e}}\\
M(t+\Delta t)&=&\sqrt{E(t+\Delta t)^{2}-\vec {P}(t+\Delta t)^{2}}\\
\vec{c}^{bh}(t+\Delta t)&=&\vec{c}^{bh}(t)-c_{i}\\
Q^{bh}(t+\Delta t)&=&Q^{bh}(t)-Q_i
\end{eqnarray}
Here $\vec{c}^{bh}$ is the color 3-vector of the black-hole and can
have arbitrary integer entries.

Due to the recoil from the emitted particle, 
the black-hole will acquire a velocity $\vec {v}$ 
and move to a position $\vec {x}$:
\begin{eqnarray}
\vec {v}(t)&=&\vec {P}(t)/E\\
\vec {x}(t+\Delta t)&=&\vec{x}(t)+\vec {v}(t) \Delta t
\end{eqnarray}

Since fermions are constrained to live 
on the 3+1-dimensional regular brane, 
the recoil from fermions is not important. 
Only the emission of vector fields, scalar fields 
and gravitons gives a black hole momentum in extradimension. 
Once a black hole gains momentum in extradimension, 
it is able to leave the regular brane if it carries no gauge charge.
In the split fermion case, it can move within the mini-bulk even if
it carries gauge charge.
In the case of rotating black holes, because the gray-body
factor for gravitons is not yet known, graviton emission is turned off
in the generator and the black holes experience no bulk recoil.  

Recoil can in principle change the radiation spectrum of the black-hole in two 
ways.
First, the spectrum will not be perfectly thermal or spherically symmetric in 
the laboratory frame, but rather boosted due to the motion of the black-hole. 
However, as we shall see, the black-hole never becomes highly relativistic,
so the recoil does not significantly affect the shape of the spectrum. 

As the lifetime of a small black-hole is relatively short,
and its recoil velocity non-relativistic, it does not move 
far from its point of creation. 
However, even a recoil of the order of one fundamental length $\sim$ $M_*^{-1}$ 
in the bulk direction could dislocate the black-hole from the 
brane\footnote{This is very unlikely because most of the black-holes have gauge 
charges.}. 
In single-brane models this would result in apparent missing energy 
for an observer located on the brane able to detect only Standard Model 
particles. 
In the split-fermion model, as the black-hole moves off or on particular
fermion branes, the decay channels open to it will change.

\subsection{Rotation}
\label{subsec:rotation}

Since two colliding particles always define a single plane of rotation, 
rotating black-holes are formed with a single rotational parameter. 
For two particles colliding along the $z$-axis, 
there should be only one rotation axis perpendicular to the $z$-axis. 
However, due to angular-momentum loss both in the formation process,
and subsequently in the black-hole decay, three things can happen:
i) the amount of rotation can change, 
ii) the rotation axis can be altered,
and 
iii) more rotation axes can emerge,because there are more than three spatial 
dimensions. 
Also, if the colliding particles have a non-zero impact parameter 
in bulk directions\footnote{Due to the finite thickness of the single brane 
or splitting between the quark branes.} the plane of rotation 
will not lie entirely in the brane direction.
Because solutions do not exist for rotating black-holes with more than one
rotation axis, we forbid the emergence of secondary rotation axes.
We do, on the other hand, allow the single rotation axis of the black-hole 
to evolve. However, no gray-body factors are known if the single rotation axis
acquires components in the extra dimensions, therefore we
limit the rotation axis to the brane dimensions.
Relaxing these limitations is a subject for future research.

We next must determine the rotational axis of the black-hole.
The rotation parameter of a black-hole with angular-momentum 
quantum numbers $(j,j)$ is taken to be 
\begin{equation}
a=\frac{J}{M}\frac{n+2}{2},
\end{equation}
where $J=\sqrt{j(j+1)}\hbar$.
The direction of the black-hole angular momentum is taken to be
\begin{equation}
\vec{J}=j\hbar\hat{\omega}+\sqrt{j}\hbar\hat{l}_{\perp},
\end{equation}
where $\hat{l}_{\perp}$ is a unit vector in the plane perpendicular
to $\hat{\omega}$. We chose the direction of $\hat{l}_{\perp}$randomly.

When the black-hole emits a particle with angular-momentum quantum numbers 
$(l,m)$, 
there are several possible final states in which the black-hole can end up. 
We use Clebsch-Gordan coefficients to find the probability of each state.

\begin{equation}
|j,j\!\!>=\!\!\!\!\!\!\sum_{j'\leq j+l}^{|j-l|} 
\!\!\!\!\!\!C(j,j;l,m,j',j\!\!-m)|l,m\!\!>|j',j\!\!-m>
\end{equation}

We use $|C(j,j;l,m,j',j-m)|^{2}$ as the probability that the 
new angular-momentum quantum numbers of the black-hole will be $(j',j-m)$.
From angular-momentum conservation $\vec{J}=\vec{L}+\vec{J'}$, 
we can calculate the tilt angle of the black-hole rotation axis as:
\begin{equation}
cos \theta =\frac{j(j+1)+j'(j'+1)-l(l+1)}{2\sqrt{j(j+1)j'(j'+1)}} .
\end{equation}
We randomly choose a direction with the tilt angle $\theta$ 
as a new rotation axis and change quantum numbers to $(j',j')$. 

In calculating the gray-body factors,
the black hole is always treated  as a fixed unchanging background. 
The power spectrum of emitted particles can be calculated from 
\begin{equation}
\label{eqn:dEdt}
\frac{dE}{dt}=\sum_{l,m}|A_{l,m}|^2\frac{\omega}{exp((\omega-m\Omega)/T_{H})\mp 1}\frac{d\omega}{2\pi} .
\end{equation}
Here $l$ and $m$ are angular momentum quantum numbers. 
$\omega$ is the energy of the emitted particle. 
$\Omega$ is definied by
\begin{equation}
\Omega=\frac{a_*}{(1+a_*^2)r_h} .
\end{equation}

The exponential factor in the denominator of  (\ref{eqn:dEdt})
causes the black hole to prefer to emit high angular momentum particles. 
However, since the TeV black holes are quantum black holes, 
the gray-body factors should really depend on both the initial
and final black-hole parameters. The calculation of the gray-body spectra 
on a fixed background can cause some problems.
In particular, in the current case,  the angular momentum of the 
emitted particle (as indeed the energy) may well be comparable 
to that of the black hole itself.  There should be a
suppression of particle  emission processes in which the black hole 
final state is very different from the initial state. 
We therefore introduce a new phenomenological suppression factor, parameter 17, 
to reduce the probability of emission events in which the angular momentum
of the black hole changes by a large amount.

If parameter 17 is equal to 1 
(cf. section \ref{sec:input_output}), we do not take into account
the suppression of decays which increase the angular momentum 
of the black-hole.
If we are using $\Delta$Area suppression
(parameter 17 equal to 2) then
\be
\label{eqn:LsuppressiondeltaA}
F^{L}= \exp(\zeta_L (r_{h}^{bh}(t+\Delta t)^{2}/r_{h}^{bh}(t){^2}-1)) .
\ee
If we are using $J_{bh}$ suppression
(parameter 17 equal to 3) then
\be
\label{eqn:LsuppressionJ}
F^{L}= \exp(-\zeta_L |J^{bh}(t+\Delta t)|) .
\ee
If we are using $\Delta{J}_{bh}$ suppression
(parameter 17 equal to 4) then
\be
\label{eqn:LsuppressiondeltaJ}
F^{L}= \exp(-\zeta_L |{J}^{bh}(t+\Delta t)-J^{bh}(t)|) .
\ee
We might expect $\zeta_L\sim1$, however there is no
detailed theory to support this; as indeed there is no
detailed theory to choose among these three phenomenological
suppression factors. 
It is also worth noting that, while for $d=3$ and $d=4$
there is a maximum angular momentum that a black-hole
of a given mass can carry, for $d\geq 5$ there
is no such upper limit. We do impose that
$a\leq R_{s}/2$ for $d=3$ 
and $a\leq R_{s}$ for $d=4$. 

As for the charge and color suppression, 
we choose a random number $N_r$ between $0$ and $1$.
If $N_r>F^L$ then the particle emission is aborted.

The procedure described in this section is then repeated 
at each time step with each particle type, and
then successive time steps are taken
until the mass of the black-hole falls below $M_{*}$.
In practice, the time step should be set short enough that
in a given time step the probability that particles
of more than one type are emitted is small.
We set the time step to $\Delta t = 10^{-5}\,{\rm GeV}^{-1}$.

In two-body final states, 
one expects no black-hole, 
and hence no black-hole decay by emission of Hawking radiation. 
The generator therefore proceeds directly to the final burst phase.


\section{Final Burst}
\label{sec:finalburst}
\indent

In the absence of a self-consistent theory of quantum gravity, 
the last stage of the evaporation cannot be described accurately.
Once the mass of black-hole becomes close to the fundamental 
scale $M_{*}$, the classical black-hole solution can certainly not be used 
anymore.
We adopt a scenario in which the final stage of evaporation 
is a burst of particles which conserves energy, momentum 
and all of the gauge quantum numbers. 
For definiteness, we assume the remaining black-hole will decay 
into the lowest number of Standard-Model particles that conserve all quantum 
number, momentum and energy. 

A black-hole with electromagnetic charge $Q^{bh}$ 
and color-vector $\vec{c}^{\;bh}=(r^{bh},b^{bh},g^{bh})$
will be taken to emit $N_{-1/3}$ down-type quarks (i.e. d,s or b quarks), 
$N_{2/3}$ up-type quarks (u,c, or t), $N_{-1}$ charged leptons and W bosons
, $N_{gl}$ gluons, and $N_{n}$ non-charge particles (ie. $\gamma$, Z an Higgs). 
We use the following procedure to determines 
$\vec{N}_{burst}\equiv(N_{-1/3},N_{2/3},N_{-1},N_{gl},N_{n})$.

\noindent Step 1: preliminary solution:
\begin{itemize}
\item
Search all possible solutions with $N_{n}=0$. 
\item Choose the minimum number of particles as preliminary solution.
\end{itemize}

\noindent Step 2: Actual charged/colored emitted particle count:
\begin{itemize}
\item 
The preliminary $\vec{N}_{burst}\equiv(N_{-1/3},N_{2/3},N_{-1},N_{gl},N_{n})$
having now been determined. If the minimum number of solution is less 
than 2, we then add $N_{n}$ to keep the total number equal to 2. Later we choose 
one of them randomly according to the degrees of freedom of each particle.
\item After obtaining the number of emitted particles, 
we randomly assign their energies and momenta,
subject to the constraint that the total energy
and momentum equal that of the final black-hole state.
We currently neglect any bulk components of the 
final black-hole momentum.
\end{itemize}

\section{Input and Output}
\label{sec:input_output}
\indent

\begin{figure*}[ht]
\centering{
\includegraphics[width=6in]{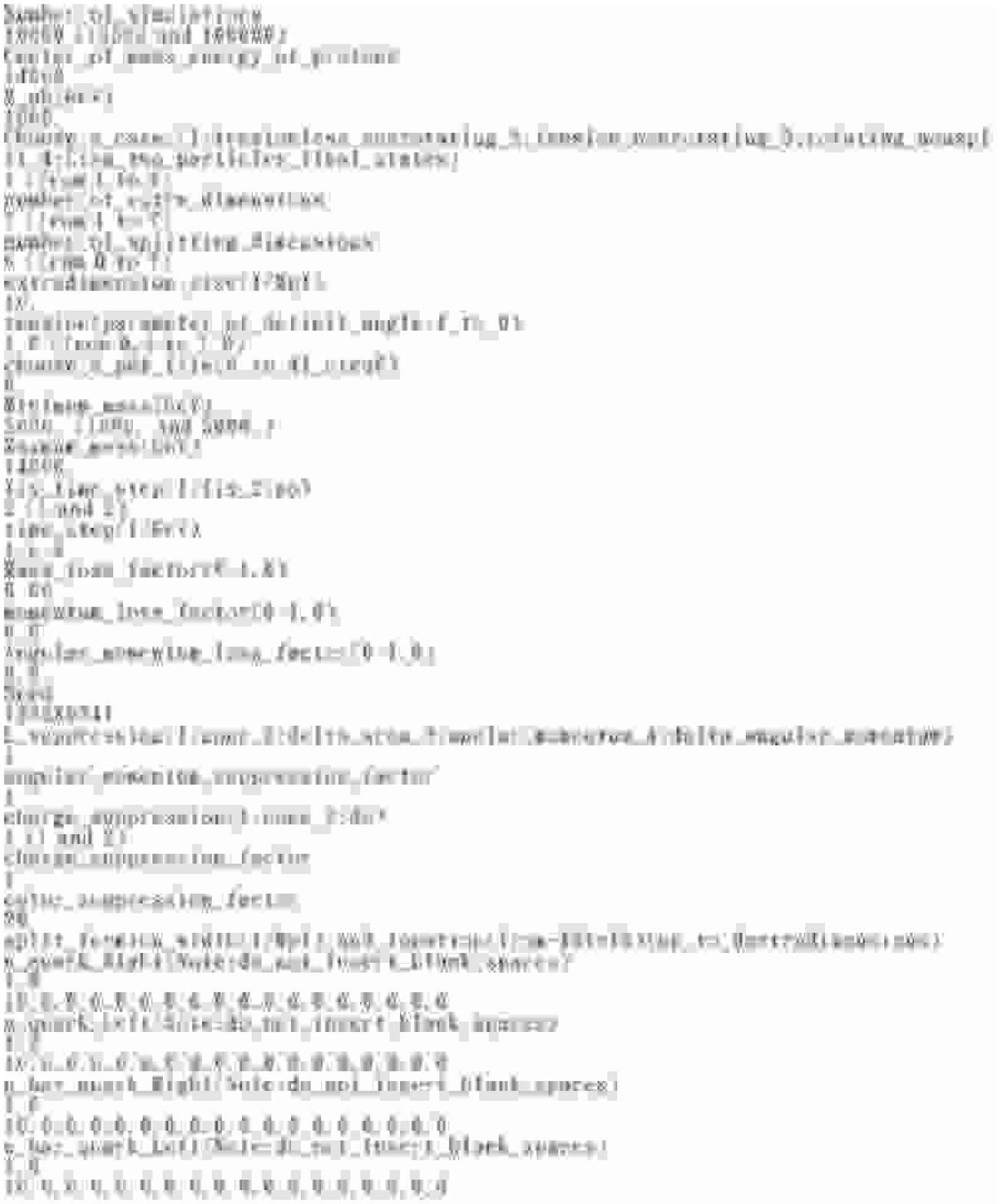} }
\caption{Parameter.txt is the input file containing the parameters that one 
can change. The words in parentheses are the parameters that are used in the paper.} \label{fig:input}
\end{figure*}

The input parameters for the generator are read from the file parameter.txt, see 
Fig.\ref{fig:input}: 
\begin{enumerate}
\item {\tt Number$\tt{\_\_}$of$\tt{\_\_}$simulations}: sets the total number of 
black-hole events to be simulated; 
\item {\tt Center$\_\_$of$\_\_$mass$\_\_$energy$\_\_$of$\_\_$protons}: sets the 
center-of-mass energy of the colliding protons in GeV;
\item {\tt M$\_\_$ph}: sets the fundamental quantum-gravity scale ($M_*$) in 
GeV;
\item {\tt Choose$\_\_$a$\_\_$case}: 
defines the extra dimension model to be simulated:
\begin{enumerate}
\item 1: non-rotating black-holes on a tensionless brane with 
possibility of fermion splitting, 
\item 2: non-rotating black-holes on a brane with non-zero positive 
tension,
\item 3: rotating black-holes on a tensionless brane with d=5,
\item 4: two-particle final-state scenario;
\end{enumerate}
\item {\tt number$\_\_$of$\_\_$extra$\_\_$dimensions}: sets the number of extra 
dimensions; this must equal 2 for brane with tension ({\tt 
Choose$\_\_$a$\_\_$case}=2);
\item {\tt number$\_\_$of$\_\_$splitting$\_\_$dimensions}: sets the number of 
extra split-fermion dimensions ({\tt Choose$\_\_$a$\_\_$case}=1);
\item {\tt extradimension$\_\_$size}: sets the size of the 
mini-bulk\footnote{This is the distance between fermion branes where only gauge 
bosons and Higgs field can propagate in split-fermion brane scenario.} in units 
of $1/\rm{TeV}$ ({\tt Choose$\_\_$a$\_\_$case}=1);
\item {\tt tension}: 
sets the deficit-angle parameter $B$~\cite{NKD,DKSS} 
({\tt Choose$\_\_$a$\_\_$case}=2); 
\item {\tt choose$\_\_$a$\_\_$pdf$\_\_$file}: 
defines which of the different 
CTEQ6 parton-distribution functions (PDF) to use;
\item {\tt Minimum$\_\_$mass}: sets the minimum mass $M_{min}$ 
in GeV of the initial black-holes;
\item {\tt fix$\_$time$\_\_$step}: If equal to 1, then code uses the next 
parameter
to determine the time interval between events; if equal to 2 then code tries
to optimize the time step, keeping the probability of emitting a particle
in any given time step below 10\%.
\item {\tt time$\_\_$step}: defines the time interval $\Delta t$ in 
$\rm{GeV}^{-1}$ which the generator will use for the black-hole evolution;
\item {\tt Mass$\_\_$loss$\_\_$factor}: 
sets the loss factor $0$
for the energy of the initial black-hole,
as defined in equation\,\ref{eqn:initlossfactors};
\item {\tt momentum$\_\_$loss$\_\_$factor}: defines the loss factor $0\leq 
f_p\leq 1$ for the momentum of initial black-holes as defined in 
equation\,\ref{eqn:initlossfactors};
\item {\tt Angular$\_\_$momentum$\_\_$loss$\_\_$factor}: sets the loss factor $0 
\leq f_L \leq 1$ for the angular momentum of initial black-holes s defined in 
equation\,\ref{eqn:initlossfactors};
\item {\tt Seed}: sets the seed for the random-number generator 
(9 digit positive integer);
\item {\tt L$\_\_$suppression}: 
chooses the model for suppressing 
the accumulation of large black-hole angular momenta 
during the evolution phase of the black-holes 
(cf. discussion surrounding equations 
\ref{eqn:LsuppressiondeltaA}-\ref{eqn:LsuppressiondeltaJ});
\begin{itemize}
\item 1: no suppression; 
\item 2: $\Delta$\,Area suppression; 
\item 3: $J_{bh}$ suppression; 
\item 4: $\Delta\,J$ suppression;
\end{itemize}
\item {\tt angular$\_\_$momentum$\_\_$suppression$\_\_$factor}: defines the 
phenomenological angular-momentum suppression factor, $\zeta_L$ (cf. discussion 
surrounding equation 
\ref{eqn:LsuppressiondeltaA}-\ref{eqn:LsuppressiondeltaJ});
\item {\tt charge$\_\_$suppression}: turns the suppression of accumulation of 
large black-hole electromagnetic and color charge during the black-hole 
evolution process on or off (cf. dicussion surrounding equation 
\ref{eqn:highqandcsuppression})
\begin{itemize}
\item 0: charge suppression turned off; 
\item 1: charge suppression turned on;
\end{itemize}
\item {\tt charge$\_\_$suppression$\_\_$factor}: 
sets the electromagnetic charge suppression factor, 
$\zeta_Q$, in \ref{eqn:highqandcsuppression};
\item {\tt color$\_\_$suppression$\_\_$factor}: 
sets the color charge suppression factor, 
$\zeta_3$ in \ref{eqn:highqandcsuppression};
\item[21-94] 
(odd entries:) 
the widths of fermion wave functions (in $M_*^{-1}$ units);
and (even entries:) 
centers of fermion wave functions (in $M_*^{-1}$ units)
in split-brane models, 
represented as 9-dimensional vectors 
(for non-split models, set all entries to 0).
\end{enumerate}

\begin{figure*}[ht]
\centering{
\includegraphics[width=6in]{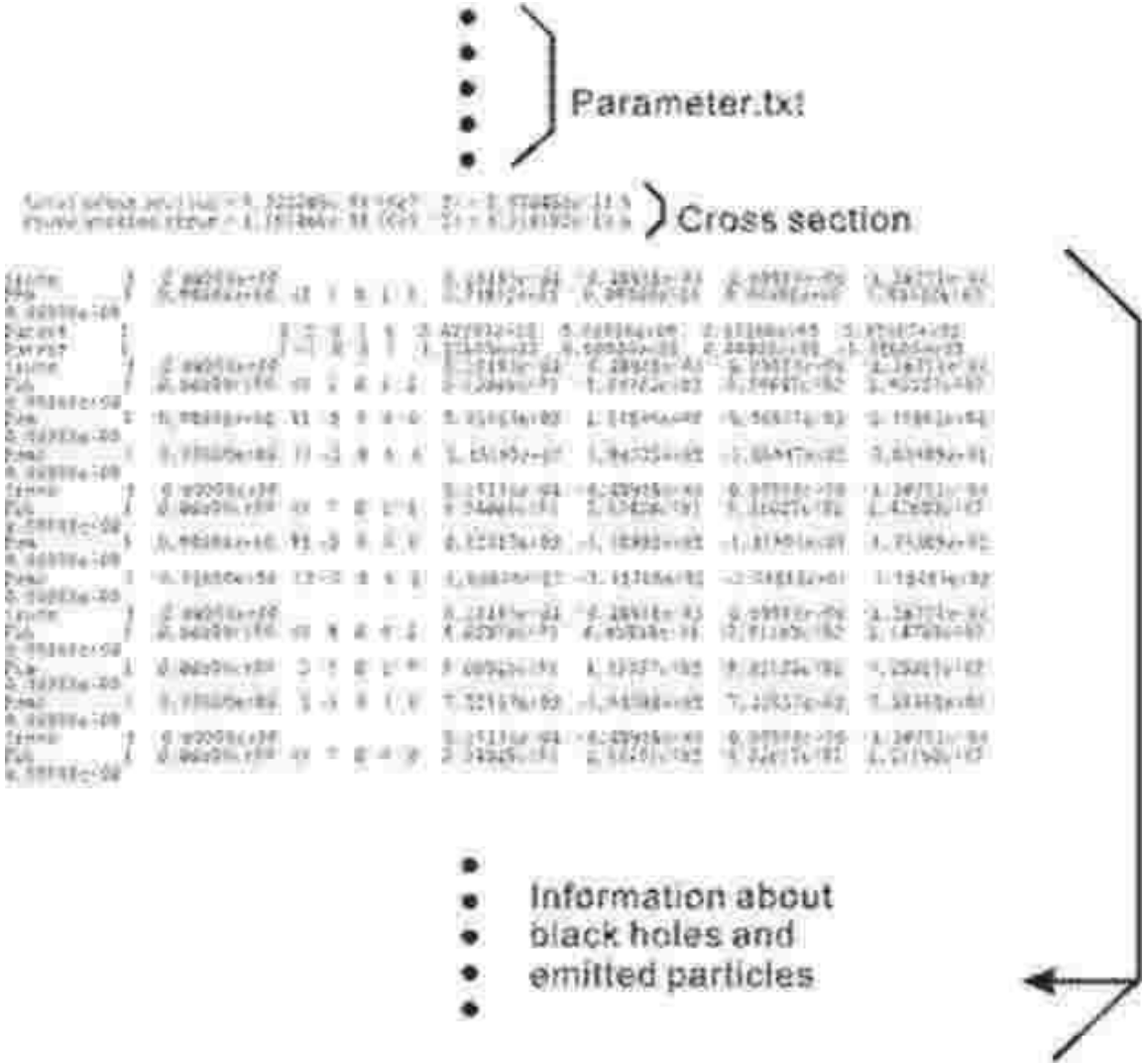} }
\caption{Output.txt: There are three parts to this file. 
The first part is a copy of parameter.txt. The second part includes 
information about the black-hole and the emitted particles. 
The first column identifies the what type of information each row is supplying 
--
parent is information about the two incoming partons; Pbh is information on the 
energy and momenta of the produced black-holes; 
trace describes the location of the black-holes;
Pem characterizes the emitted particles in the lab frame;
Pemc characterizes the emitted particles in the center-of-mass frame; 
Elast describes the final burst. \\
The third part of the file is the black-hole production cross-section 
as inferred from the events in this generator run. }
\label{fig:screenoutput} 
\end{figure*}

\begin{figure*}[ht]
\centering{
\includegraphics[width=6in]{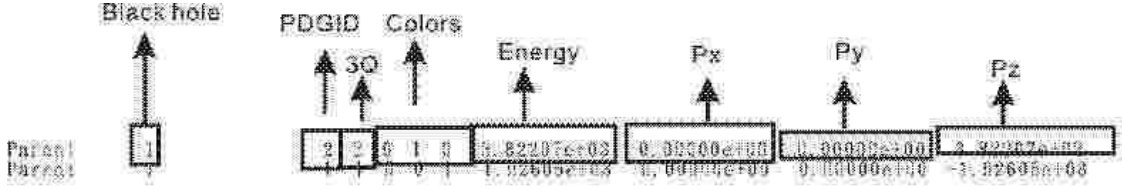} }
\caption{Lines in the output file headed by the ID = {\tt Parent} contain 
information about 
the initial partons which formed the black-hole.}
\label{fig:output1}
\end{figure*}
\begin{figure*}[ht]
\centering{
\includegraphics[width=6in]{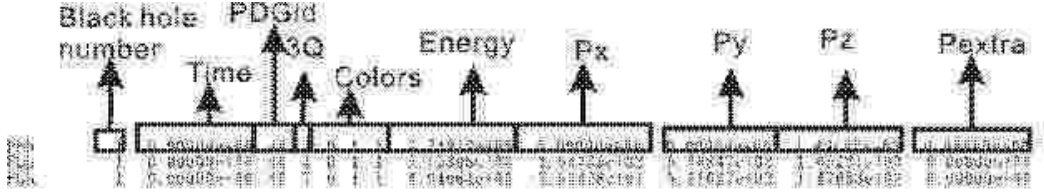} }
\caption{Lines in the output file headed by the ID = {\tt Pbh} contain the 
energies and momenta 
of the black-holes for each emission step. In case of rotating black-holes, 
the last column in the line is the angular momentum.}
\label{fig:output2}
\end{figure*}
\begin{figure*}[hb]
\centering{
\includegraphics[width=6in]{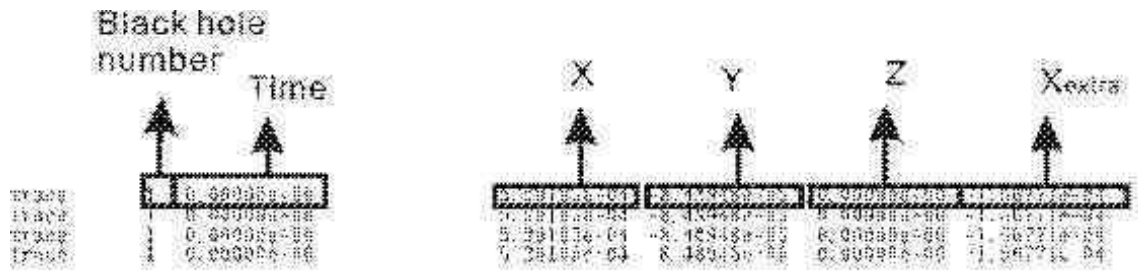} }
\caption{Lines in the output file headed by the ID = {\tt trace} contain the 
location 
of the black-hole fo reach emission step.}
\label{fig:output3}
\end{figure*}
\begin{figure*}[ht]
\centering{
\includegraphics[width=6in]{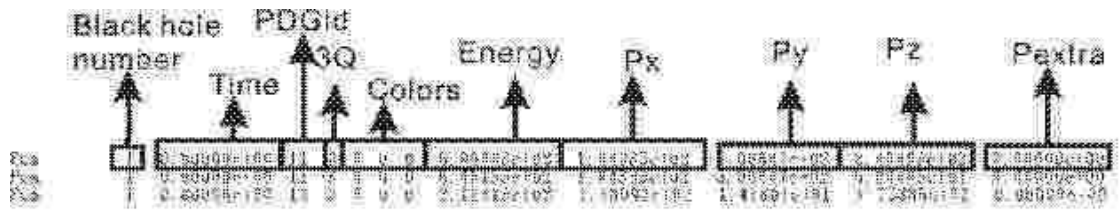} }
\caption{Lines in the output file headed by the ID = {\tt Pem} contain the 
types of the emitted particles, their energies and momenta in the lab frame 
and the times of their emission.}
\label{fig:output45}
\end{figure*}
\begin{figure*}[ht]
\centering{
\includegraphics[width=6in]{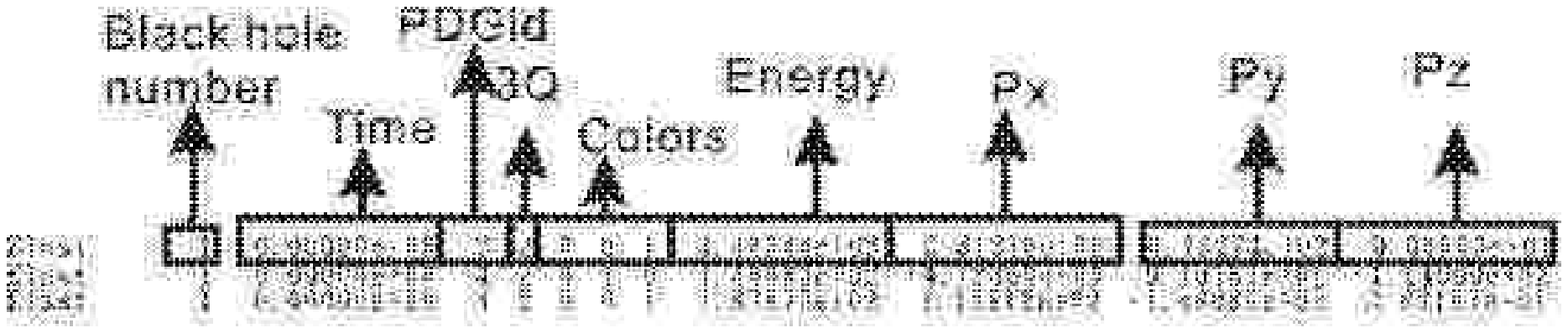} }
\caption{Lines in the output file headed by the ID = {\tt Elast} contain the 
types, energies and momenta of particles of the final burst.}
\label{fig:output6}
\end{figure*}


When the code terminates, the file output.txt with all the relevant information 
(i.e. input parameters, cross section) is output to the working directory. This 
file contains also different segments of information about the generation of 
black-holes which are labelled at the beginning of each line with an ID word 
(Parent, Pbh, trace, Pem, Pemc or Elast):

\begin{itemize}
\item {\bf Parent}: identifies the partons whose collision resulted in the 
formation of the initial black-hole (see Fig. \ref{fig:output1}).
\begin{itemize}
\item column 1: identifies the black-hole;
\item column 2: PDGID code of the parton; 
\item column 3: energy of the parton;
\item columns 4-6: brane momenta of the parton.
\end{itemize}
\item {\bf Pbh}: contains the evolution 
of the charge, color, momentum and energy of the black-holes, 
and, for rotating black-holes, their angular momentum 
(cf. Fig. \ref{fig:output2}).
\begin{itemize}
\item column 1: identifies the black-hole;
\item column 2: time at which the black-hole 
emitted a particle;
\item column 3: PDGID code of a black-hole;
\item column 4: three times the 
electromagnetic charge of the black-hole;
\item columns 5 to 7: color-charge vector 
components of the black-hole;
\item columns 8: energy of the black-hole 
in the laboratory frhame;
\item columns 9 to 11: brane components 
of the black-hole momentum in the laboratory frame;
\item columns 12 to (8+d): bulk components 
of the black-hole momentum;
\item column (9+d): angular momentum 
of the black-hole, 
in the case of rotating black-holes; 
empty otherwise.
\end{itemize}
\item {\bf trace}: contains the evolution history 
of the black-holes' positions (cf. Fig. \ref{fig:output3}):
\begin{itemize}
\item column 1: identifies the black-hole; 
\item column 2: the times at which the black-hole 
emitted a particle;
\item columns 3 to 5are the brane components 
of the black-hole position vector
when the black-hole emitted a particle;
\item columns 6 to (2+d): the bulk components 
of the black-hole position vector,
when the black-hole emitted a particle.
\end{itemize}
\item {\bf Pem}: contains a list of the black-holes,
with the history of their evolution (cf. Fig. \ref{fig:output45}):
\begin{itemize}
\item column 1: identifies the black-hole;
\item column 2: the times at which the 
black-hole emitted a particle;
\item column 3: PDGID code of the emitted particle;
\item column 4: three times 
the charge of the emitted particle;
\item columns 5 to 7: color-vector components 
of the emitted particle;
\item columns 8: energy of the emitted particle 
in the laboratory frame;
\item columns 9 to 11: brane components 
of the momentum of the emitted particle,
in the laboratory frame;
\item columns 12 to (8+d): bulk components 
of the momentum of the emitted particle.
\end{itemize}
\item {\bf Pemc}: contains the same information as Pem,
but in the center-of-mass frame of the collision.
\item {\bf Elast}: contains the same information as Pem
for the particles emitted in the final decay burst of the black-hole.
Column 12 and onwards are omitted as these particles have no bulk momentum.
\end{itemize}

\section{Results}
\label{sec:results}

\begin{figure}[ht]
\centering{
\includegraphics[width=3.2in]{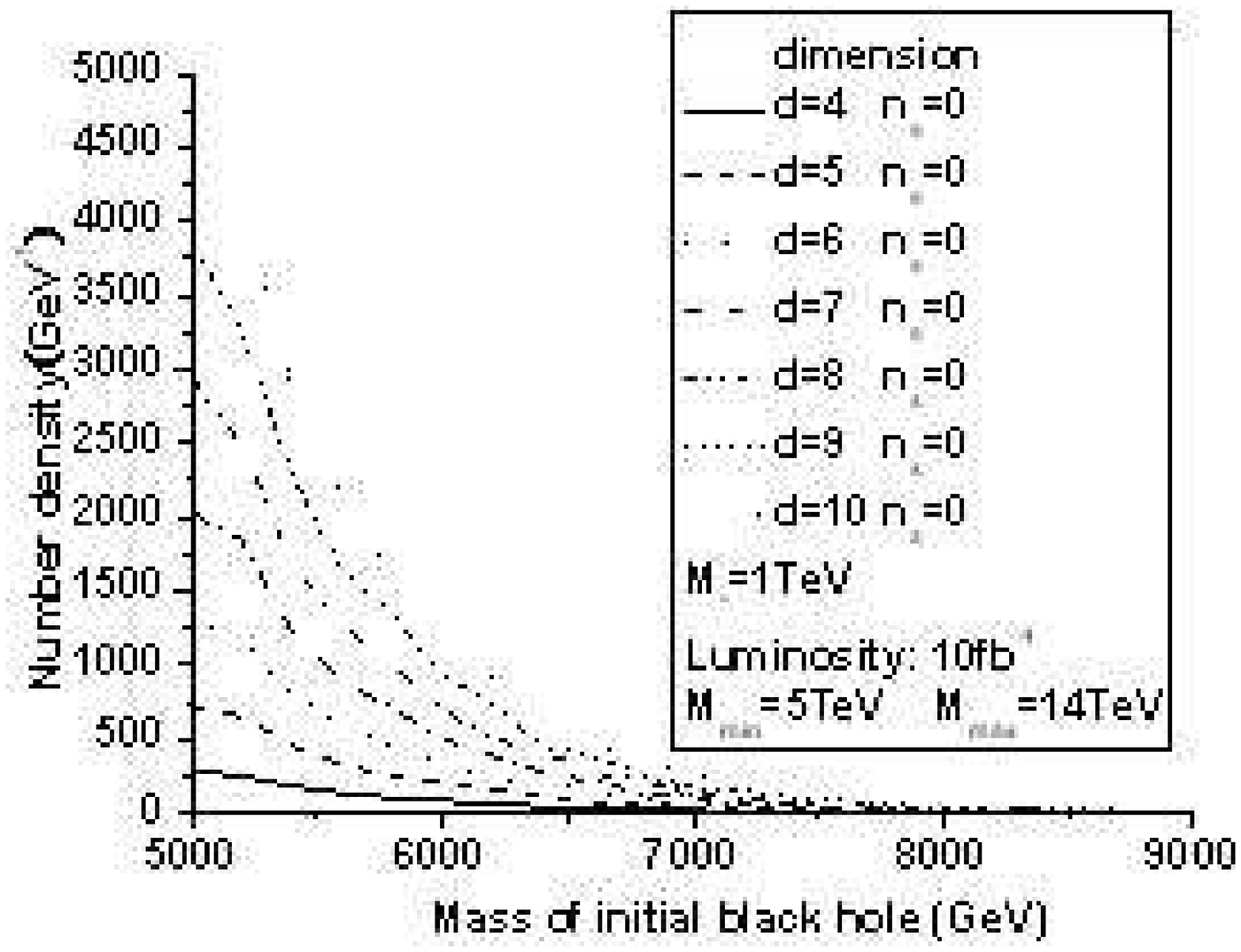} }
\caption{Mass distribution of initial black-holes (rotating and non-rotating)
on a tensionless brane for various numbers of extra dimension. }
\label{IE-brane}
\end{figure}

\begin{figure}[ht]
\centering{
\includegraphics[width=3.2in]{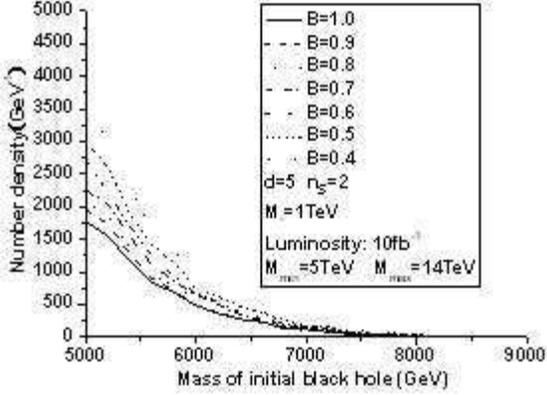} }
\caption{Mass distribution of initial (non-rotating) black-holes on a non-zero 
tension 
brane for $B=1.0$, $B=0.8$ and $B=0.6$.}
\label{IE-tension}
\end{figure}

\begin{figure}[ht]
\centering{
\includegraphics[width=3.2in]{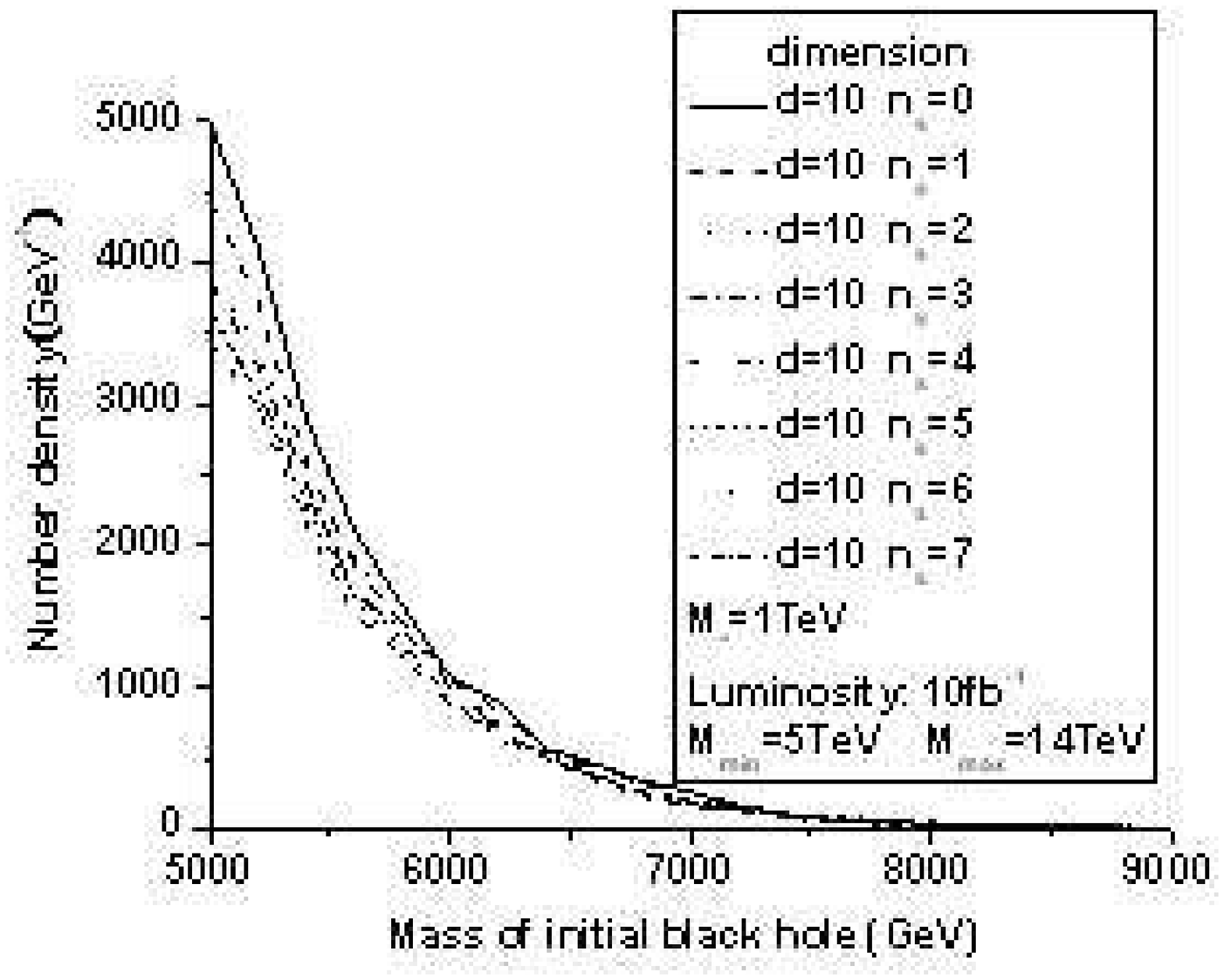} }
\caption{Mass distribution of initial (non-rotating) black-holes 
on a tensionless brane for $d=10$ and different numbers of split-fermion 
branes.}
\label{IE-split}
\end{figure}

\indent
We choose the following parameters 
for the distributions shown in this section and normalize them to an integrated 
luminosity of $10fb^{-1}$, unless otherwise stated.
The values of the parameters are chosen
to be the same as in figure \ref{fig:input}, except where
a parameter is varied to study its effect.

\begin{itemize}
\item {\tt Number$\tt{\_\_}$of$\tt{\_\_}$simulations} = 10000; 
\item {\tt 
Center$\_\_$of$\_\_$mass$\_\_$energy$\_\_$of$\_\_$protons}\,=\,14000\,Ge
V;
\item {\tt M$\_\_$ph} = 1000\,GeV;
\item {\tt extradimension$\_\_$size} = $10\,\rm{TeV}^{-1}$;
\item {\tt choose$\_\_$a$\_\_$pdf$\_\_$file}=0;
\item {\tt Minimum$\_\_$mass}=5000\,GeV;
\item {\tt time$\_\_$step}=$10^{-5}\,\rm{GeV}^{-1}$;
\item {\tt Mass$\_\_$loss$\_\_$factor}=0.0;
\item {\tt momentum$\_\_$loss$\_\_$factor}=0.0;
\item {\tt Angular$\_\_$momentum$\_\_$loss$\_\_$factor}=0.2;
\item {\tt Seed}=123589341;
\item {\tt L$\_\_$suppression}=1;
\item {\tt charge$\_\_$suppression}=1;
\item {\tt charge$\_\_$suppression$\_\_$factor} 
irrelevant since {\tt charge$\_\_$suppression}=1;
\item {\tt color$\_\_$suppression$\_\_$factor}
irrelevant since {\tt charge$\_\_$suppression}=1;
\item widths of fermion wave functions = 1\,$M_*^{-1}$ ;
\item centers of quark wave functions: $(10^{-2}/3,0,0,....)({\rm GeV}^{-1})$ 
(i.e. all quarks have Gaussian wave functions centered on a point 
displaced from the origin by $10^{-2}/3{\rm GeV}^{-1}$ in the first splitting 
direction);
\item distribution center of leptons: $(-10^{-2}/3,0,0,....)({\rm GeV}^{-1})$.
\end{itemize}

In this section, we present some distributions 
of properties of the initial and evolving black-holes 
and of the particles which are emitted by them during the 
Hawking radiation and final-burst phases. 

\subsection{Mass of the Initial Black-Holes}
\label{subsec:initialmass}

Figures \ref{IE-brane}, \ref{IE-tension} and \ref{IE-split} show the the initial 
black-hole mass distribution for three different extra dimension scenarios: 
non-rotating black-holes on a tensionless brane, non-rotating black-holes on a 
non-zero tension brane and non-rotating black-holes with split fermion branes 
respectively. 
Because we chose $5\,\rm{TeV}$ as the minimum mass of the initial black-hole,
the distributions have a cut off at $5\,\rm{TeV}$.

\subsection{Movement of Black-Holes in the Bulk}
\begin{figure}[ht]
\centering
\includegraphics[width=3.2in]{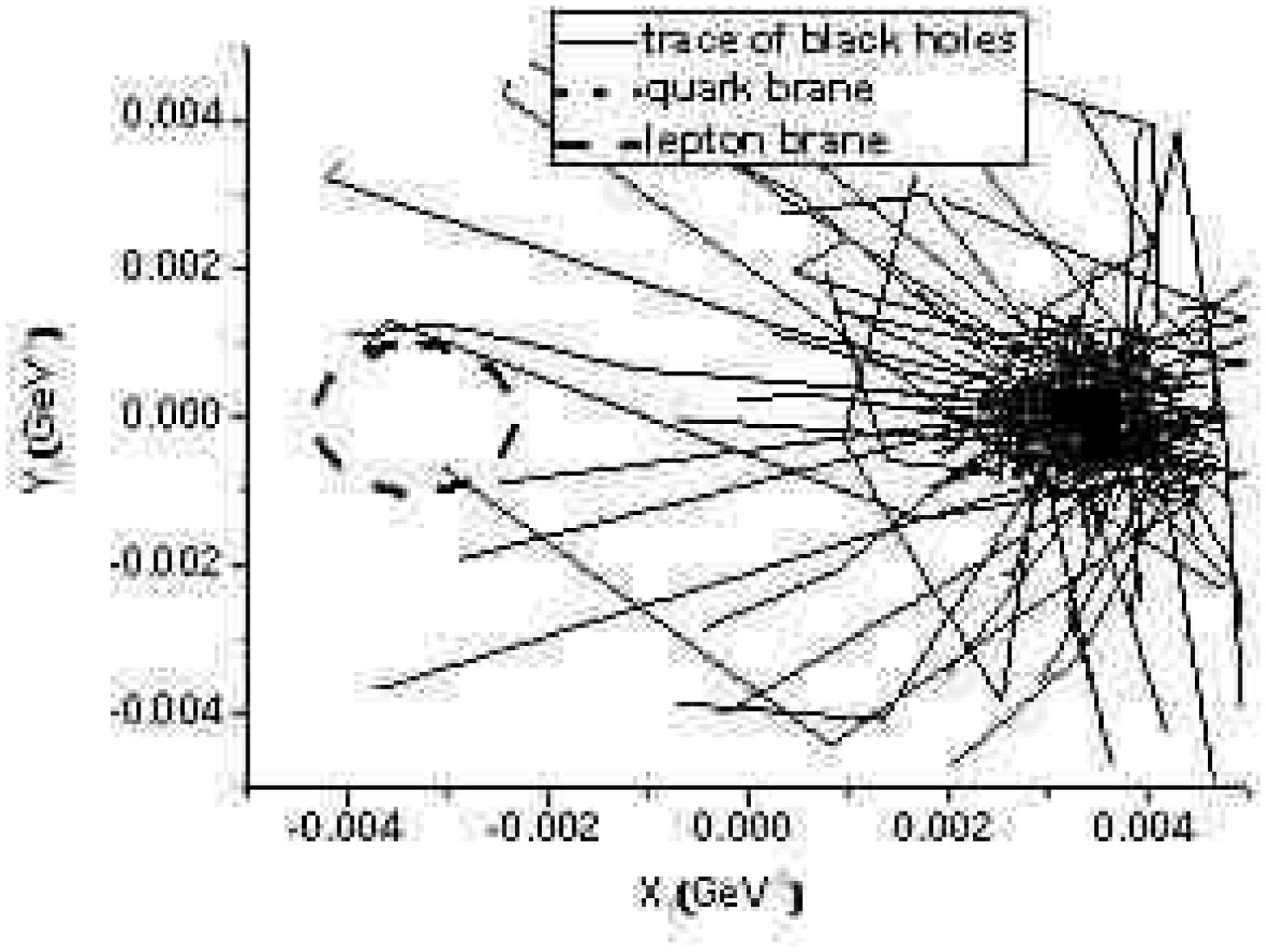}
\caption{black-hole movement in the mini-bulk due to recoil. $X_{1}$, 
$Y_{1}$ are coordinates in two extra dimensions. The red circle indicates the 
width of the quark brane. The blue circle indicates the width of the lepton 
brane. The black lines are black-holes traces. The size of the mini-bulk is 
$10\,\rm{TeV}^{-1} \times 10\,\rm{TeV}^{-1}$. black-holes with non-zero standard 
model gauge charges bounce back from the wall of the mini-bulk. black-holes with 
zero Standard Model gauge charges can leave the mini-bulk.}
\label{fig:bh-trace}
\end{figure}

The generator includes recoil of black-holes due to Hawking radiation. 
The recoil modifies the spectrum due to the Doppler effect. Even if the effect 
is small, the high energy tail of the emitted particle's energy spectrum is 
longer than for pure Hawking radiation. 
Fig. \ref{fig:bh-trace} shows the random motion in the mini-bulk for $10,000$ 
black-holes as consequence of recoil. While most of the black-holes remain on 
the 
brane where they were formed, a significant number of them are capable of 
drifting 
all the way to the lepton brane. There are also a few events where the black 
holes leave the mini-bulk completely. Since Standard-Model charges are 
confined to the mini-bulk, a black-hole needs to carry zero charge in order to 
be 
able to leave the mini-bulk. Once out of the mini-bulk, a black-hole cannot 
emit Standard-Model particles anymore.
Models with an additional bulk $Z_2$ symmetry (e.g. Randall Sundrum 
models) do not allow for a black-hole recoil from the brane 
\cite{Stojkovic:2004hp}.
Unfortunately, the number of black-holes which escape the mini-bulk is so small 
that experimentally we are unlikely to be able to distinguish between 
models on this basis. 

\subsection{Initial Black-Hole Charge Distribution}
\label{subsec:initchargedis}

\begin{figure}[ht]
\centering{
\includegraphics[width=3.2in]{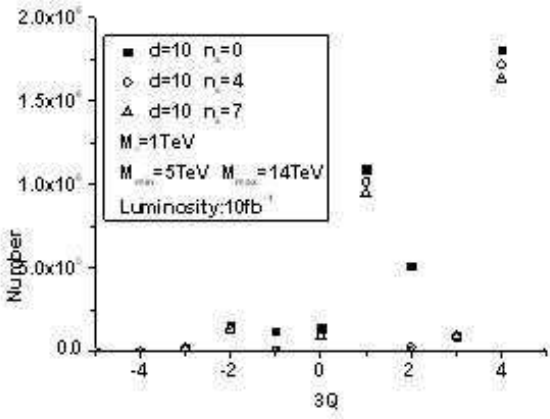} }
\caption{Electromagnetic charge distribution of the initial black-holes.}
\label{fig:black-charge-in}
\end{figure}

Most of the initial black-holes are created by u and d quarks. 
Denote by $N_{3Q}$ the number of black-holes that
have electromagnetic charge $Q$.
Fig. \ref{fig:black-charge-in} is a histogram of $3Q$
for $n_{s}=0$, $n_{s}=4$, and $n_{s}=7$ in $d=10$ space.
Since at these parton momenta,
there are roughly twice as many u-quarks in a proton as d-quarks, 
we expect that most of the black-holes have $3Q=4$.
{\it i.e.} are made of two u-quarks ($f_{uu}$). 
One does indeed see a large peak at $3Q=4$ in figure \ref{fig:black-charge-in}.
A second peak at $3Q=1$ 
corresponds to black-holes made of one d and one u-quark,
or from one anti-d and one gluon. 
Since there are only a few gluons or anti-quarks at these momenta, 
$f_{1}\simeq f_{ud}$. 

Similarly, the small peak at $3Q=-2$ is predominantly $dd$
and not ${\bar {\rm u}}$-gluon.

We expect that $f_{ud}\simeq2\sqrt{f_{uu}f_{dd}}$, and thus 
$f_{1}\simeq2\sqrt{f_{4}f_{-2}}$.
This relation is roughly satisfied in Fig. \ref{fig:black-charge-in}. 
In the split-fermion case, 
since gluons can move in the mini-bulk, 
there is a further suppression of the gluon contribution 
due to the wave-function-overlap suppression between the gluons and fermions. 
In particular $3Q=2$ and $3Q=-1$, 
which are dominated by gluon-quark collisions,
are suppressed, as can be seen in Fig.\ref{fig:black-charge-in}. 
For a large number of split dimensions, 
there are almost no gluon-gluon or gluon-quark black-holes.
The decline in the gluon-quark configurations
accounts for the simultaneous rise in the fraction of
quark-only configurations({\it i.e.} $3Q=4,1,-2$).

\subsection{Initial Black-Hole Color Distribution}
\label{subsec:initdis}

\begin{figure}[ht]
\centering{\includegraphics[width=3.2in]{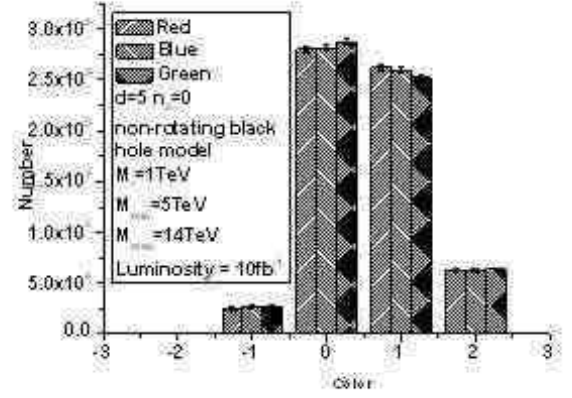} }
\caption{Initial color distribution of the created black-holes. The 
vertical lines are error bar. }
\label{fig:black-color-in}
\end{figure}

The colliding partons that form the black-hole carry gauge charges, 
in particular color and electromagnetic charge. 
From the PDFs \cite{pdfs} we see that, at the relevant parton momentum,
most of the partons are u and d type quarks -- 
essentially the valence quarks.
Contributions from ``the sea'' -- 
other quarks, antiquarks, gluons and other partons -- are subdominant. 
We therefore estimate the distribution of the 
colors of the initial black-holes to be
\be
\label{eqn:colordistnestimate}
N_i(0):N_i(1):N_i(2)=4:4:1 .
\ee
Here $N_i(p)$ is the number of black-holes 
whose i-th color-vector component (i=1 is red,
i=2 blue, i=3 green) has the value $p$.
This agrees very well with the graph in Fig. \ref{fig:black-color-in}. 
$N_i(-1)$ and $N_i(-2)$ refer to black-holes created from 
collisions involving gluons or anti-quarks.
Their numbers are hard to estimate, 
but we expect that $N_i(-2) \ll f_i(-1) \ll N_i(0\leq p \leq 2)$, 
again consistent with Fig. \ref{fig:black-color-in}. 

At energy scales accessible at the LHC, 
the color distribution of black-holes in different brane world models 
does not differ from each other significantly.
However, were $M_{min}$ significantly lower (at or below $1$\,TeV),
then black-hole production by gluon-gluon scattering would be more important,
significantly altering the color distribution,
and making it more sensitive to fermion brane-splitting
(which lowers the gluon-gluon contribution).

\begin{figure}[ht]
\centering{
\includegraphics[width=3.2in]{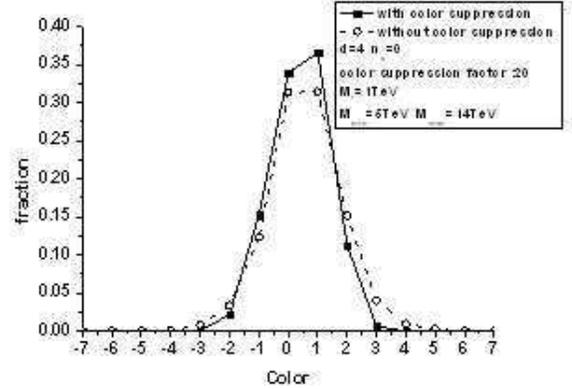} }
\caption{Cumulative color distribution for non-rotating black-holes on a 
tensionless 
brane with $d=4$ and no fermion brane splitting.
Histogram with the black squares (open circles) is with (without) color 
suppression.}
\label{fig:charge-suppress}
\end{figure}

\subsection{Evolution of Black-Hole Color and Charge during the Hawking 
Radiation Phase}
\label{subsec:evolutioncolorcharge}

Figure \ref{fig:charge-suppress} shows the color distribution of the black-holes 
which they accumulate during the evaporation phase. 
From equation \ref{eqn:colordistnestimate},
the expected average initial color of the black-holes is $2/3$.
Since the colors of emitted particles are assigned
randomly, we expect the cumulative color distribution (CCD)
to be symmetric around $2/3$ and peaked at the value.
This is indeed what we find.

The width of the CCD depends on the total number of particles 
emitted by the black-hole during its evaporation phase. 

As discussed above, we allow for the possibility of suppressing
particle emission which increases the charge, color or angular
momentum of the black-hole excessively (cf. discussion around 
equations \ref{eqn:highqandcsuppression}, 
and \ref{eqn:LsuppressiondeltaA}-\ref{eqn:LsuppressiondeltaJ}).
Figure \ref{fig:charge-suppress} shows also the cumulative 
black-hole color distribution
where we suppressed the accumulation of large color charges during the 
evaporation phase. 
In order to amplify the effect of color suppression,
we have set $f_3=20$ instead of the expected $f_3\simeq0.1$.
We see that the number of black-holes with a color charge larger 
than $1$ is decreased.

\subsection{Number of emitted particles}
\label{subsec:numberempart}
\begin{figure}[ht]
\centering{
\includegraphics[width=3.2in]{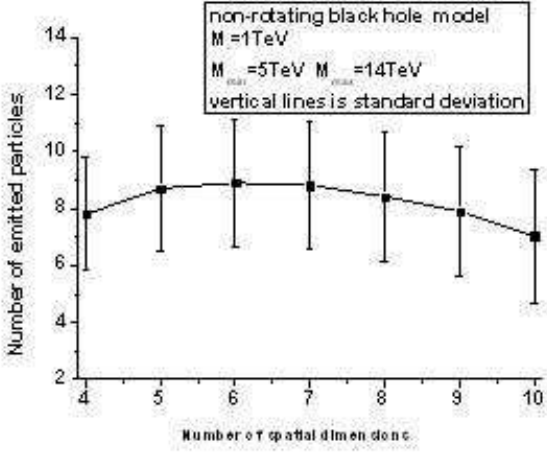} }
\caption{Number of particles emitted by a non-rotating 
black-hole on a tensionless brane prior to the ``final burst'' as function 
of number of 
extra dimension. Here $n_{s}=0$. The error bars denote one standard deviation 
range.}
\label{fig:step-brane}
\end{figure}

\begin{figure}[ht]
\centering{
\includegraphics[width=3.2in]{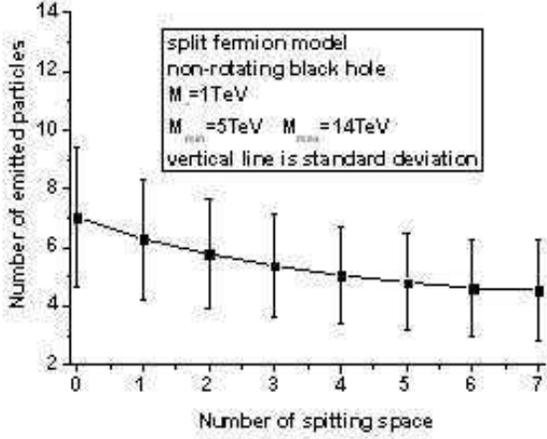} }
\caption{Number of particles emitted by a non-rotating 
black-hole in the split-fermion model prior to the ``final burst'' 
as a function of the number of brane-splitting dimensions, with $d=10$.
Error bars denote one standard deviation range.}
\label{fig:step-splitting}
\end{figure}

\begin{figure}[ht]
\centering{
\includegraphics[width=3.2in]{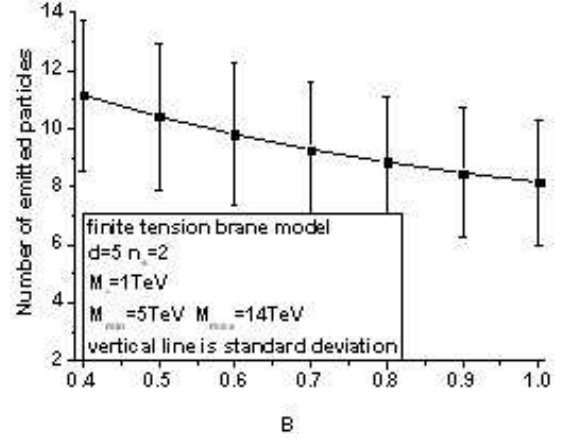} }
\caption{Number of particles emitted by a non-rotating black-hole on a 
non-zero tension brane prior to the ``final burst'' as function of defizit angle 
parameter $B$ with $d=5$ and $n_{s}=2$. Error bars denote one standard deviation 
range.}
\label{fig:step-tension}
\end{figure}

\begin{figure}[ht]
\centering
\includegraphics[width=3.2in]{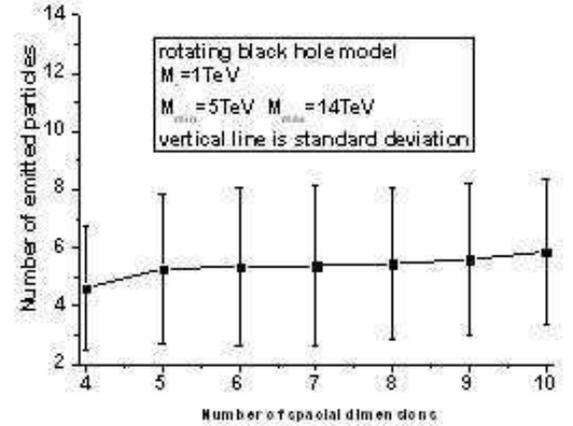}
\caption{Number of particles emitted by a rotating black-hole prior to the 
``final burst'' as function of number of extra dimension with $n_{s}=0$.
Error bars denote one standard deviation range.}
\label{fig:step-rotation}
\end{figure}

\begin{figure}[ht]
\centering
\includegraphics[width=3.2in]{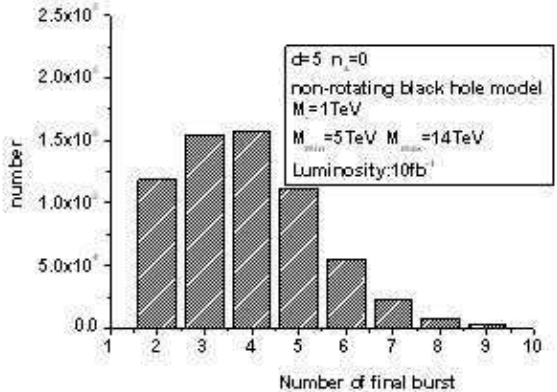}
\caption{Number of particles emitted at the final burst: the average number 
of final burst is about 3.4.}
\label{fig:number-final}
\end{figure}

Figures \ref{fig:step-brane} through \ref{fig:step-rotation} 
show the number of particles that are emitted by a microscopic black-hole 
during the decay process before its final burst for a variety of models. 

In the single tensionless brane model (Fig. \ref{fig:step-brane}), 
the number of emitted particles first increases with 
the number of dimensions for a non-rotating black-hole, but then decreases. 
This behaviour is a result of the complicated interplay
of a number of effects: the horizon size of a black-hole of a given
mass as a function of $d$, and its effect on the particle emission
spectra; the dependence of the Hawking temperature on $r_h^{(d)}$;
the existence due to energy-momentum conservation 
of an upper limit of $M^{bh}/2$ on the energy of an emitted particle.
The location of the peak will shift as a function of the
input parameter $M_{min}$, the minimum initial mass of a black-hole.

In the split fermion model (Fig. \ref{fig:step-splitting}),
the number of emitted particles decreases with the number of extra dimensions. 
This is because, even for a fixed Hawking temperature,
the average energies of emitted gauge bosons and scalar fields
increase as $n_s$ increases. 

In the model with a finite-tension brane (Fig. \ref{fig:step-tension}),
the number of particles decreases as the parameter $B$ increases,
{\it i.e.} with decreasing tension.
As the tension increases, $B$ gets smaller but
the horizon radius of the black-hole increases. 
The Hawking temperature therefore decreases, and,
as a consequence, the average energy of emitted particles falls.
More particles will therefore be emitted in the evolution of the black-hole.

For rotating black-holes (Fig. \ref{fig:step-rotation})
(on a tensionless, unsplit brane),
the number of emitted particles first increases, then decreases,
and finally reaches a plateau. This is due to similar reasons as for 
non-rotating black-holes.
Compared to non-rotating black-holes of the same mass,
rotation shifts the energy of emitted particles 
to higher values because it decreases the horizon radius
and increases the emission of higher angular-momentum modes.
This decreases the total number of emitted particles.
It also means that the effect of the upper kinematic limit of $M^{bh}/2$
on the emitted particle's energy is magnified. 

Figure \ref{fig:number-final} shows the number of particles that are emitted at 
the final burst stage for a non-rotating black-hole on a tensionless brane with 
$d=5$ and $n_s = 0$. The average number of emitted particles is about $3$. 
During the Hawking radiation phase a black-hole emits about 10 particles, 
so approximately 30\% of the emitted particles will be from the final burst 
stage. 
In the case examined in Figure \ref{fig:number-final}, 
we did not include suppression of large black-hole color or electric charge. 
Thus some black-holes acquire large color and electric charges by
the end of the Hawking radiation phase. These black-holes then
must decay into a large number of particles ($>5$) in the final burst.

\subsection{Energy Distributions of the Emitted Particles}
\label{subsec:energy}

Once formed, the black-holes decay by emission of Hawking radiation,
a process which continues until the mass of the black-hole falls to 
the fundamental quantum-gravity scale. At this stage we chose the black-holes to 
burst into a set of Standard-Model particles as described in 
section\,\ref{sec:finalburst}. 
The observable signatures of the decay will depend on the
distributions of energy, momentum and particle types of the emitted particles.

\begin{figure}[ht]
\centering
\includegraphics[width=3.2in]{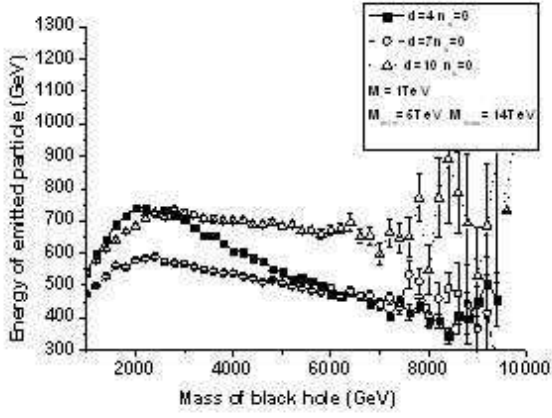}
\caption{
Average energy of the particles emitted 
by (non-rotating) black-holes on unsplit branes
versus the mass of the black-hole at the time of emission.
Here $d=4$, $d=7$ and $d=10$.
Note that the final burst is not included,
because it occurs when the mass of the black-hole 
is less than 1 TeV. }
\label{fig:Energy-Mass-brane}
\end{figure}

\begin{figure}[ht]
\centering
\includegraphics[width=3.2in]{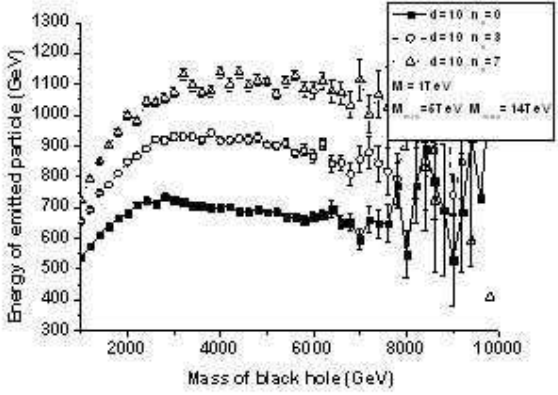}
\caption{Average energy of the particles emitted 
by (non-rotating) black-holes on split branes
versus the mass of the black-hole at the time of emission.
Here $d=10$ and $n_s=0$, $n_s=3$, $n_s=7$. }
\label{fig:Energy-Mass-splitting}
\end{figure}

\begin{figure}[ht]
\centering
\includegraphics[width=3.2in]{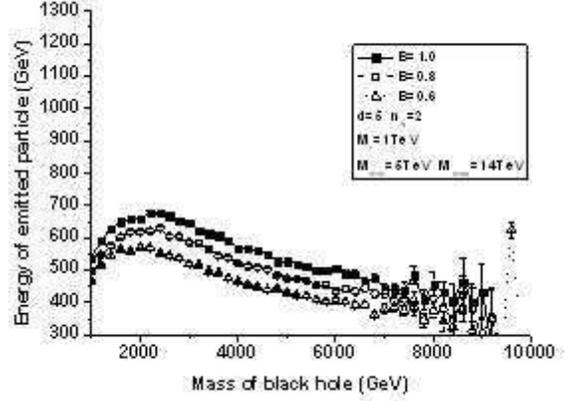}
\caption{Average energy of the particles emitted
by (non-rotating) black-holes on a brane with tension,
versus the mass of the black-hole at the time of emission.
Here $d=5$, $n_s=2$ and $B=1$, $B=0.8$, $B=0.6$. }
\label{fig:Energy-Mass-tension}
\end{figure}

\begin{figure}[ht]
\centering
\includegraphics[width=3.2in]{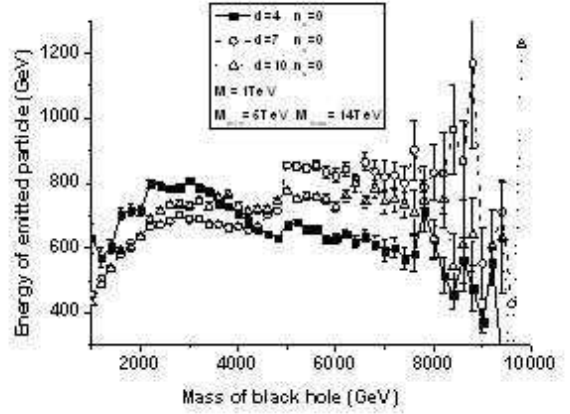}
\caption{Average energy of the particles emitted
by (possibly rotating) black-holes 
versus the mass of the black-hole at the time of emission. 
Here $d=4$, $d=7$ and $d=10$, with a tensionless unsplit brane.
}
\label{fig:Energy-Mass-rotation}
\end{figure}

Figures \ref{fig:Energy-Mass-brane} through \ref{fig:Energy-Mass-rotation} 
show the relation between the mass of the evolving black-hole and the
average energy of emitted particles for different extra dimension models. 
The error bars denote $1/\sqrt{N}$ times the standard deviation of the mean 
energy.
The minimum mass of the initial black-hole is taken to be $M_{min}=5$\,TeV. 

We see from figure \ref{fig:Energy-Mass-brane} that, 
for a single (i.e. unsplit) brane, when $M_{bh} >> M_*$. 
a black-hole in $d=10$ emits higher energy particles 
than a black-hole in lower dimensions.
For black-hole masses closer to $M_*$, 
the highest energy particles are emitted when the dimensionality
of space is low, i.e. $d=4$. 
This reversal can be understood from figure \ref{fig:THawk}. 
In the LHC energy range, 
the curves of Hawking temperature as a function of black-hole mass 
for different dimensions cross.
At high mass, high $d$ exhibits the highest Hawking temperature;
at low mass, low $d$ exhibits the highest Hawking temperature.
It is easy from this figure 
to estimate the number of emitted particles,
and to roughly reproduce the results shown 
in figure \ref{fig:step-brane}.

The main difference between the curves for different $d$,
comes from the changing size of the black-hole horizon.
For low $d$, 
the horizon radius increases more quickly with the mass 
than for higher $d$, as seen in figure \ref{fig:rhorizon}. 
The Hawking temperature of the black-hole 
is inversely proportional to $r_h$. 
So long as the Hawking temperature remains well below
the black-hole mass (here for $M_{bh}\gapp 2$TeV),
the energy of emitted particles decreases as the mass increases. 
However as equations (\ref{eqn:horizon2}-\ref{eqn:horizon3}) show, 
this change is slow for high $d$.
By $d=10$,
the energy of emitted particles is almost constant 
from $2TeV$ to $5TeV$.

The increase of the energy of emitted particles stops at about $2$\,TeV 
and then a decrease begins. 
The reason, as stated above (equation (\ref{eqn:M-max})), is that by 
energy-momentum conservation,
a black-hole can only emit particles 
with less than half of its mass. 

In the split-fermion model (Fig. \ref{fig:Energy-Mass-splitting}),
the average energy of the emitted particles increases
as the number of dimensions in the mini-bulk increases. 
The energy shift comes from the gauge bosons and scalar fields, 
which access the higher-dimensional phase space of the mini-bulk 
(cf. appendix A: figure \ref{spectra:s-2} through \ref{spectra:ga-7}).

For the brane with non-zero tension (Fig. \ref{fig:Energy-Mass-tension}), 
the radius of the black-hole increases with tension 
hence the energy of emitted particles decreases with tension. 

For a rotating black-hole (Fig. \ref{fig:Energy-Mass-rotation}), 
angular momentum decreases the size of the horizon. 
Thus since black-holes are typically formed with 
some initial angular momentum, 
they emit higher energy particles than non-rotating black-holes
of the same mass.
However, the black-hole tends to shed its angular momentum rapidly
as it emits particles. This increases the horizon size,
lowers the Hawking temperature, and lowers the average
energy of the emitted particles. The rapid shedding
of angular momentum thus leads to a drop in the 
average emitted particle energy around $M_{\rm min}$.

If one compares rotating with non-rotating black-holes, 
one finds that the energy of the emitted particles is 
always larger for the rotating black-holes. 
By the time the mass of the black-hole has 
dropped well below $M_{\rm min}$
(here to approximately $1-2$\,TeV), 
almost all of the angular momentum has been lost and 
the difference between the rotating and 
non-rotating black-holes is small.

\begin{figure}[ht]
\centering
\includegraphics[width=3.2in]{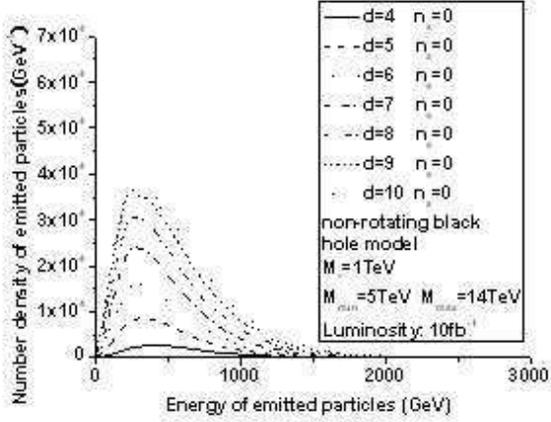}
\caption{Energy distribution of emitted particles in the Hawking radiation 
step for single-brane non-rotating black-hole.}
\label{fig:E-emitted-brane-without}
\end{figure}
\begin{figure}[ht]
\centering
\includegraphics[width=3.2in]{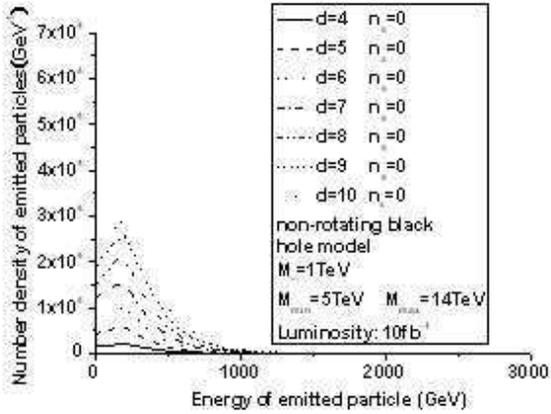}
\caption{Energy distribution of emitted particles at the final burst step 
for single-brane non-rotating black-hole.}
\label{fig:E-emitted-brane-final}
\end{figure}

As a Standard-Model particle is emitted from a black-hole, 
this particle may carry some momentum in the extra dimensional
directions. 
As an observer in 3-space observes this particle, 
he/she will find the apparent mass of the particle is 
\begin{equation}
M_{ob}=\sqrt{M_{p}^{2}+\sum_{e}P_{e}^{2}}
\end{equation}
Here $M_{ob}$ is the observed mass, 
$M_{p}$ is the true mass of the particle,
and $P_{e}$ is the particles extra-dimensional momentum. 
Clearly $M_{ob} \geq M_{p}$. 
A Standard-Model particle, however, 
cannot leave the Standard Model brane. 
Its extra-dimensional momentum 
must therefore be absorbed by the brane
or carried away by bulk particles (such as gravitons).
We therefore calculate an emitted particle's 
``energy on the brane'' according to:
\begin{equation}
E_{b}=\sqrt{M_{p}^{2}+\sum_{i=1}^3P_{i}^{2}}
\end{equation}
where $P_{i}$ is the regular 3-momentum.
We will assume that the shedding of extra-dimensional
momentum is rapid, and {\bf henceforth} we
will refer to $E_{b}$ (rather than the initial emission
energy) as the energy of the emitted particle.

Figure \ref{fig:E-emitted-brane-without} shows the energy distribution of the 
particles from Hawking radiation in the single-brane model. 
The cross section for black-hole production increases with $d$ (figure \ref{fig:cross-d}).
The area under the curves also increase with $d$. 
The peaks of the curves are around $200$GeV to $400$GeV. 
One can compare, for example, the energy distributions for $d=9$ and $d=10$. 
A black-hole in high r $d$ tends to emit particles with higher energy, 
so the curve for $d=10$ has a longer higher energy tail than for $d=9$.

Figure \ref{fig:E-emitted-brane-final} shows the energy distribution of the 
particles from the final burst in the single-brane model. 
The energy these  particles share is much smaller than the energy in the
earlier Hawking radiation phase. 
The peak in the energy distribution of these particle is
around $200$GeV to $300$GeV. 
The tails extend just to $1$TeV, 
which is the mass at which the  black hole is taken to be 
unstable and undergo its final burst. 

\begin{figure}[ht]
\centering
\includegraphics[width=3.2in]{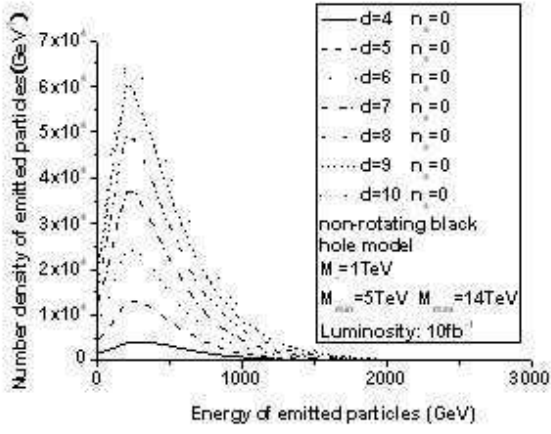}
\caption{Energy density distribution of emitted particles from non-rotating 
black-holes on a tensionless brane with no splitted fermion branes. Shown are 
the distributions for $4 \le d \le 10$. The spectra include the final burst 
particles.}
\label{fig:E-emitted-brane}
\end{figure}

\begin{figure}[ht]
\centering
\includegraphics[width=3.2in]{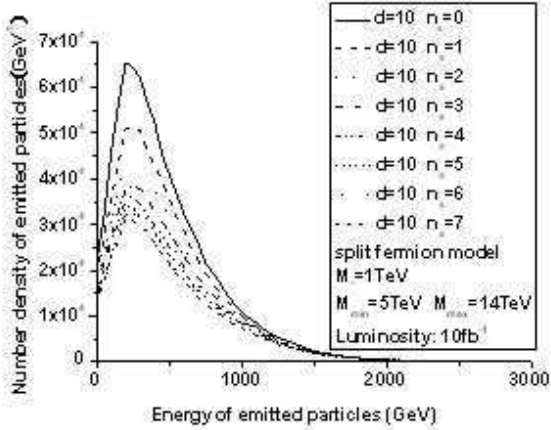}
\caption{Energy density distribution of emitted particles from non-rotating 
black-holes on a tensionless brane with fermion brane splitting. Shown are the 
distributions for $d=10$ and $0 \le n_s \le 7$. The spectra include the final 
burst particles.}
\label{fig:E-emitted-splitting}
\end{figure}

\begin{figure}[ht]
\centering
\includegraphics[width=3.2in]{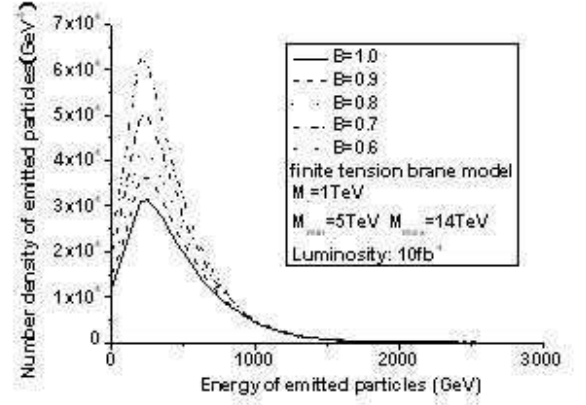}
\caption{Energy distribution of emitted particles for non-rotating black 
holes on a non-zero tension brane with $B=1$, $B=0.8$ and $B=0.6$. The spectra 
include the final burst particles.}
\label{fig:E-emitted-tension}
\end{figure}

\begin{figure}[ht]
\centering
\includegraphics[width=3.2in]{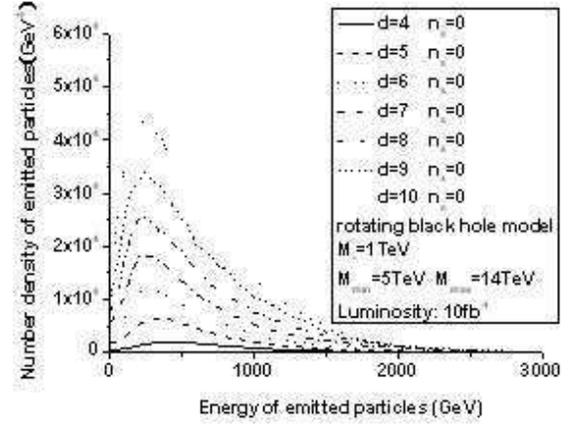}
\caption{Energy distribution of emitted particles for rotating black-holes 
for $4 \le d \le 10$.}
\label{fig:E-emitted-rotation}
\end{figure}

Figures \ref{fig:E-emitted-brane} through \ref{fig:E-emitted-rotation} 
show the energy distribution of emitted particles 
(including final burst particles) in the various 
models that BlackMax can simulate. 

Figure \ref{fig:E-emitted-brane} shows the energy distribution in the single-brane model. 
The cross section of black-hole increases with $d$ (figure \ref{fig:cross-d}).
The area under the curves also increases with $d$. 
The peaks of the curves are around $200\,$GeV to $400\,$. 
Again comparing $d=9$ and $d=10$, a black-hole in higher $d$ 
tends to emit particles with higher energy, 
so the curve for $d=10$ has a longer high energy tail than for $d=9$.

Figure \ref{fig:E-emitted-splitting} illustrates the split-fermion model.
In this figure, we keep the total number of dimensions $d$ fixed
but change the dimensionality $n_{s}$ of the mini-bulk.
This affects the spectra of only the gauge boson and scalar fields,
as only their propagation is affected by the the mini-bulk's dimensionality. 
(Gravitons propagate in the full bulk; other Standard-Model particles
propagate only on the brane.)
As explained above, these spectra will shift to higher energies 
as the number of splitting dimensions is increased.
One can see that the curve in $n_{s}=7$ has the longest high energy tail.

Figure \ref{fig:E-emitted-tension} illustrates the brane with tension model. 
The energies of the emitted particles shift to lower energy as B decreases. 

The energy distribution of emitted particles for rotating black-holes 
(figure \ref{fig:E-emitted-rotation}) 
has the same general characteristics 
as the distribution for non-rotating black-holes. 
Angular momentum causes a black-hole to tend to emit higher energy particles than 
a non-rotating black-hole. The curves have longer higher energy tails than 
the non-rotating black-holes.


\begin{figure}[ht]
\centering
\includegraphics[width=3.2in]{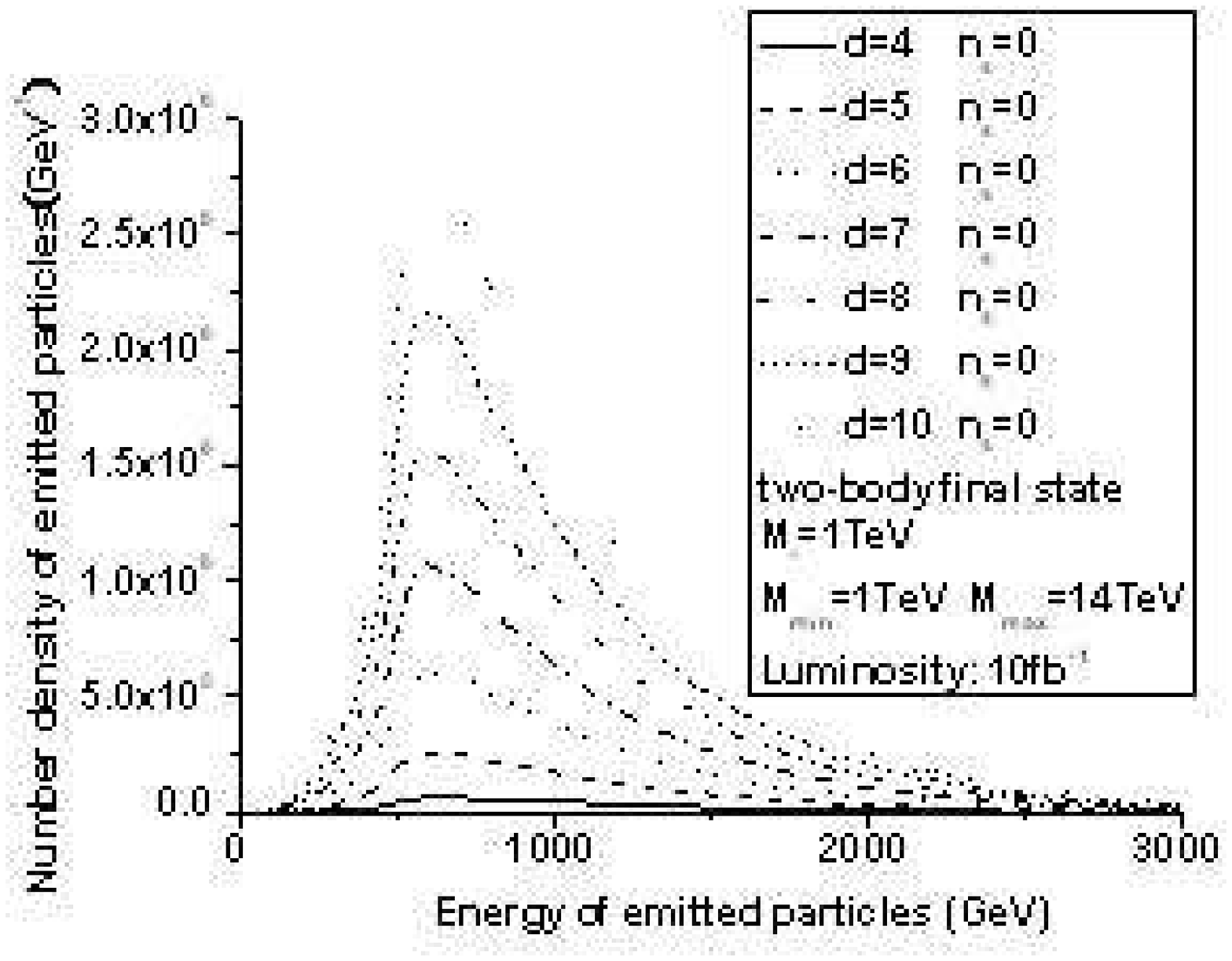}
\caption{Energy distribution of emitted particles for two-body final 
states.}
\label{fig:spectrum-lisa}
\end{figure}

Figure \ref{fig:spectrum-lisa} shows the Energy distribution of emitted 
particles from two-body final-states scenario. 
The energy of the emitted particles is about the half of the incoming partons. 

\begin{figure}[ht]
\centering
\includegraphics[width=3.2in]{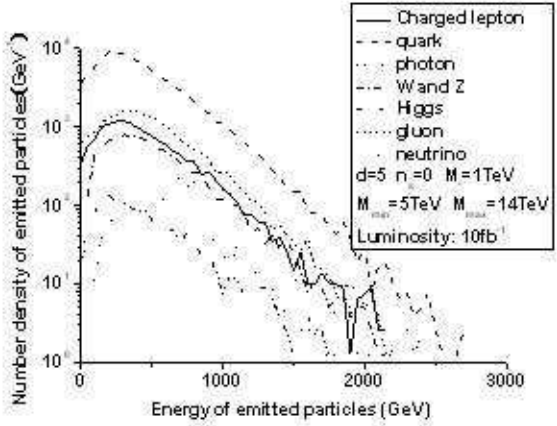}
\caption{Energy distribution of each particle type in the $d=5$ single 
brane model.}
\label{fig:E-emitted-type}
\end{figure}

In figure \ref{fig:E-emitted-type} we show the energy distribution of 
different types of particles in the $d=5$ single-brane model.
The area of each curve is dependent on the degree of freedom of each particle 
and its power spectrum. One can compare the ratio of the same type of particles. 
For example, the area of gluons should be 8 times as large as the are a of 
photons. It is roughly the same as what the figure shows.

%
\subsection{Pseudorapidity Distributions of the Emitted Particles}
\label{subsec:eta}

\begin{figure}[ht]
\centering
\includegraphics[width=3.2in]{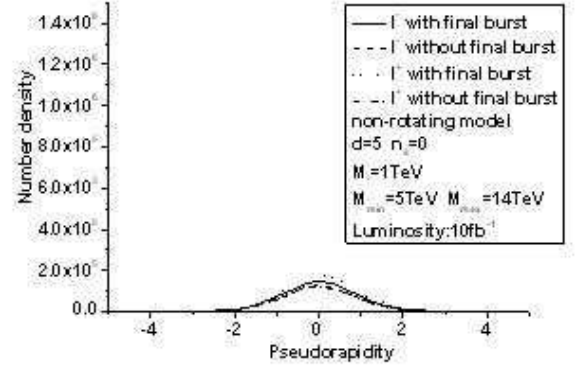}
\caption{Pseudorapidity distribution of charge leptons and anti leptons for 
non-rotating black-holes on a tensionless brane with and without the final burst 
particles; $d=5$.}
\label{fig:eta-e-brane}
\end{figure}

\begin{figure}[ht]
\centering
\includegraphics[width=3.2in]{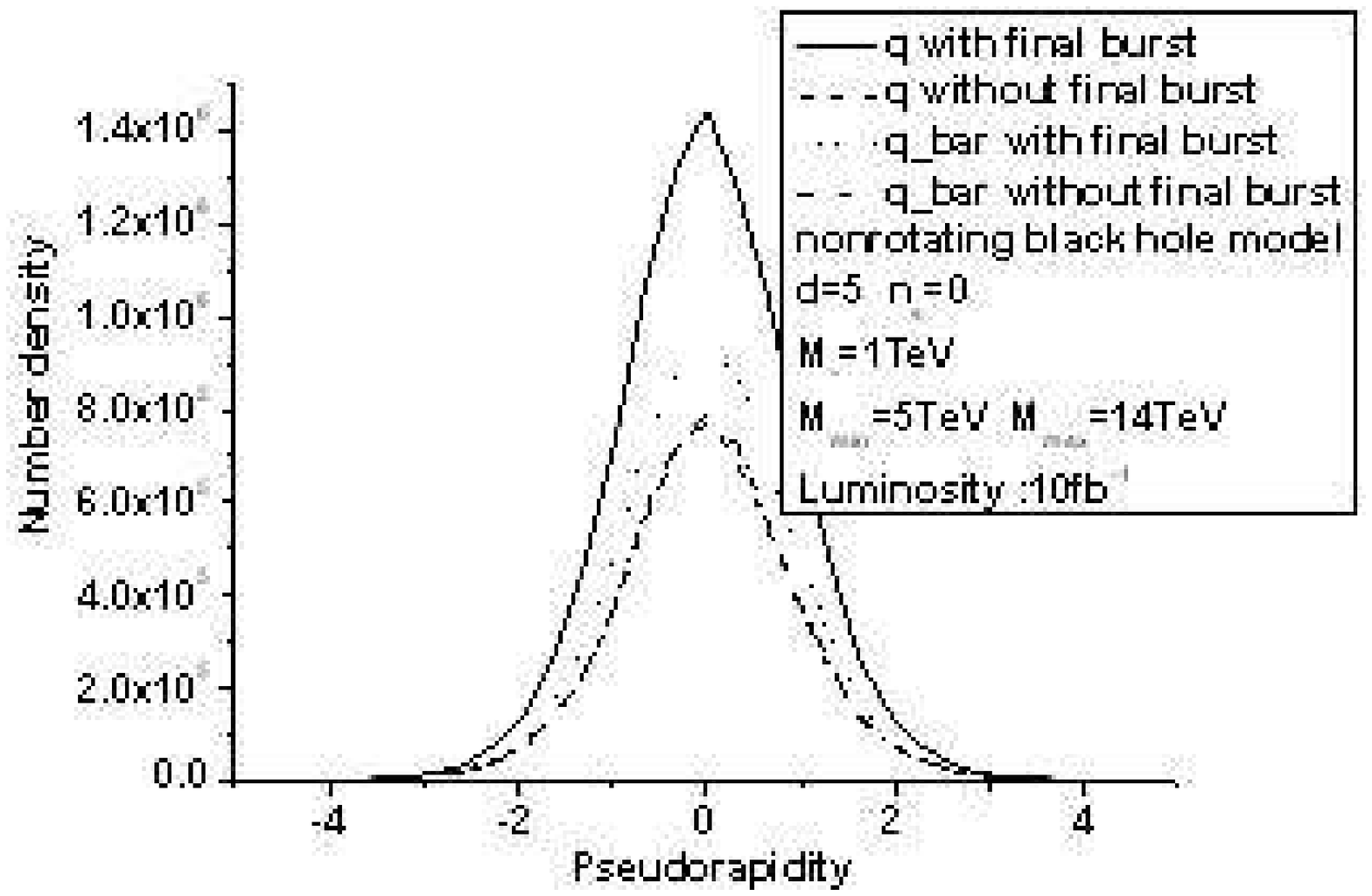}
\caption{Pseudorapidity distribution of quarks and anti quarks for 
non-rotating black-holes on a tensionless brane with and without the final burst 
particles; $d=5$.}
\label{fig:eta-q-brane}
\end{figure}

\begin{figure}[ht]
\centering
\includegraphics[width=3.2in]{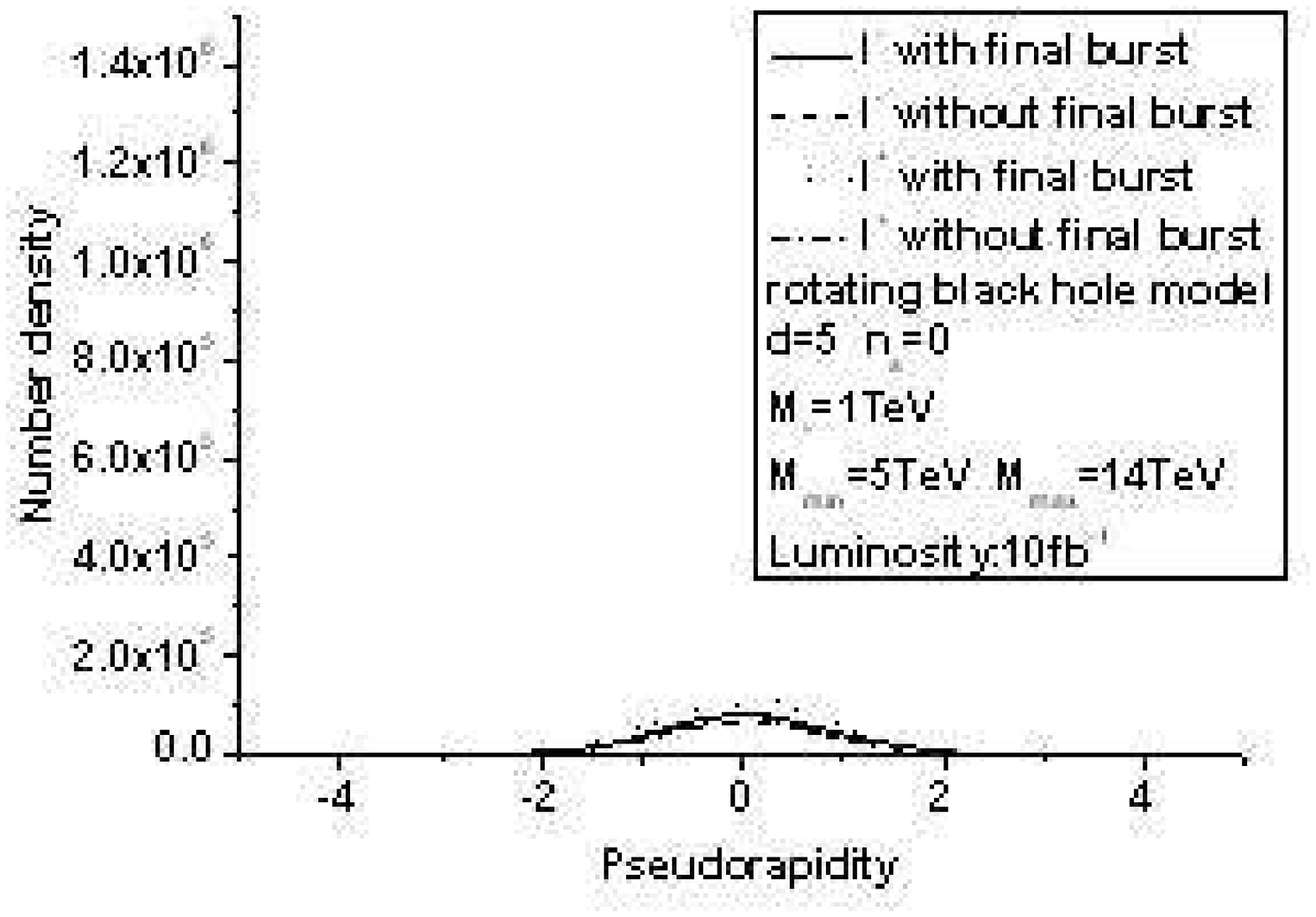}
\caption{Pseudorapidity distribution of charged leptons and anti-leptons 
for rotating black-holes, on a tensionless brane,
with and without the final burst particles. Here $d=5$.}
\label{fig:eta-e-rotation}
\end{figure}

\begin{figure}[ht]
\centering
\includegraphics[width=3.2in]{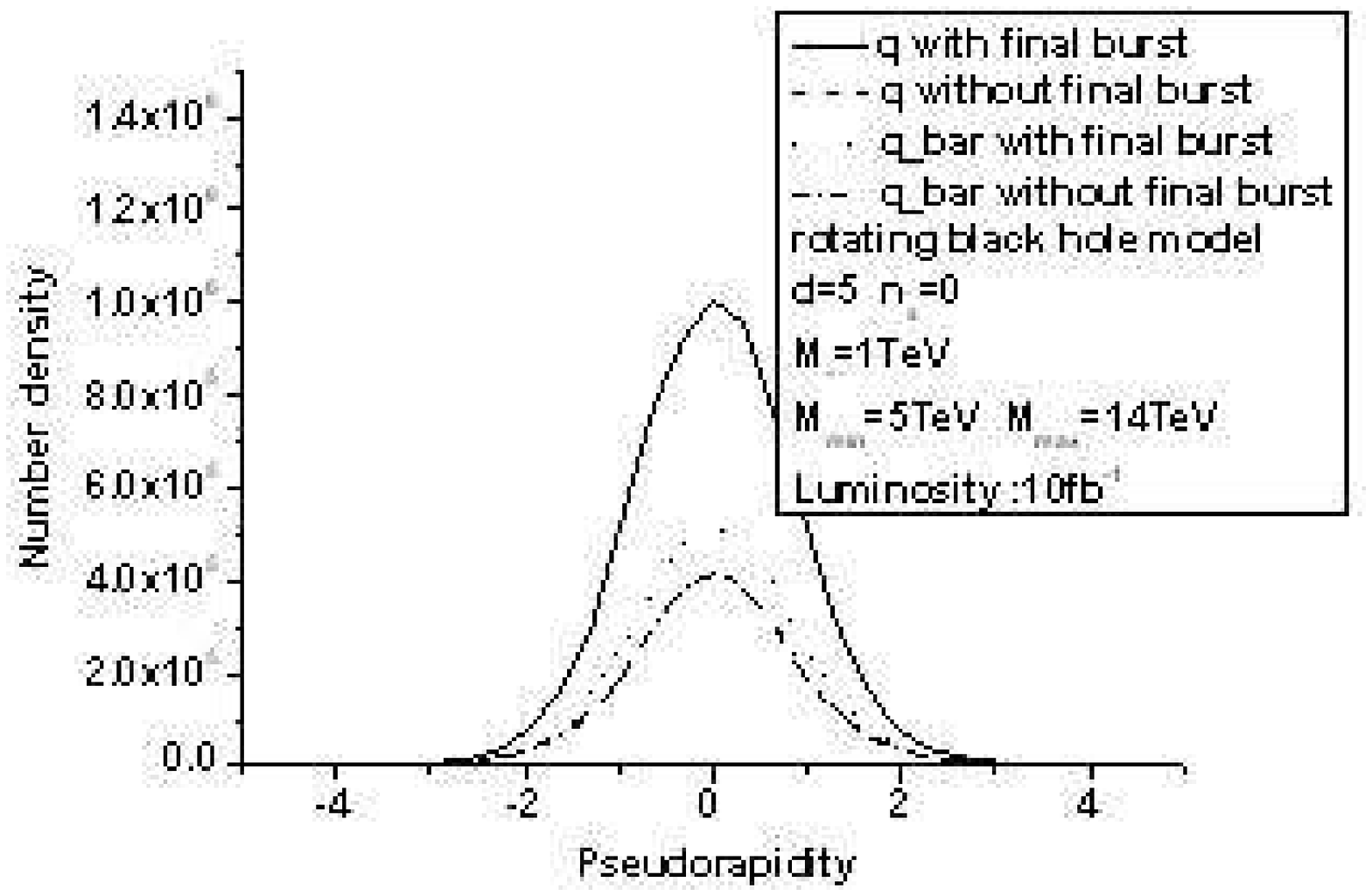}
\caption{Pseudorapidity distribution of quarks and anti-quarks 
for rotating black-holes, on a tensionless brane,
with and without the final burst particles. Here $d=5$.}
\label{fig:eta-q-rotation}
\end{figure}

Figures\,\ref{fig:eta-e-brane} to \ref{fig:eta-q-rotation} show the 
pseudorapidity distributions of the emitted particles for different extra 
dimension scenarios. 


Most of black-holes are made of two $u$ quarks and have charge $3/4$. 
For the shown figures we did not include charge suppression, because of that the 
black-holes tend to emit the same number 
of particles and antiparticles during Hawking radiation phase. This can be seen 
from the curves without final burst in figure \ref{fig:eta-e-brane} through 
figure \ref{fig:eta-q-rotation}. 
The majority of the black-holes are positive charged and will tend to emit 
positive particles in the final burst. That is why there are more positrons than 
electrons for the distributions which include the final burst particles. 

\begin{figure}[ht]
\centering
\includegraphics[width=3.2in]{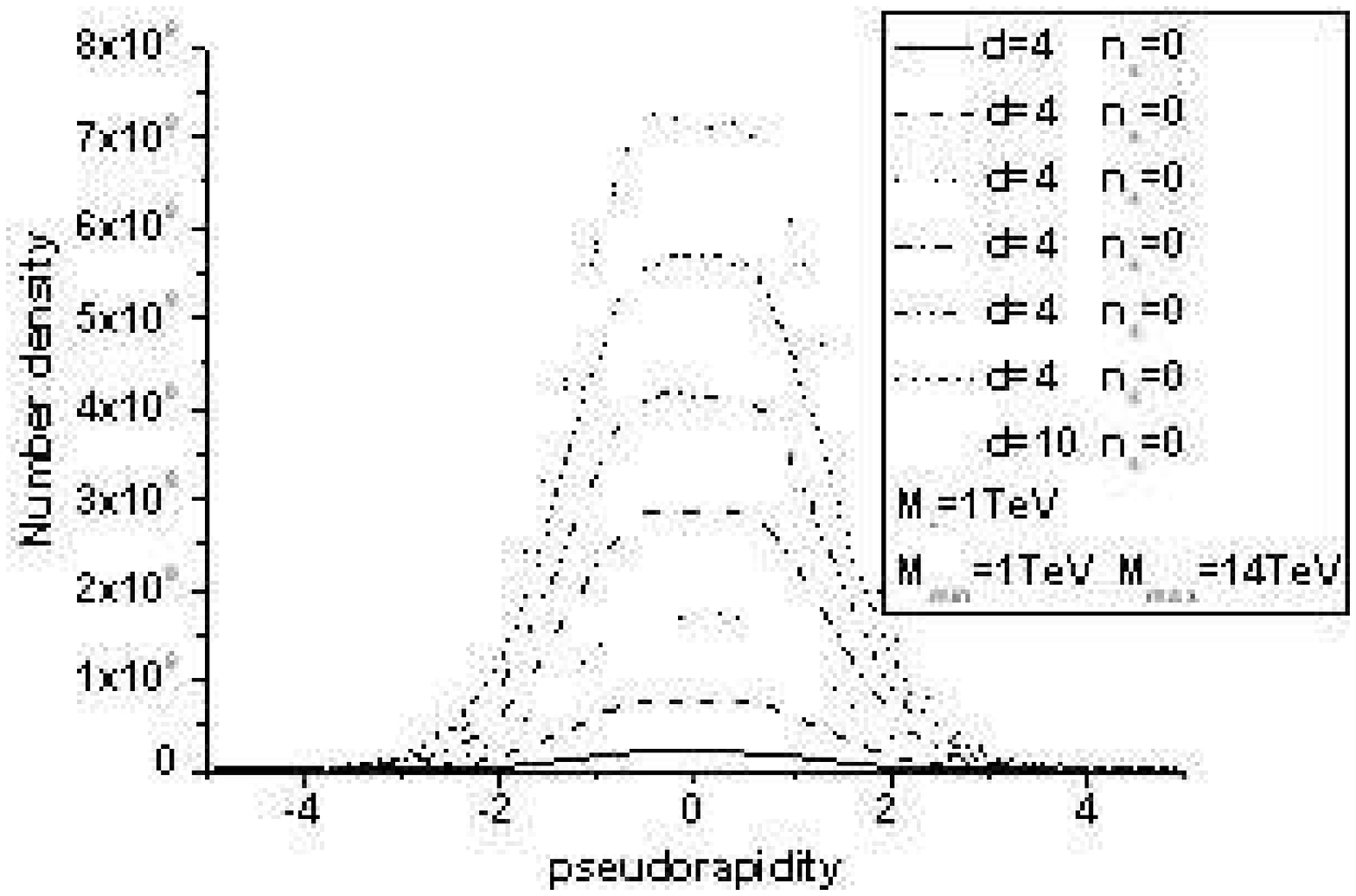}
\caption{Pseudorapidity distribution 
for the two-body final-state scenario. 
The distribution in 2-body final states 
is much flatter near $\eta =0$ 
than emission due to Hawking radiation.}
\label{fig:eta-lisa}
\end{figure}

Figure \ref{fig:eta-lisa} shows the pseudorapidity distribution in the two-body 
final-state scenario. The distribution is much 
wider than the equivalent distribution from Hawking radiation. 
In this model we do not consider the angular momentum 
of the black hole. We therefore take the decay process in the two-body final state 
scenario to be isotropic in the ordinary spatial directions,
just as in other models without angular momentum.
In the center of mass frame, particles are therefore emitted in directions
uncorrelated with the beam direction. Nevertheless, because the threshold
energy of the two-body final-state model is much lower than that of other models,
the intermediate  state tends to have a higher velocity down the beam-pipe.  
The decay products  are therefore emitted with larger pseudorapidity. 
In truth, one may expect that the intermediate state of the
two-body final-state scenario has non-negligible angular momentum.
However, since the intermediate state is not a real black hole, 
it is unclear exactly what role the angular momentum of the intermediate 
state plays.

In the final state scenario,
the momenta of the two emitted particles are 
correlated with the initial parton momenta,
and hence the pseudorapidity distribution of the emitted particles
reflects that of the initial partons.
The ratio of the number of events for pseudorapidity between 
$0$ and $0.5$ divided by that for pseudorapidity between 
$0.5$ and $1$ is about $1.1$. 
This is much higher than the asymptotic QCD value of 0.6, 
as predicted by \cite{Meade,newref}. 
If the ratio is found not to equal $0.6$, 
then this would suggest new physics beyond the Standard Model. 

\subsection{Emitted Particle Types}
\label{subsec:parttypes}

Table \ref{tab:particle-ratio} shows, for a variety 
of representative extra-dimension scenarios the fraction of emitted
particles which are of each possible type -- quarks, gluons,
(charged) leptons, (weak) gauge bosons, neutrinos, gravitons, 
Higgs bosons and photons.
One notable feature is that 
the intensity of gravitons relative to other particles 
increases with the number of extra dimensions. 
Note that the absence of gravitons in the case of a rotating
black-hole is {\em not} physical, but rather reflects our
ignorance of the correct gray-body factor.

\section{Conclusion}
\label{sec:conclusion}

Hitherto, black-hole generators for the large-extra-dimension
searches at the LHC have made many simplifying assumptions 
regarding the model of both our three-dimensional 
space and the extra-dimensional space, and simplifying
assumptions regarding the properties of the black-holes
that are produced. In this paper we have discussed a 
new generator for black-holes at the LHC, BlackMax, which 
removes many of these assumptions. With regard to the
extra-dimensional model it allows for brane tension, 
and brane splitting. With regard to the black-hole,
it allows for black-hole rotation, charge (both electro-magnetic
and color) and bulk recoil. It also introduces
the possibility of a two-body final state that is
not a black-hole.

Although BlackMax represents a major step forward,
there remain important deficiencies that will need
to be addressed in the future. BlackMax continues
to insist on a flat geometry for the bulk space,
whereas there is considerable interest in a warped
geometry \cite{RS} or in a compact hyperbolic
geometry \cite{CHMADD}. While BlackMax allows
for black-hole rotation, the absence of either
an analytic or a numerical gray-body factor for the
graviton in more than three space dimensions
for rotating black-holes is a {\em serious} shortcoming
that can be expected to materially change the 
signature of black-hole decay for rotating black-holes.
Other issues include how to properly account 
for the likely suppression of decays that
cause a black-hole to acquire very ``large'' color,
charge or angular momentum. (As oppposed to the
somewhat contrived phenomenological approach
currently taken.) These are but a few of the
fundamental issues that remain to be clarified.

Despite these (and no doubt other) shortcomings, 
we expect that BlackMax will allow for a much improved
understanding of the signatures of black-holes
at the LHC. Work in progress focuses on using
BlackMax to explore the consequences for the ATLAS 
experiment of more realistic black-hole and 
extra-dimension models.

We thank Nicholas Brett for discussions in the early stages of this paper.
We thank Daisuke Ida, Kin-ya Oda, and Seong Chan Park for providing some 
of the spectra of rotating black-holes.
DCD, DS and GDS thank Oxford's Atlas group for its hospitality at various
stages of this project. DCD, DS and GDS have been supported in part by
a grant from the US DOE; GDS was supported in part by the John Simon Guggenheim
Memorial Foundation and by Oxford's Beecroft Institute for Particle Astrophysics
and Cosmology.

\vskip 1 in
\appendix{Appendix A: Particle Emission Spectra}
\label{app:PES}

\begin{figure}[h]
\centering
\includegraphics[width=3.2in]{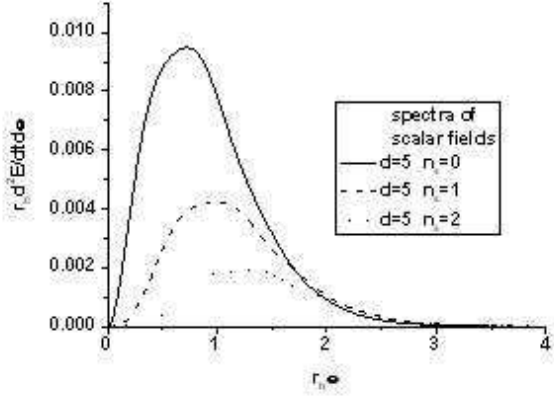}
\caption{Spectra of scalar fields in d=5 space.}
\label{spectra:s-2}
\end{figure}
\begin{figure}[h]
\centering
\includegraphics[width=3.2in]{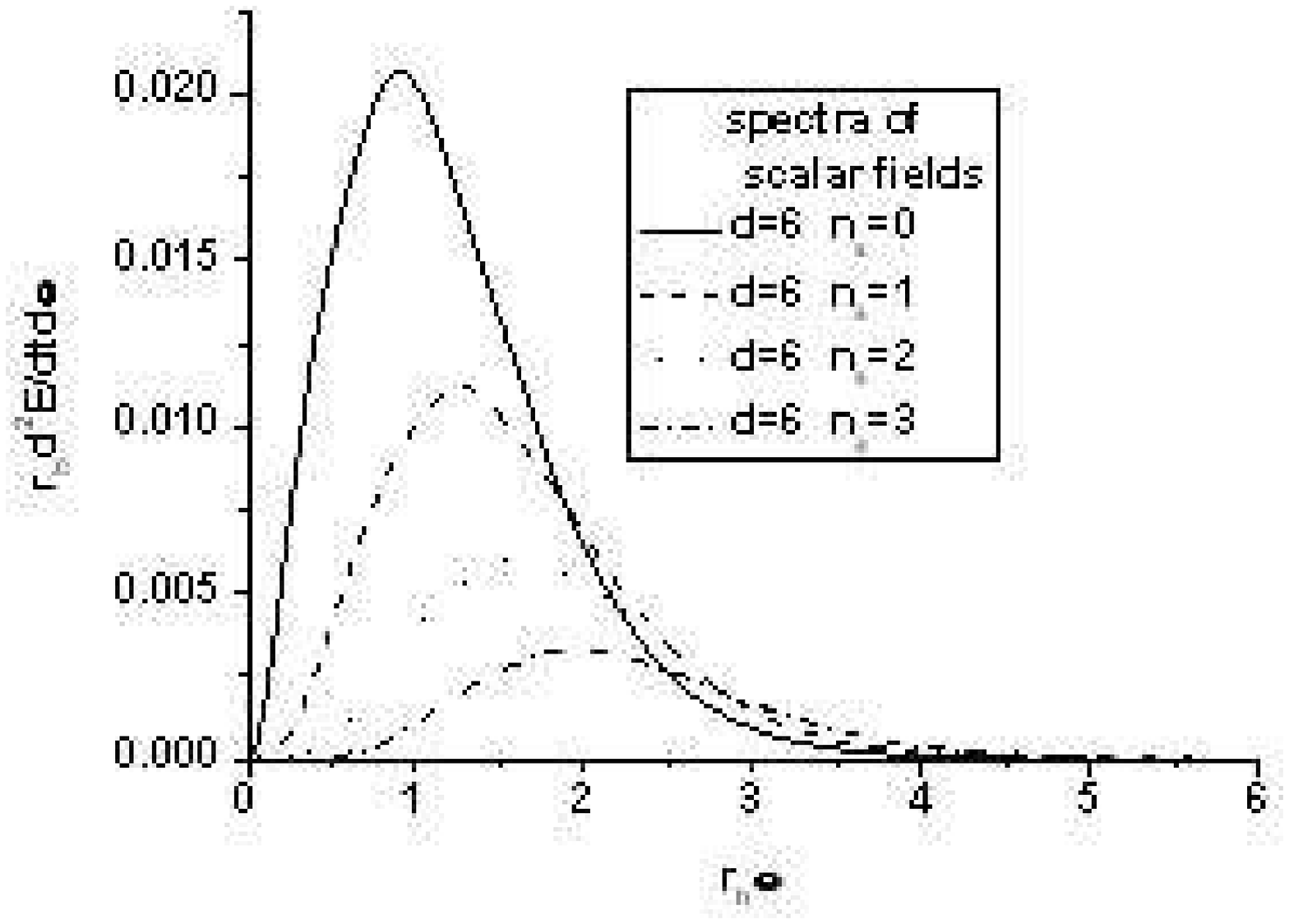}
\caption{Spectra of scalar fields in d=6 space.}
\label{spectra:s-3}
\end{figure}
\begin{figure}[h]
\centering
\includegraphics[width=3.2in]{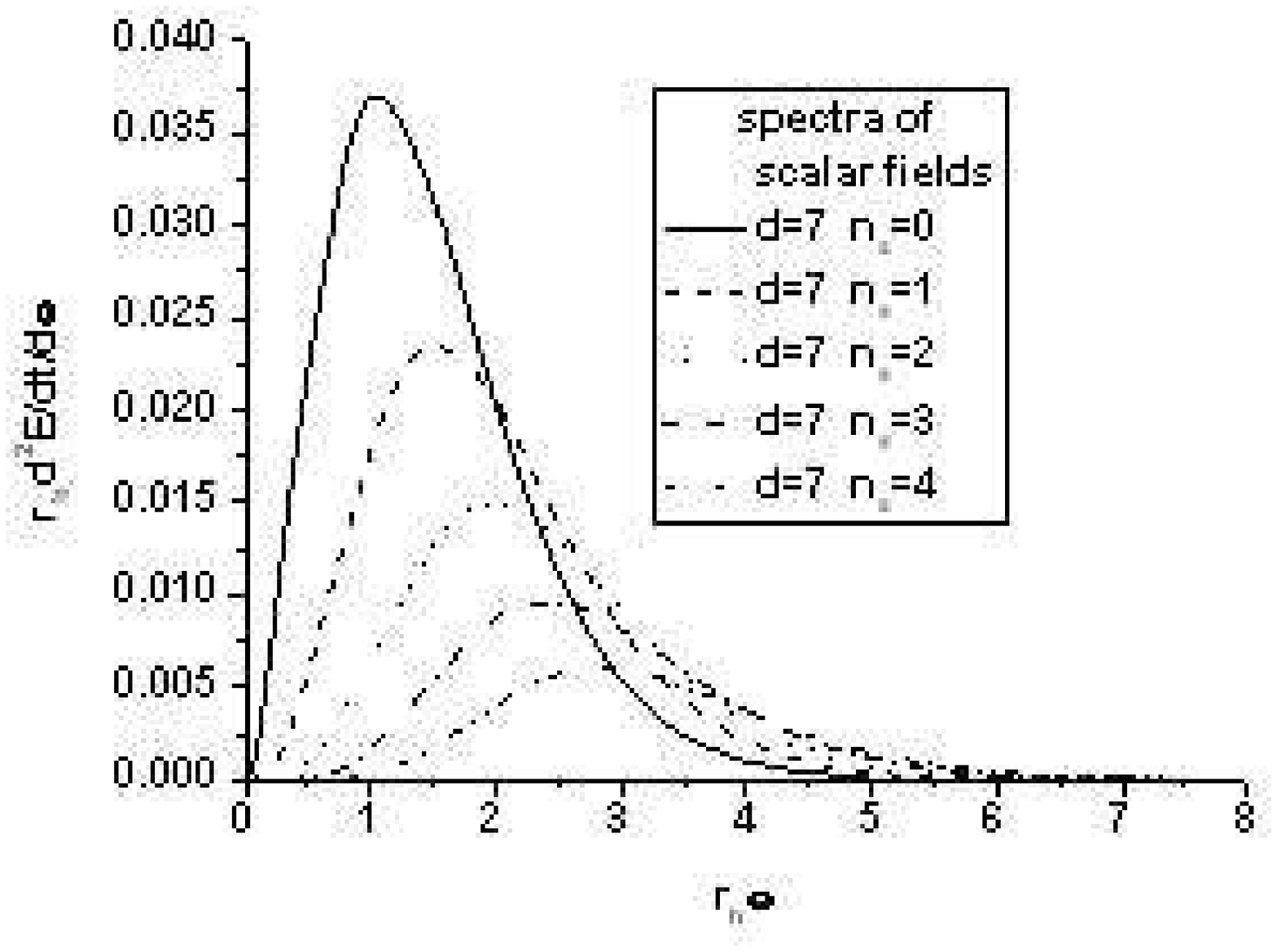}
\caption{Spectra of scalar fields in d=7 space.}
\label{spectra:s-4}
\end{figure}
\begin{figure}[h]
\centering
\includegraphics[width=3.2in]{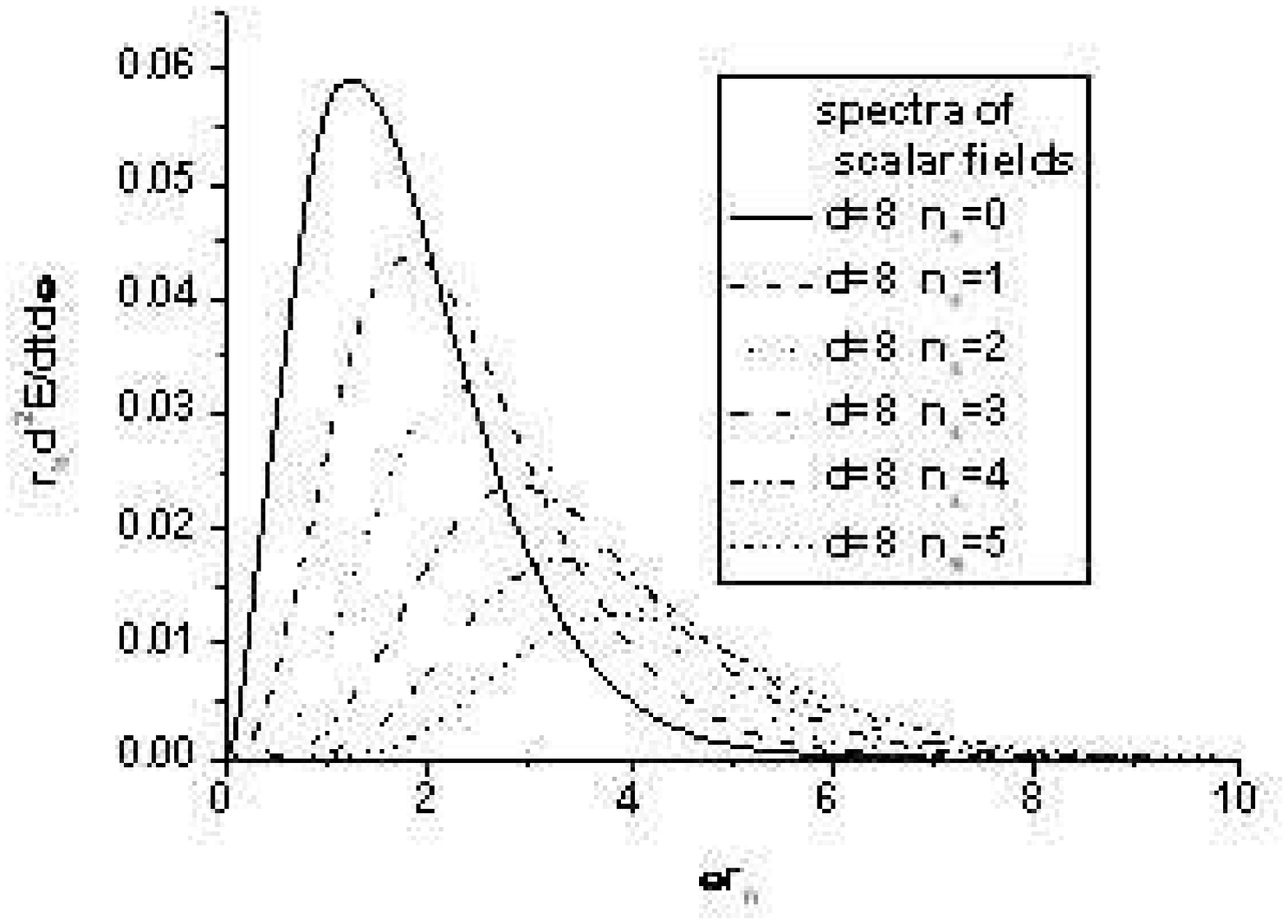}
\caption{Spectra of scalar fields in d=8 space.}
\label{spectra:s-5}
\end{figure}
\begin{figure}[h]
\centering
\includegraphics[width=3.2in]{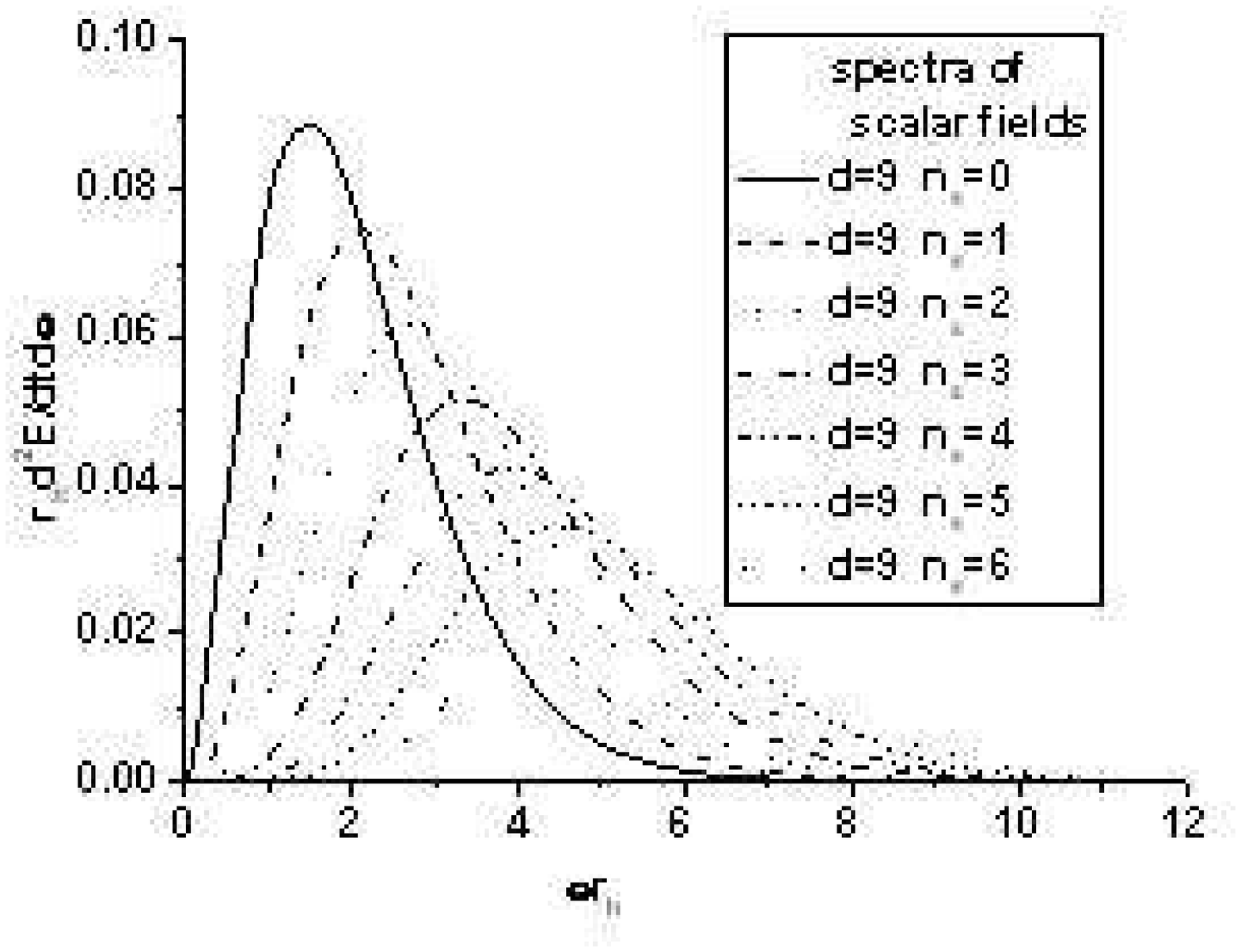}
\caption{Spectra of scalar fields in d=9 space.}
\label{spectra:s-6}
\end{figure}
\begin{figure}[h]
\centering
\includegraphics[width=3.2in]{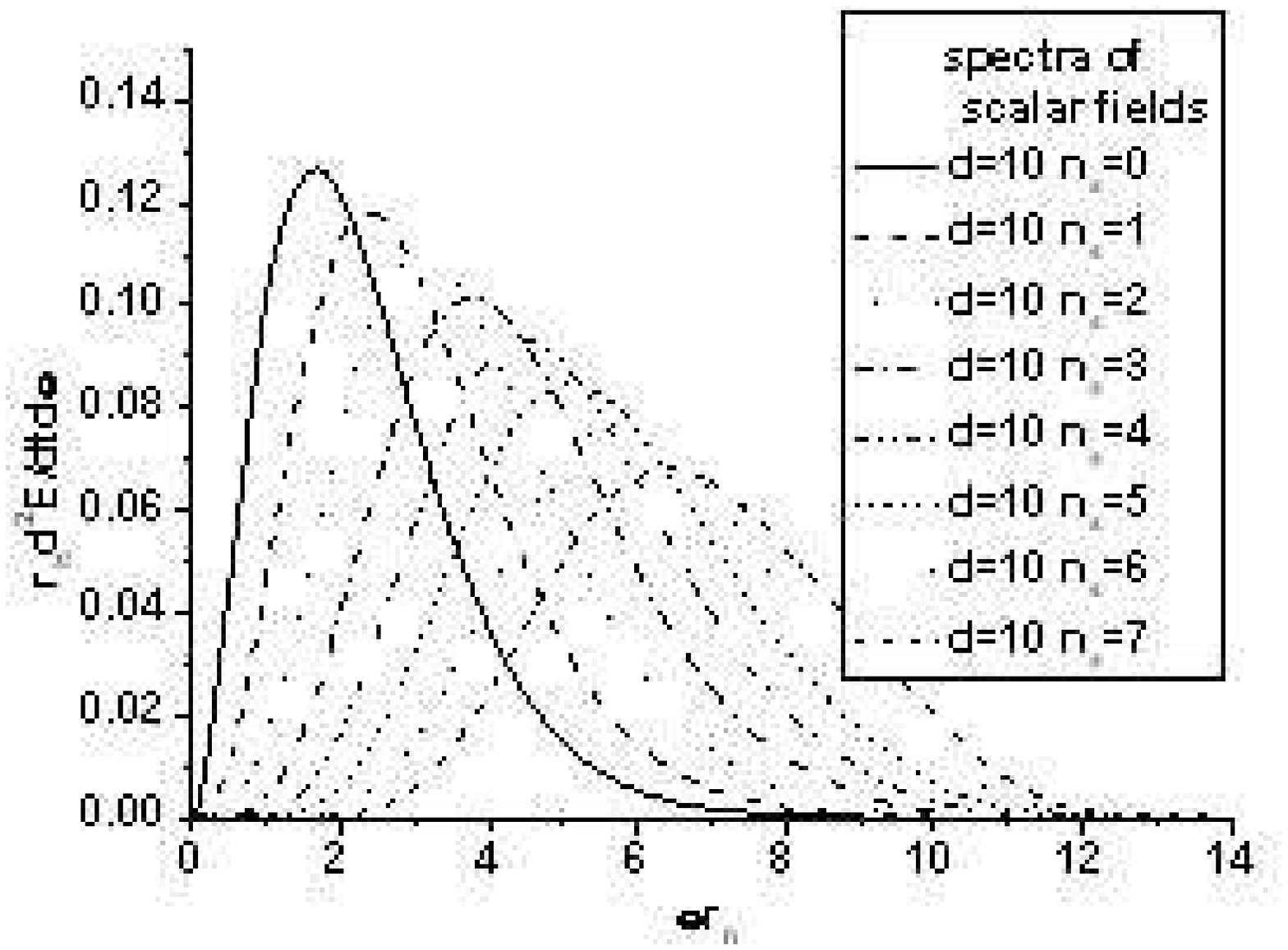}
\caption{Spectra of scalar fields in d=10 space.}
\label{spectra:s-7}
\end{figure}
\begin{figure}[h]
\centering
\includegraphics[width=3.2in]{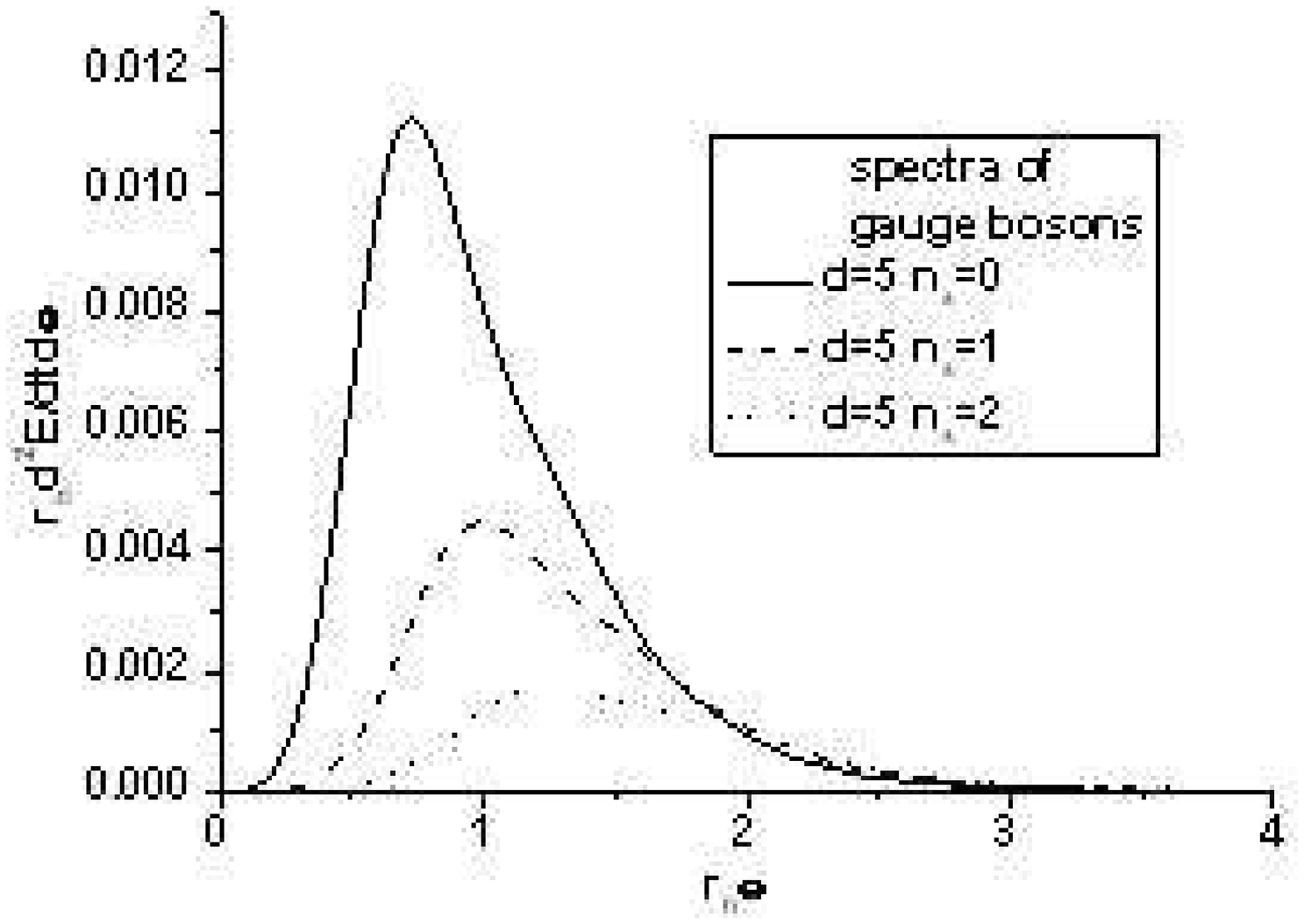}
\caption{Spectra of gauge bosons in d=5 space.}
\label{spectra:ga-2}
\end{figure}
\begin{figure}[h]
\centering
\includegraphics[width=3.2in]{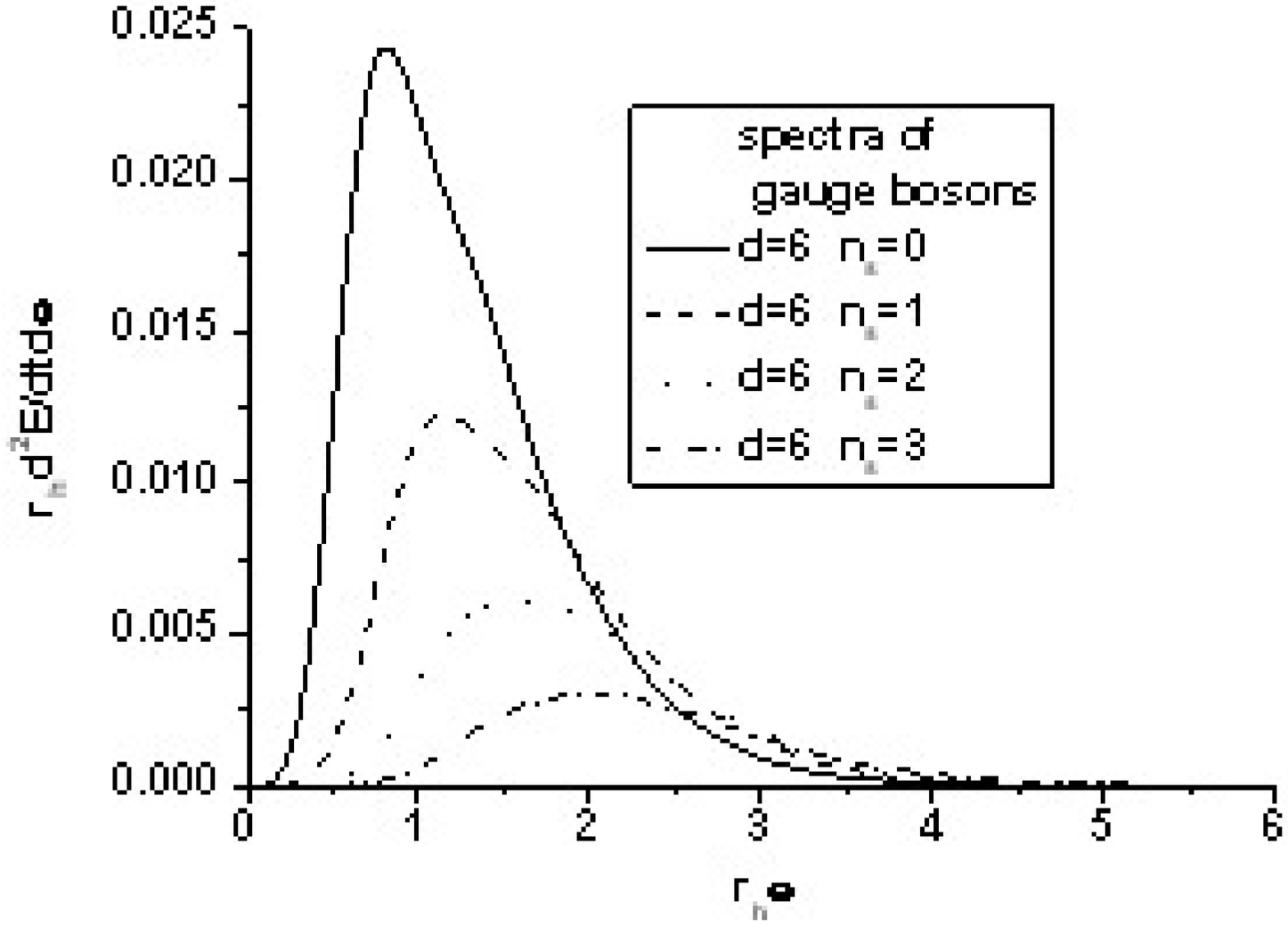}
\caption{Spectra of gauge bosons in d=6 space.}
\label{spectra:ga-3}
\end{figure}
\begin{figure}[h]
\centering
\includegraphics[width=3.2in]{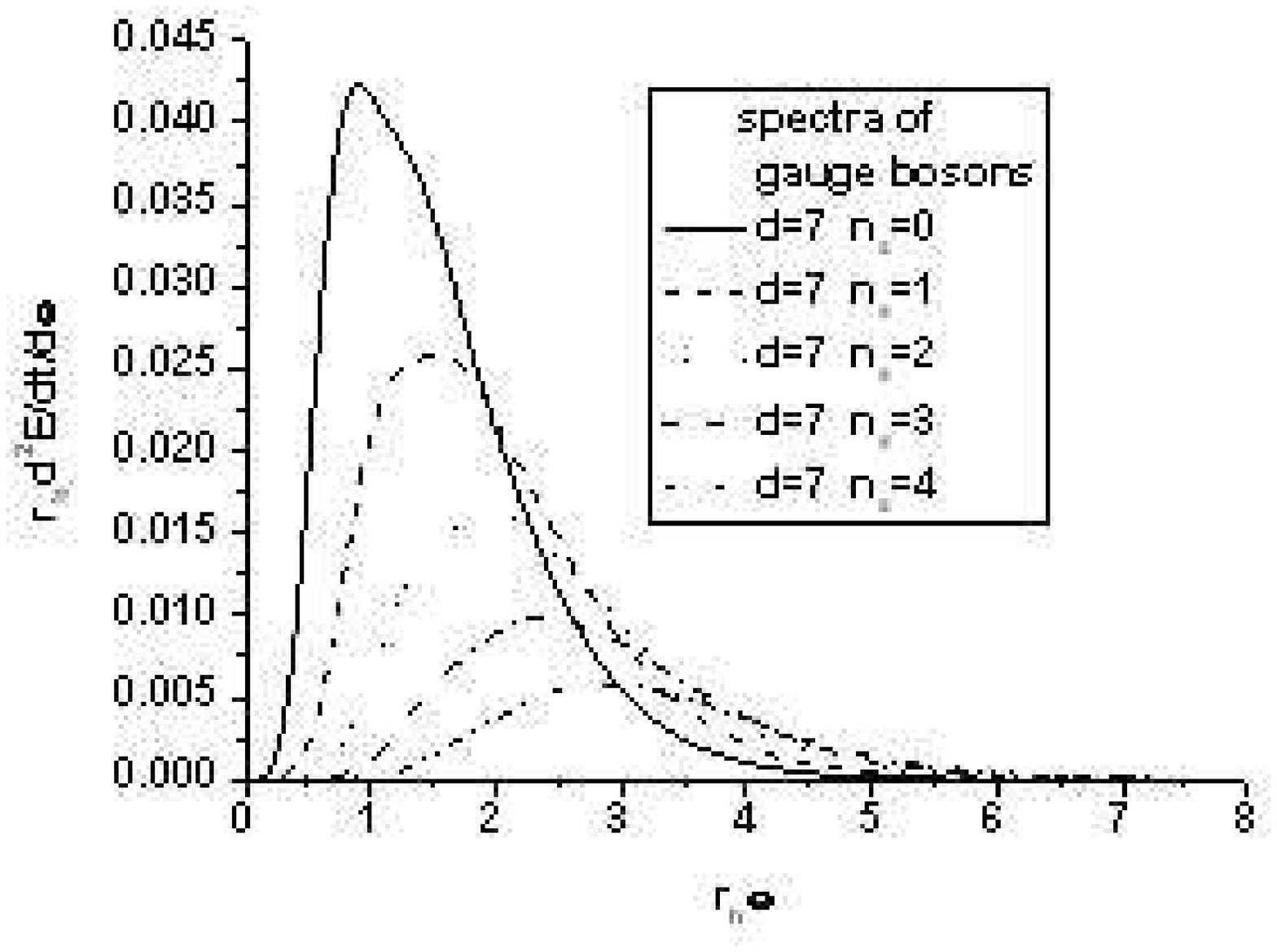}
\caption{Spectra of gauge bosons in d=7 space.}
\label{spectra:ga-4}
\end{figure}
\begin{figure}[h]
\centering
\includegraphics[width=3.2in]{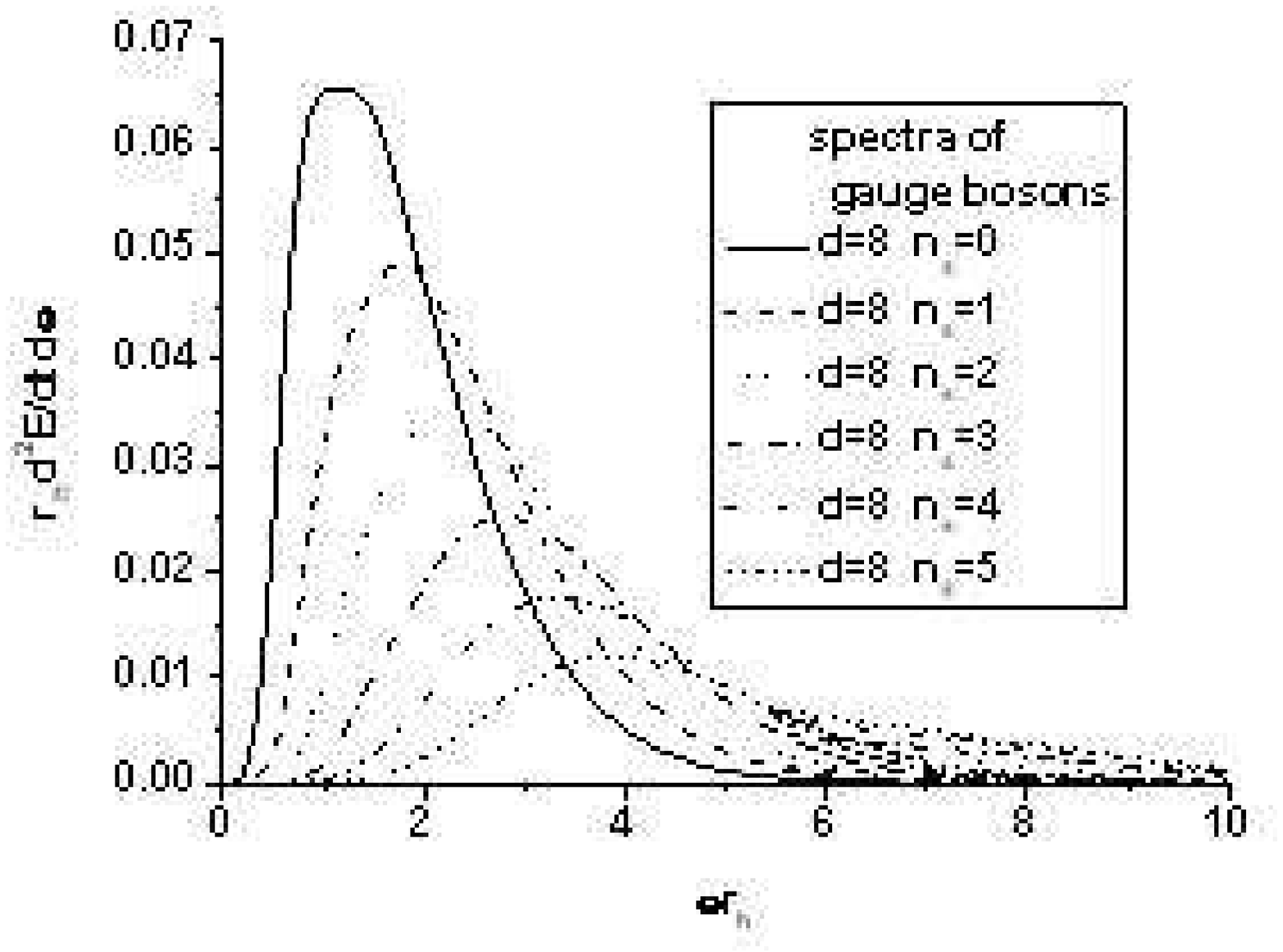}
\caption{Spectra of gauge bosons in d=8 space.}
\label{spectra:ga-5}
\end{figure}
\begin{figure}[h]
\centering
\includegraphics[width=3.2in]{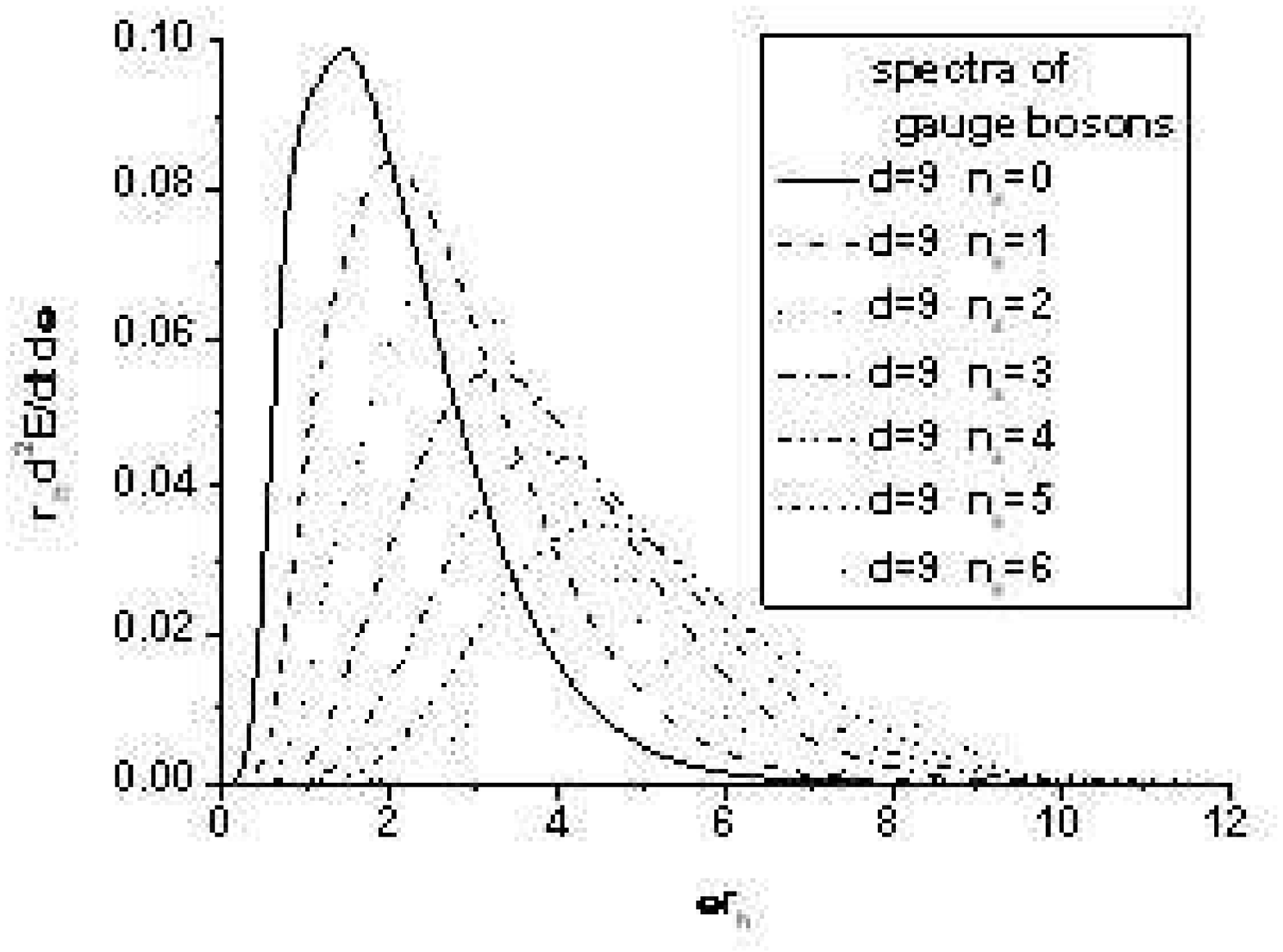}
\caption{Spectra of gauge bosons in d=9 space.}
\label{spectra:ga-6}
\end{figure}
\begin{figure}[h]
\centering
\includegraphics[width=3.2in]{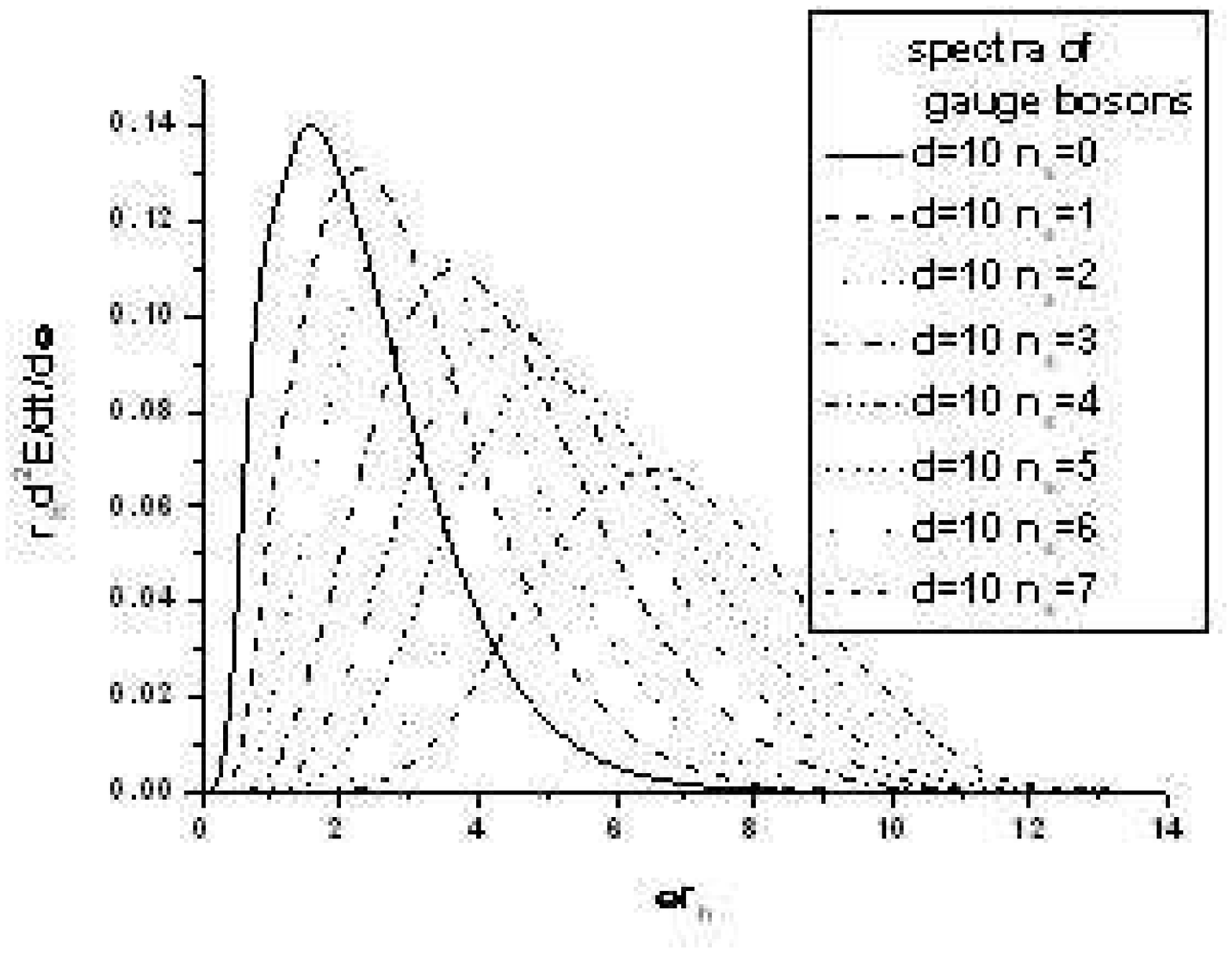}
\caption{Spectra of gauge bosons in d=10 space.}
\label{spectra:ga-7}
\end{figure}

Nowadays gray-body factors can be found in many papers. We collect the 
relevant papers in table \ref{tab:spectra}. 
We follow these papers to calculate energy power spectra in our database except 
for the split-fermions model (although we independently confirm the results
as well). We perform an original calculation of  
the spectra of gauge bosons and scalar fields in the split-fermions model 
as a function of the number of dimensions $n_{s}$
in which the fermion branes are split.
These are shown in figures \ref{spectra:s-2} to \ref{spectra:ga-7}. 

Anyone who has a better spectrum can upgrade our database. 
For example, we only calculate the spectra up to the $l=9$ mode. 


\clearpage
\begin{table*}[h]
\caption{Literature sources for particle emission spectra}
\label{tab:spectra}
\begin{tabular}{|l|l|l|l|}
\hline
Type of particle & Type of black hole & Brane model & References\\
\hline
\hline
Standard-Model particles & non-rotating & unsplit; tensionless 
&\cite{SM-brane}\cite{SM-brane-1}\\
\hline
gravitons & non-rotating & split/unsplit; tensionless & 
\cite{Cardoso-graviton}\\
\hline
Standard-Model particles & non-rotating & split/unsplit; with tension 
&\cite{DKSS}\\
\hline
gravitons & non-rotating & split/unsplit; with tension &\cite{DKSS}\\
\hline
scalars and gauge bosons & non-rotating & split; tensionless& figures 
\ref{spectra:s-2}-\ref{spectra:ga-7}\\
\hline
fermions &rotating & unsplit; tensionless 
&\cite{kant-ro-fer}\cite{Ida1}\\
\hline
gauge bosons &rotating & unsplit; tensionless 
&\cite{kant-ro-b}\cite{Ida1}\\
\hline 
scalar fields &rotating & unsplit; tensionless &\cite{SM-brane} 
\cite{kant-ro-s}\cite{kant-ro-s-1}\cite{Ida1}\\
\hline
\end{tabular}
\end{table*}

\begin{table*}
\caption{ Degrees of freedom of Standard-Model particles 
which are emitted from a black hole. For gravitons, 
the table shows $1$, because the appropriate growth in the
number of degrees of freedom is included explicitly in the graviton
emission spectrum. 
$n_{s}$ is the number of extra dimensions in which vector and scalar fields 
can propagate.}
\label{tab:degenecy1}
\begin{tabular}{|l|l|l|l|r|}
\hline
particle type & $d_{0}$ & $d_{1/2}$ & $d_{1}$ & $d_{2}$\\
\hline
Quarks & 0 & 6 & 0 & 0\\
\hline
Charged leptons & 0 & 2 & 0 & 0\\
\hline
Neutrinos & 0 & 2 & 0 & 0\\
\hline
Photons or gluons& 0 & 0 & $2+n_{s}$ & 0\\
\hline
$Z^{0}$ & 1 & 0 & $2+n_{s}$ & 0\\
\hline
$W^{+}$ and $W^{-}$ & 2 & 0 & $2(2+n_{s})$ & 0\\
\hline
Higgs boson & 1 & 0 & 0 & 0\\
\hline
Graviton & 0 & 0 & 0 & 1\\
\hline
\end{tabular}
\end{table*}

\begin{table*}
\caption{The fraction of emitted particles 
of different types (including final burst particles)
in a variety of extra dimension scenarios. 
$^*$Note that the absence of gravitons in the case of a rotating black hole 
is due exclusively to our current ignorance of the correct gray-body factor.
}
\label{tab:particle-ratio}
\begin{tabular}{|l|l|l|l|l|l|l|l|l|}
\hline
Scenario & quarks & gluons & leptons & gauge bosons& neutrinos & gravitons &
Higgs bosons & photons\\
\hline
$d=4$ $n_{s}=0$ non-rotating black 
hole&68.21&10.79&9.45&5.72&3.87&2.00e-01&8.99e-01&8.61e-01\\
\hline
$d=5$ $n_{s}=0$ non-rotating black 
hole&65.37&13.29&9.04&6.12&3.76&4.60e-01&8.26e-01&1.13\\
\hline
$d=6$ $n_{s}=0$ non-rotating black 
hole&63.63&14.51&8.93&6.58&3.51&7.59e-01&7.76e-01&1.30\\
\hline
$d=7$ $n_{s}=0$ non-rotating black 
hole&61.25&15.94&8.75&7.17&3.45&1.20&7.89e-01&1.44\\
\hline
$d=8$ $n_{s}=0$ non-rotating black 
hole&60.99&15.94&8.56&6.99&3.35&1.93&7.63e-01&1.47\\
\hline
$d=9$ $n_{s}=0$ non-rotating black 
hole&59.40&16.26&8.26&7.08&3.26&3.48&7.45e-01&1.51\\
\hline
$d=10$ $n_{s}=0$ non-rotating black 
hole&57.56&16.15&7.68&6.82&3.17&6.46&6.97e-01&1.46\\
\hline
$d=10$ $n_{s}=1$ split fermions 
model&61.58&19.76&1.64&7.30&3.83e-01&6.95&5.88e-01&1.80\\
\hline
$d=10$ $n_{s}=2$ split fermions 
model&61.28&20.40&1.66&7.09&3.49e-01&6.91&4.56e-01&1.86\\
\hline
$d=10$ $n_{s}=3$ split fermions 
model&62.10&20.33&1.65&6.76&4.46e-01&6.43&4.15e-01&1.87\\
\hline
$d=10$ $n_{s}=4$ split fermions 
model&62.70&19.76&1.72&6.60&4.66e-01&6.62&3.28e-01&1.80\\
\hline
$d=10$ $n_{s}=5$ split fermions 
model&63.67&19.34&1.73&6.14&5.17e-01&6.50&2.65e-01&1.83\\
\hline
$d=10$ $n_{s}=6$ split fermions 
model&64.96&18.24&1.82&5.85&5.56e-01&6.69&2.35e-01&1.65\\
\hline
$d=10$ $n_{s}=7$ split fermions 
model&66.38&17.23&1.83&5.44&5.50e-01&6.68&2.51e-01&1.64\\
\hline
$d=2$ $n_{s}=2$ $B=1.0$ tension brane 
model&84.13&9.39&1.66&2.80&2.02e-01&9.23e-01&2.51e-01&6.30e-01\\
\hline
$d=2$ $n_{s}=2$ $B=0.9$ tension brane 
model&85.33&8.88&1.65&2.49&1.63e-01&7.55e-01&1.90e-01&5.47e-01\\
\hline
$d=2$ $n_{s}=2$ $B=0.8$ tension brane 
model&86.42&8.32&1.54&2.32&1.79e-01&5.70e-01&1.99e-01&4.49e-01\\
\hline
$d=2$ $n_{s}=2$ $B=0.7$ tension brane 
model&87.06&7.93&1.63&2.08&1.75e-01&5.19e-01&1.98e-01&4.00e-01\\
\hline
$d=2$ $n_{s}=2$ $B=0.6$ tension brane 
model&87.92&7.39&1.60&1.94&1.59e-01&4.46e-01&1.88e-01&3.59e-01\\
\hline
$d=2$ $n_{s}=2$ $B=0.5$ tension brane 
model&88.51&7.11&1.46&1.84&1.33e-01&3.96e-01&1.86e-01&3.67e-01\\
\hline
$d=2$ $n_{s}=2$ $B=0.4$ tension brane 
model&89.16&6.58&1.58&1.72&1.49e-01&3.54e-01&1.66e-01&2.90e-01\\
\hline
$d=4$ $n_{s}=0$ rotating black 
hole&64.82&15.41&7.94&6.25&3.50&0.00$^*$&6.06e-01&1.48\\
\hline
$d=5$ $n_{s}=0$ rotating black 
hole&61.38&18.11&7.89&7.06&3.30&0.00$^*$&5.56e-01&1.70\\
\hline
$d=6$ $n_{s}=0$ rotating black 
hole&59.21&20.38&7.27&7.59&3.08&0.00$^*$&5.85e-01&1.89\\
\hline
$d=7$ $n_{s}=0$ rotating black 
hole&57.52&21.82&7.08&8.08&2.87&0.00$^*$&5.77e-01&2.05\\
\hline
$d=8$ $n_{s}=0$ rotating black 
hole&54.41&24.02&6.75&9.09&2.77&0.00$^*$&6.07e-01&2.36\\
\hline
$d=9$ $n_{s}=0$ rotating black 
hole&51.88&24.05&7.98&9.27&3.55&0.00$^*$&7.15e-01&2.55\\
\hline
$d=10$ $n_{s}=0$ rotating black 
hole&52.43&25.67&6.45&9.62&2.52&0.00$^*$&6.79e-01&2.63\\
\hline
$d=4$ $n_{s}=0$ two-body final 
states&58.88&16.89&1.38&6.08&11.14&0.00e+00&1.91&3.71\\
\hline
$d=5$ $n_{s}=0$ two-body final 
states&57.71&17.46&1.37&6.22&11.53&0.00e+00&1.91&3.80\\
\hline
$d=6$ $n_{s}=0$ two-body final 
states&57.16&17.72&1.39&6.27&11.59&0.00e+00&1.93&3.94\\
\hline
$d=7$ $n_{s}=0$ two-body final 
states&57.02&17.91&1.37&6.22&11.59&0.00e+00&1.98&3.91\\
\hline
$d=8$ $n_{s}=0$ two-body final 
states&56.58&18.08&1.35&6.31&11.79&0.00e+00&1.92&3.97\\
\hline
$d=9$ $n_{s}=0$ two-body final 
states&56.51&18.02&1.41&6.40&11.75&0.00e+00&1.96&3.95\\
\hline
$d=10$ $n_{s}=0$ two-body final 
states&56.66&18.06&1.43&6.28&11.74&0.00e+00&1.92&3.90\\
\hline
\end{tabular}
\end{table*}


\begin{thebibliography}{99}
\bibitem{code} 
{\tiny http://www-pnp.physics.ox.ac.uk/$\sim$issever/BlackMax/blackmax.html}


\bibitem{ADD}
Nima Arkani-Hamed, Savas Dimopoulos, G.R. Dvali, 
Phys. Rev. D59:086004, 1999. hep-ph/9807344.

Ignatios Antoniadis, Nima Arkani-Hamed, Savas Dimopoulos, G.R. Dvali,
Phys. Lett. B436:257-263, 1998. hep-ph/9804398.

Nima Arkani-Hamed, Savas Dimopoulos, G.R. Dvali, 
Phys. Lett. B429:263-272, 1998. hep-ph/9803315.

\bibitem{RS}
Lisa Randall, Raman Sundrum, Phys.Rev.Lett.83:4690-4693,1999. hep-th/9906064.
\bibitem{CHMADD} 
N.~Kaloper, J.~March-Russell, G.~D.~Starkman and M.~Trodden,
Phys.\ Rev.\ Lett.\ {\bf 85}, 928 (2000)
[arXiv:hep-ph/0002001];
G.~D.~Starkman, D.~Stojkovic and M.~Trodden,
Phys.\ Rev.\ Lett.\ {\bf 87}, 231303 (2001)
[arXiv:hep-th/0106143].
G.~D.~Starkman, D.~Stojkovic and M.~Trodden,
Phys.\ Rev.\  D {\bf 63}, 103511 (2001)
[arXiv:hep-th/0012226].

\bibitem{UED} 
I.~Antoniadis,
Phys.\ Lett.\ B {\bf 246}, 377 (1990);
K.~R.~Dienes, E.~Dudas and T.~Gherghetta,
Phys.\ Lett.\ B {\bf 436}, 55 (1998)
[arXiv:hep-ph/9803466].

\bibitem{BHacc} T. Banks, W. Fischler, {\it hep-th/9906038 };
S. Dimopoulos, G. Landsberg, Phys. Rev. Lett. {\bf 87} 161602 (2001)
; S. B. Giddings and S. Thomas, Phys. Rev. {\bf D65} 056010 (2002)


\bibitem{hoop}
K. S. Thorne, Nonspherical gravitational collapse: A short review. In J R 
Klauder, Magic Without Magic, San Francisco 1972, 231-258;
D. Ida and K.-i. Nakao, Phys. Rev. D66 (2002)064026, [gr-qc/0204082]; 
H. Yoshino and Y. Nambu, Phys. Rev. D66 (2002)065004, [gr-qc/0204060]. 





\bibitem{Vachaspati:2006ki}
T.~Vachaspati, D.~Stojkovic and L.~M.~Krauss,
Phys.\ Rev.\ D {\bf 76}, 024005 (2007)
[arXiv:gr-qc/0609024].
T.~Vachaspati and D.~Stojkovic,
arXiv:gr-qc/0701096.


\bibitem{othergenerators}
TRUENOIR: Savas Dimopoulos, Greg L. Landsberg, 
Phys. Rev. Lett. 87:161602,2001. hep-ph/0106295

CHARYBDIS: C.M. Harris, P. Richardson, B.R. Webber, JHEP 0308:033, 2003. 
hep-ph/0307305.

Catfish: M. Cavaglia, R. Godang, L. Cremaldi, D. Summers, hep-ph/0609001.

\bibitem{Dai:2006dz}
D.~C.~Dai, G.~D.~Starkman and D.~Stojkovic,
Phys.\ Rev.\ D {\bf 73}, 104037 (2006)
[arXiv:hep-ph/0605085];
  D.~Stojkovic and G.~D.~Starkman,D.~C.~Dai,
  Phys.\ Rev.\ Lett.\  {\bf 96}, 041303 (2006)
  [arXiv:hep-ph/0505112]

\bibitem{Stojkovic:2005zq}
D.~Stojkovic, F.~C.~Adams and G.~D.~Starkman,
Int.\ J.\ Mod.\ Phys.\ D {\bf 14}, 2293 (2005)
[arXiv:gr-qc/0604072];
C.~Bambi, A.~D.~Dolgov and K.~Freese,
Nucl.\ Phys.\ B {\bf 763}, 91 (2007)
[arXiv:hep-ph/0606321]; arXiv:hep-ph/0612018.



\bibitem{splitfermions} N. Arkani-Hamed, M. Schmaltz, Phys. Rev. {\bf
D61} 033005 (2000)

\bibitem{superradiance}
Ya. B. Zel'dovich, Pis'ma v Zh. Eksp. Teor. Fiz. 12, 443;
Ya. B. Zel'dovich, Sov. Phys. JETP Lett. 14, 180;
Ya. B. Zel'dovich, Sov. Phys. JETP 35, 1085;
C. W. Misner, Phys. Rev. Lett. 28, 994 (1972);
V. P. Frolov, D. Stojkovic, Phys. Rev. {\bf D67} 084004 (2003);
Phys. Rev. {\bf D68} 064011 (2003)

\bibitem{NKD}
Nemanja Kaloper, Derrick Kiley, JHEP 0603:077,2006. hep-th/0601110

\bibitem{DKSS}
De-Chang Dai, Nemanja Kaloper, Glenn D. Starkman, Dejan Stojkovic. Phys. Rev. 
D75:024043,2007. hep-th/0611184

\bibitem{Meade}
Patrick Meade, Lisa Randall, arXiv:0708.3017 [hep-ph] 



%
%
\bibitem{Cardoso-graviton}
Vitor Cardoso, Marco Cavaglia, Leonardo Gualtieri. JHEP 0602:021,2006. 
hep-th/0512116
\bibitem{SM-brane}
Christopher Michael Harris. Ph.D. thesis. hep-ph/0502005
\bibitem{SM-brane-1}
Chris M. Harris, Panagiota Kanti. JHEP 0310:014,2003. hep-ph/0309054
\bibitem{kant-ro-s}
G. Duffy, C. Harris, P. Kanti, E. Winstanley. JHEP 0509:049,2005. hep-th/0507274
\bibitem{kant-ro-s-1}
C.M. Harris, P. Kanti. Phys.Lett.B633:106-110,2006. hep-th/0503010
\bibitem{kant-ro-b}
M. Casals, P. Kanti, E. Winstanley. JHEP 0602:051,2006. hep-th/0511163
\bibitem{kant-ro-fer}
M. Casals, S.R. Dolan, P. Kanti, E. Winstanley. JHEP 0703:019,2007. 
hep-th/0608193
\bibitem{Ida1}
Daisuke Ida, Kin-ya Oda, and Seong Chan Park, Phys. \ Rev. \ {\bf D73}, 124022 
(2006), 
hep-th/0602188

\bibitem{Creek:2007sy}
S.~Creek, O.~Efthimiou, P.~Kanti and K.~Tamvakis,
Phys.\ Rev.\ D {\bf 75}, 084043 (2007)
[arXiv:hep-th/0701288];arXiv:0707.1768 [hep-th]
T.~Harmark, J.~Natario and R.~Schiappa,
arXiv:0708.0017 [hep-th].

\bibitem{newref}
B.~Abbott {\it et al.} [D0 Collaboration],
Phys.\ Rev.\ Lett.\ {\bf 82}, 2457 (1999)
[arXiv:hep-ex/9807014].

\bibitem{Stojkovic:2004hp}
D.~Stojkovic,
Phys.\ Rev.\ Lett.\ {\bf 94}, 011603 (2005)
[arXiv:hep-ph/0409124].

\bibitem{Anchordoqui}
Luis A. Anchordoqui, Jonathan L. Feng, Haim Goldberg, Alfred D. Shapere, 
Phys.Lett.B594:363-367,2004. hep-ph/0311365 


\bibitem{recoil}
V. Frolov, D. Stojkovic, Phys. Rev. Lett. {\bf 89} 151302
(2002); Phys. Rev. {\bf D66} 084002 (2002); D. Stojkovic,
Phys.Rev.Lett. {\bf 94} 011603 (2005)

\bibitem{flux} D. Stojkovic, JHEP 0409:061 (2004)




\bibitem{ida} D. Ida, K. Oda, S. C. Park, Phys.Rev.{\bf D67} 064025
(2003) Erratum-ibid.{\bf D69}
049901 (2004)



\bibitem{pdfs}
J. Pumplin et al. JHEP 0207, 012 (2002)[hep-ph/0201195]; D. Stump et
al., JHEP 0310, 046 (2003)[hep-ph/0303013].



\bibitem{1} A. Flachi, O. Pujolas, M. Sasaki, T. Tanaka,
hep-th/0601174;
A. Flachi, T. Tanaka, Phys. Rev. Lett. {\bf 95} 161302 (2005)


\bibitem{2} T. G. Rizzo, hep-ph/0601029; hep-ph/0510420; JHEP {\bf
0501} 028 (2005)


\bibitem{3} D.K. Park, hep-th/0512021; hep-th/0511159

\bibitem{4} M. Casals, P. Kanti, E. Winstanley, hep-th/0511163

\bibitem{5} R. da Rocha, C. H. Coimbra-Araujo, JCAP {\bf 0512} 009
(2005)

\bibitem{6} A.S. Cornell, W. Naylor, M. Sasaki; hep-th/0510009

\bibitem{7} J. Grain, A. Barrau, P. Kanti, Phys.Rev.{\bf D72} 104016
(2005)


\bibitem{9} H. Yoshino, T. Shiromizu, M. Shibata, Phys.Rev. {\bf D72}
084020 (2005)


\bibitem{10} G. Duffy, C. Harris, P. Kanti, E. Winstanley, JHEP {\bf
0509} 049 (2005)

\bibitem{11} E. Jung, D.K. Park, Nucl.Phys. {\bf B731} 171 (2005)


\bibitem{12} L. Lonnblad, M. Sjodahl, T. Akesson, JHEP {\bf 0509} 019
(2005)


\bibitem{13} J.I. Illana, M. Masip, D. Meloni, Phys.Rev.{\bf D72}
024003 (2005)

\bibitem{14} E. Jung, S. Kim, D.K. Park, Phys.Lett. {\bf B619} 347
(2005);
Phys.Lett. {\bf B614} 78 (2005)

\bibitem{15} H. Yoshino, V. S. Rychkov, Phys.Rev. {\bf D71} 104028
(2005)

\bibitem{16} A.S. Majumdar, N. Mukherjee, Int.J.Mod.Phys. {\bf D14}
1095 (2005); astro-ph/0403405;
  Mod.\ Phys.\ Lett.\  A {\bf 20}, 2487 (2005).

\bibitem{17} A. Perez-Lorenzana, hep-ph/0503177


\bibitem{18} D. Ida, K. Oda, S. C. Park, Phys.Rev. {\bf D71} 124039
(2005);hep-th/0602188

\bibitem{19} V. P. Frolov, D. V. Fursaev, D.
Stojkovic, Class. Quant. Grav. {\bf 21} 3483 (2004); JHEP
{\bf 0406} 057 (2004);


\bibitem{20} A.N. Aliev, A.E. Gumrukcuoglu, Phys.Rev. {\bf D71} 104027
(2005);
A.N. Aliev, gr-qc/0505003;
A.N. Aliev, V. P. Frolov, Phys.Rev.{\bf D69} 084022 (2004)

\bibitem{21} P. Kanti, J. Grain, A. Barrau, Phys.Rev. {\bf D71} 104002
(2005);
P. Kanti, K. Tamvakis, Phys.Rev. {\bf D65} 084010 (2002);
P. Kanti, J. March-Russell, Phys.Rev.{\bf D67} 104019 (2003);
P. Kanti, Int.J.Mod.Phys.{\bf A19} 4899 (2004)

\bibitem{22} H. Yoshino, Y. Nambu, Phys.Rev. {\bf D70} 084036 (2004)

\bibitem{23} E. F. Eiroa, gr-qc/0511004; Phys.Rev. {\bf D71} 083010
(2005)


\bibitem{24} P. S. Apostolopoulos, N. Brouzakis, E. N. Saridakis, N.
Tetradis Phys.Rev. {\bf D72} 044013 (2005)


\bibitem{25} S. Creek, O. Efthimiou, P. Kanti, K. Tamvakis,
hep-th/0601126

\bibitem{26} D. Stojkovic, Phys.Rev.{\bf D67} 045012 (2003)


\bibitem{27} E. Berti, V. Cardoso, M. Casals, Phys.Rev. {\bf D73}
024013 (2006)


\bibitem{28} M. Vasudevan, K. A. Stevens, Phys.Rev. {\bf D72} 124008
(2005)

\bibitem{29} B. M.N. Carter, I. P. Neupane, Phys.Rev. {\bf D72} 043534
(2005)


\bibitem{30} L. Lonnblad, M. Sjodahl, T. Akesson, JHEP {\bf 0509} 019
(2005)

\bibitem{31} V. Frolov, M. Snajdr and D. Stojkovic,
Phys. Rev. {\bf D68} 044002 (2003);

\bibitem{32} M. Nozawa, K. Maeda, Phys.Rev. {\bf D71} 084028 (2005)

\bibitem{33} C. M. Harris, hep-ph/0502005

\bibitem{34} Y. Morisawa, D. Ida, Phys.Rev.{\bf D71} 044022 (2005)


\bibitem{35} 
V. Cardoso, O.J.C. Dias, J. P.S. Lemos, Phys.Rev.{\bf D67} 064026
(2003); V. Cardoso, O. J.C. Dias, J. L. Hovdebo, R. C. Myers, hep-th/0512277


\bibitem{36} A. Cafarella, C. Coriano, T.N. Tomaras, hep-ph/0412037

\bibitem{37} A. Chamblin, F. Cooper, G. C. Nayak, Phys.Rev. {\bf D70}
075018 (2004); Phys.Rev. {\bf D69} 065010 (2004)



\bibitem{38} D. Ida, Y. Uchida, Y. Morisawa, Phys.Rev.{\bf D67} 084019
(2003)

\bibitem{39} E.J. Ahn, M. Cavaglia, hep-ph/0511159

\bibitem{40} M. Cavaglia, S. Das, Class.Quant.Grav. {\bf 21} 4511
(2004);
M. Cavaglia, S. Das, R. Maartens, Class.Quant.Grav. {\bf 20} L205
(2003)

\bibitem{41} V. Cardoso, M. Cavaglia, L. Gualtieri, hep-th/0512116

\bibitem{42} L. A. Anchordoqui, J. L. Feng, H. Goldberg, A. D. Shapere,
Phys.Lett.{\bf B594} 363 (2004); L. Anchordoqui, H. Goldberg, Phys.Rev.{\bf D67} 
064010 (2003);
L. Anchordoqui, T. Han, D. Hooper, S. Sarkar, hep-ph/0508312

\bibitem{43} A. Casanova, E. Spallucci, Class.Quant.Grav. {\bf 23}
R45 (2006)

\bibitem{44} J. E. Aman, N. Pidokrajt, Phys.Rev.{\bf D73} 024017 (2006)


\bibitem{45} U. Harbach, M. Bleicher, hep-ph/0601121


\bibitem{46} B. Koch, M. Bleicher, S. Hossenfelder, JHEP {\bf 0510} 053
(2005); S. Hossenfelder, Phys.Lett. {\bf B598} 92 (2004)

\bibitem{47} M. Fairbairn, hep-ph/0509191

\bibitem{48} G.L. Alberghi, R. Casadio, D. Galli, D. Gregori, A. Tronconi,
V. Vagnoni, hep-ph/0601243

\bibitem{49} P. Davis, hep-th/0602118; P. Krtous, J. Podolsky,
Class.Quant.Grav. {\bf 23} 1603 (2006)

\bibitem{50} K.A. Bronnikov, S.A. Kononogov, V.N. Melnikov, gr-qc/0601114



\bibitem{51}
B.~Koch, M.~Bleicher and H.~Stoecker,
arXiv:hep-ph/0702187.

\bibitem{52}
S.~Chen, B.~Wang and R.~K.~Su,
Phys.\ Lett.\ B {\bf 647}, 282 (2007)
[arXiv:hep-th/0701209].

\bibitem{53}
H.~T.~Cho, A.~S.~Cornell, J.~Doukas and W.~Naylor,
Phys.\ Rev.\ D {\bf 75}, 104005 (2007)
[arXiv:hep-th/0701193].

\bibitem{54}
L.~h.~Liu, B.~Wang and G.~h.~Yang,
arXiv:hep-th/0701166.

\bibitem{55}
B.~Koch,
arXiv:0707.4644 [hep-ph].


\bibitem{56}
H.~Stoecker,
J.\ Phys.\ G {\bf 32}, S429 (2006).

\bibitem{57}
B.~Betz, M.~Bleicher, U.~Harbach, T.~Humanic, B.~Koch and H.~Stoecker,
arXiv:hep-ph/0606193.

\bibitem{58}
A.~N.~Aliev,
Phys.\ Rev.\ D {\bf 75}, 084041 (2007)
[arXiv:hep-th/0702129].

\bibitem{59}
T.~G.~Rizzo,
Phys.\ Lett.\ B {\bf 647}, 43 (2007)
[arXiv:hep-ph/0611224].


\bibitem{60}
D.~M.~Gingrich,
Int.\ J.\ Mod.\ Phys.\ A {\bf 21}, 6653 (2006)
[arXiv:hep-ph/0609055];arXiv:0706.0623 [hep-ph].

\bibitem{61}
E.~Abdalla, C.~B.~M.~Chirenti and A.~Saa,
arXiv:gr-qc/0703071.

\bibitem{62}
G.~L.~Landsberg,
J.\ Phys.\ G {\bf 32}, R337 (2006)
[arXiv:hep-ph/0607297].

\bibitem{63}
H.~Yoshino and R.~B.~Mann,
Phys.\ Rev.\ D {\bf 74}, 044003 (2006)
[arXiv:gr-qc/0605131].

\bibitem{64}
Y.~i.~Takamizu, H.~Kudoh and K.~i.~Maeda,
Phys.\ Rev.\ D {\bf 75}, 061304 (2007)
[arXiv:gr-qc/0702138].

\bibitem{65}
R.~da Rocha and C.~H.~Coimbra-Araujo,
Phys.\ Rev.\ D {\bf 74}, 055006 (2006)
[arXiv:hep-ph/0607027].


\bibitem{66}
S.~Creek, R.~Gregory, P.~Kanti and B.~Mistry,
Class.\ Quant.\ Grav.\ {\bf 23}, 6633 (2006)
[arXiv:hep-th/0606006].


\bibitem{67}
A.~Perez-Lorenzana,
J.\ Phys.\ Conf.\ Ser.\ {\bf 18}, 224 (2005)
[arXiv:hep-ph/0503177].

\bibitem{68}
A.~Lopez-Ortega,
Gen.\ Rel.\ Grav.\ {\bf 39}, 1011 (2007)
[arXiv:0704.2468 [gr-qc]].


\bibitem{69}
H.~K.~Kunduri, J.~Lucietti and H.~S.~Reall,
Phys.\ Rev.\ D {\bf 74}, 084021 (2006)
[arXiv:hep-th/0606076].


\bibitem{70}
  H.~T.~Cho, A.~S.~Cornell, J.~Doukas and W.~Naylor,
  Phys.\ Rev.\  D {\bf 75}, 104005 (2007)
  [arXiv:hep-th/0701193].

\bibitem{71}
  S.~B.~Giddings,
  arXiv:0709.1107 [hep-ph].

\bibitem{72}
  S.~Creek, O.~Efthimiou, P.~Kanti and K.~Tamvakis,
  arXiv:0709.0241 [hep-th].

\bibitem{73}
  S.~Chen, B.~Wang and R.~K.~Su,
  arXiv:0710.3240 [hep-th].

\bibitem{74}
  A.~Barrau, J.~Grain and C.~Weydert,
  arXiv:0710.1998 [hep-th].


\bibitem{75}
  U.~A.~al-Binni and G.~Siopsis,
  arXiv:0708.3363 [hep-th].

\bibitem{76}
  D.~Kiley,
  arXiv:0708.1016 [hep-th].

\end{thebibliography}
\end{document}